\documentclass[twocolumn]{aastex631}

\renewcommand{\thefootnote}{\arabic{footnote}}
\shorttitle{The LSST AGN Data Challenge: Selection methods}
\shortauthors{Savić et al.}

\usepackage[T1]{fontenc}
\usepackage{natbib}
\usepackage{booktabs}
\usepackage{float}
\usepackage{hyperref}
\hypersetup{colorlinks, linkcolor={red}, citecolor={blue}, urlcolor={magenta}}
\usepackage{hyphenat}
\usepackage{savesym}
\savesymbol{tablenum}
\usepackage[separate-uncertainty = true,multi-part-units=single]{siunitx}
\restoresymbol{SIX}{tablenum}\DeclareSIUnit\Jy{\ensuremath{\mathrm{Jy}}}
\DeclareSIUnit\erg{\ensuremath{\mathrm{erg}}}
\DeclareSIUnit\ev{\ensuremath{\mathrm{eV}}}
\DeclareSIUnit\pix{\ensuremath{\mathrm{pixel}}}
\DeclareSIUnit\byte{\ensuremath{\mathrm{B}}}
\usepackage{amsmath}
\usepackage{multirow}
\usepackage{footmisc}
\usepackage{xcolor}
\defcitealias{2011ApJ...733...10R}{R11}
\usepackage{textcomp}

\begin{document}

\title{The LSST AGN Data Challenge: Selection methods}

\author[0000-0003-0880-8963]{Đorđe V.\,Savić}
\affiliation{Institut d’Astrophysique et de Géophysique, Université de Liège \\ Allée du 6 Août 19c, 4000 Liège, Belgium}
\affiliation{Astronomical Observatory, Volgina 7, 11000 Belgrade, Serbia}

\author[0000-0002-0431-2357]{Isidora Jankov}
\affiliation{University of Belgrade - Faculty of Mathematics, Department of astronomy, Studentski trg 16 Belgrade, Serbia}

\author[0000-0003-1262-2897]{Weixiang Yu}
\affiliation{Drexel University, Department of Physics, 32 S. 32nd Street,
Philadelphia, PA 19104, USA}

\author[0000-0002-3078-856X]{Vincenzo Petrecca}
\affiliation{Department of Physics, University of Napoli “Federico II”, via Cinthia 9, 80126 Napoli, Italy}
\affiliation{INAF - Osservatorio Astronomico di Capodimonte, via Moiariello 16, 80131 Napoli, Italy}

\author[0000-0001-8433-550X]{Matthew J. Temple}\thanks{Fondecyt fellow}
\affiliation{Instituto de Estudios Astrof\'{\i}sicos, Universidad Diego Portales, Av. Ej\'ercito Libertador 441, Santiago, Chile}

\author[0000-0002-8577-2717]{Qingling Ni}
\affiliation{Max-Planck-Institut f\"{u}r extraterrestrische Physik (MPE), Gie{\ss}enbachstra{\ss}e 1, D-85748 Garching bei M\"unchen, Germany}

\author[0000-0002-1114-0135]{Raphael Shirley}
\affiliation{Astronomy Centre, Department of Physics and Astronomy, University of
Southampton, Southampton SO17 1BJ, UK}
\affiliation{Institute of Astronomy, University of Cambridge, Madingley Road, Cambridge CB3 0HA, UK}



\author[0000-0002-0786-7307]{Andjelka B. Kovačević}
\affiliation{University of Belgrade - Faculty of Mathematics, Department of astronomy, Studentski trg 16 Belgrade, Serbia}
\affiliation{PIFI Research Fellow, Key Laboratory for Particle Astrophysics, Institute of High Energy Physics, Chinese Academy of Sciences,19B Yuquan Road, 100049 Beijing, China}

\author[0000-0003-3105-7037]{Mladen Nikolić}
\affiliation{University of Belgrade - Faculty of Mathematics, Department of astronomy, Studentski trg 16 Belgrade, Serbia}

\author[0000-0002-1134-4015]{Dragana Ilić}
\affiliation{University of Belgrade - Faculty of Mathematics, Department of astronomy, Studentski trg 16 Belgrade, Serbia}
\affiliation{Humboldt Research Fellow, Hamburger Sternwarte, Universit{\"a}t Hamburg, Gojenbergsweg 112, 21029 Hamburg, Germany}

\author[0000-0003-2398-7664]{Luka Č.\,Popović}
\affiliation{Astronomical Observatory, Volgina 7, 11000 Belgrade, Serbia}
\affiliation{University of Belgrade - Faculty of Mathematics, Department of astronomy, Studentski trg 16 Belgrade, Serbia}

\author[0000-0003-4210-7693]{Maurizio Paolillo}
\affiliation{Department of Physics, University of Napoli “Federico II”, via Cinthia 9, 80126 Napoli, Italy}
\affiliation{INAF - Osservatorio Astronomico di Capodimonte, via Moiariello 16, 80131 Napoli, Italy}

\author[0000-0002-5854-7426]{Swayamtrupta Panda}\thanks{CNPq fellow}
\affiliation{Laborat\'orio Nacional de Astrof\'isica - MCTI, R. dos Estados Unidos, 154 - Na\c{c}\~oes, Itajub\'a - MG, 37504-364, Brazil}
\affiliation{Center for Theoretical Physics, Polish Academy of Sciences, Al. Lotnik{\'o}w 32/46, 02-668 Warsaw, Poland}

\author[0000-0003-1281-7192]{Aleksandra Ćiprijanović}
\affiliation{Fermi National Accelerator Laboratory, P.O. Box 500, Batavia, IL 60510, USA}

\author[0000-0002-1061-1804]{Gordon T.\,Richards}
\affiliation{Drexel University, Department of Physics, 32 S. 32nd Street, Philadelphia, PA 19104, USA}



\begin{abstract}
Development of the Rubin Observatory Legacy Survey of Space and Time (LSST) includes a series of Data Challenges (DC) arranged by various LSST Scientific Collaborations (SC) that are taking place during the project's preoperational phase. The AGN Science Collaboration Data Challenge (AGNSC-DC) is a partial prototype of the expected LSST AGN data, aimed at validating machine learning approaches for AGN selection and characterization in large surveys like LSST. The AGNSC-DC took part in 2021 focusing on accuracy, robustness, and scalability. The training and the blinded datasets were constructed to mimic the future LSST release catalogs using the data from the Sloan Digital Sky Survey Stripe 82 region and the XMM-Newton Large Scale Structure Survey region.  Data features were divided into astrometry, photometry, color, morphology, redshift and class label with the addition of variability features and images. We present the results of four DC submitted solutions using both classical and machine learning methods. We systematically test the performance of supervised (support vector machine, random forest, extreme gradient boosting, artificial neural network, convolutional neural network) and unsupervised (deep embedding clustering) models when applied to the problem of classifying/clustering sources as stars, galaxies or AGNs. We obtained classification accuracy \SI{97.5}{\percent} for supervised and clustering accuracy \SI{96.0}{\percent} for unsupervised models and \SI{95.0}{\percent} with a classic approach for a blinded dataset. We find that variability features significantly improve the accuracy of the trained models and correlation analysis among different bands enables a fast and inexpensive first order selection of quasar candidates. 
\end{abstract}



\section{Introduction} \label{sec:intro}
A few percent of galaxies show enhanced emission from the nucleus that typically surpasses the stellar emission from the rest of the galaxy \citep[e.g.,][]{2019ApJ...874...54M}; such sources are known as active galactic nuclei (AGNs). Emission from AGNs is produced by an accretion disk and ionized clouds surrounding a central super-massive black hole \citep{1964ApJ...140..796S,1964SPhD....9..246Z,1993ARA&A..31..473A,2015ARA&A..53..365N}. AGNs emit across the whole electromagnetic spectrum \citep{2017A&ARv..25....2P} and are readily observed at large distances due to their high luminosity, with potential to be used as probes of cosmology \citep{2019NatAs...3..272R, 2019FrASS...6...75P, 2022arXiv220906563C}.  AGNs have profound effects on the life and evolution of their entire host galaxy \citep{2000ApJ...539L...9F, 2000ApJ...543L...5G, 2013ARA&A..51..511K}. Outflows and jets interact with the local environment and release a large amount of energy capable of driving away the nearby gas, hence terminating star formation \citep{2012ARA&A..50..455F}. Moreover, AGNs also have an impact on the surrounding hot intergalactic medium and play an active role in the evolution of the host galaxy clusters \citep{2021Univ....7..142E}. Therefore, each successful detection and observation of AGNs and measuring their physical properties is crucial for many areas of modern astrophysics and cosmology. 

Ongoing and forthcoming large-scale photometric surveys (e.g.\,Zwicky Transient Facility - \citealt{2014htu..conf...27B}; Pan-STARRS - \citealt{Chambers16}; Gaia -  \citealt{2016A&A...595A...1G}) will produce catalogs for a vast number of sources which brings the astronomy to the new era of ``big data''. The Vera C.\,Rubin Observatory Legacy Survey of Space and Time (LSST) is designed to address the main challenges for probing dark energy and dark matter, solar system exploration, exploring the transient optical sky and mapping the Milky Way \citep{2017arXiv170804058L, 2019ApJ...873..111I}. With a state-of-the-art 3.2 Gigapixel flat-focal array camera mounted on an 8.4m telescope, LSST will cover the whole observable sky every $\sim$4 nights in the optical/near-infrared $ugrizy$-bands. The expected data volume of LSST is $\sim\SI{300}{\peta\byte}$ of raw data and $\sim\SI{4d10}{}$ objects after 10 years of planned survey \citep{2019ApJ...873..111I}. Every night, LSST will monitor tens of millions of AGNs over $\sim$\SI{18000}{deg\tothe{2}} area \citep{2017ApJS..228....2L,2021A&A...645A.103D}. Although the actual number of AGNs that will be detected is a subject of the optimal observing strategy \citep{2022ApJS..258....1B}, LSST  will produce an AGN sample that supersedes the largest current AGN samples by more than an order of magnitude. This present work is a preparatory step towards producing a high-purity AGN sample with LSST.

To identify AGNs within LSST, the main challenge is separating AGNs from normal galaxies and stars. Construction of LSST’s AGN census will build upon a considerable volume of past work, making use of colors, proper motion, variability and image morphologies.
The idea of performing a multi-faceted quasar selection (ie., combining information from multiple observables) has long been proposed \citep{1986PASP...98..285K}. However, the quality, quantity, and type of data of LSST will allow for a more complete AGN selection and thus these approaches should be considered from scratch.

Color selection has been widely used as the gold standard for identification of unobscured AGNs since their discovery \citep{1982A&A...105..107K,1991ApJS...76....1W,2002AJ....123.2945R}, but we expect AGN colors to change as a function of luminosity as we probe fainter towards LSST-like depths \citep[e.g.,][]{2021MNRAS.508..737T}. While an application of modern statistical techniques to color data could be used to select AGNs, we expect the addition of multi-parameter data to result in a purer and more complete AGN selection function.

As some AGNs and stars have similar colors, the fact that AGNs lack proper motions (unlike Galactic stars) has long been used as a discriminant \citep{1967ApJ...148..767S,1981PASP...93..397K}. LSST's use of astrometric data will be no different in that regard. One way that LSST will be unique is in its ability to take advantage of differential chromatic refraction (DCR) of AGN \citep{2009AJ....138...19K, 2020RNAAS...4..252Y} which makes use of the astrometric offset of an emission-line object from that expected (in the astrometric solution) for a power-law source--to break degeneracies in photometric redshifts of luminous AGNs (henceforth quasars or QSOs).

Selection of AGNs via time-series due to their variability is another well known method \citep{1979A&AS...35..391B,1989AJ.....98..108T,2011AJ....141...93B,2019A&A...627A..33D,2020A&A...634A..50P}. As quasars display higher fractional variability in their brightness than the average star and with different characteristics than the typical variable star, variability will be a cornerstone of AGN classification for LSST \citep{2021ApJ...907...96S}. For luminous quasars, it has been shown that variability combined with colors works better for selection than variability alone \citep{2015ApJ...811...95P}. Lower-luminosity AGNs are expected to have the most variable nuclei, however increased contamination from the host galaxy could compromise variability-selection methods if insufficient care is taken. The implementation of Difference Image Analysis (DIA) in the LSST reduction pipelines will completely revolutionize the detection of AGNs through variability by removing the contribution from the host galaxy \citep{2001AcA....51..317Z,2008MNRAS.386L..77B,2010ApJ...716..530K,2016ApJ...817..119K}. 
Finally, the addition of high quality resolution images is expected to considerably increase the performance of the selection methods \citep{2022A&A...666A.171D}. 

Several data challenges (DCs) have been created in the past to facilitate the preparation of LSST, by other groups and for science use cases other than the study of AGN~\citep[e.g.,][]{Sanchez2020, Hlozek2020}. In 2021, the LSST AGN Science Collaboration (AGNSC\footnote{\url{https://agn.science.lsst.org/}}) organized a DC to get more people involved in the work needed for the AGN science with the upcoming LSST data. The main goal of the LSST AGN DC was to address the problem of AGN selection.

Unlike the previous data challenges, which relied on simulated datasets, this AGN DC utilizes real observational data. Major tasks also include establishing public training/test sets that will be used as a benchmark to test different machine learning (ML) algorithms. There were five proposed solutions submitted to the DC: one using a classical approach; and four applying ML-based AGN selection. We will present the solution using classical approach and three ML-based solutions, while the one remaining ML-based solution is addressed by \citet{2022A&A...666A.171D}.

The paper is organized as follows: in Section \ref{sec:dataconstruction} we address the data retrieval and the construction of training and blinded datasets. We elaborate on applied ML methods in Section \ref{sec:ML}. The results obtained from the various methodologies employed in this DC are presented in Section \ref{sec:results}. We discuss further issues relevant to our work and summarize our findings in Section \ref{sec:summary}.


\section{Datasets construction}\label{sec:dataconstruction}
LSST will deliver three levels of data products and services: prompt data products that are computed and released within 24 hours of observation, data release data products that are computed during annual processing campaigns, and user generated data products. The Data Products Definitions Document (DPDD\footnote{\url{https://docushare.lsst.org/docushare/dsweb/Get/LSE-163}}) is the ultimate reference for descriptions of the planned LSST data products and pipelines \cite[see also][]{2019ApJ...873..111I}. 

The input data for the AGNSC-DC were modified such that the column names and units used for different measurements (e.g., flux) comply with the DPDD standards for data release catalogs (DPDD, Section 4.3). We refer to distinct astrophysical bodies that emit light detected as ``Objects'' and individual instances~(detection) of those objects as ``Sources''. Observations from a specific point in time will appear in the {\tt Source} tables in the data releases, while ``co-added'' (averaging/summing over time) information will appear in the {\tt Object} tables.  So-called ``light curves'' (brightness as a function of time) will appear in the {\tt ForcedSource} tables (with summary statistics in the {\tt Object} tables). ``Simulated'' training data will attempt to heel to this data structure as closely as possible. 

The datasets released in this data challenge are pulled from different sources (public archives) and put together to mimic the architecture of future LSST data release catalogs as much as possible, but without taking into account the expected number of objects that will appear in certain regions of LSST sky. Details on how the tables are constructed can be found under the `docs' folder in the main github repository~\citep{yu2021}. Here, we provide a brief overview of the datasets. 

\begin{figure*}[httb]
    \centering
    \includegraphics[width=0.85\hsize]{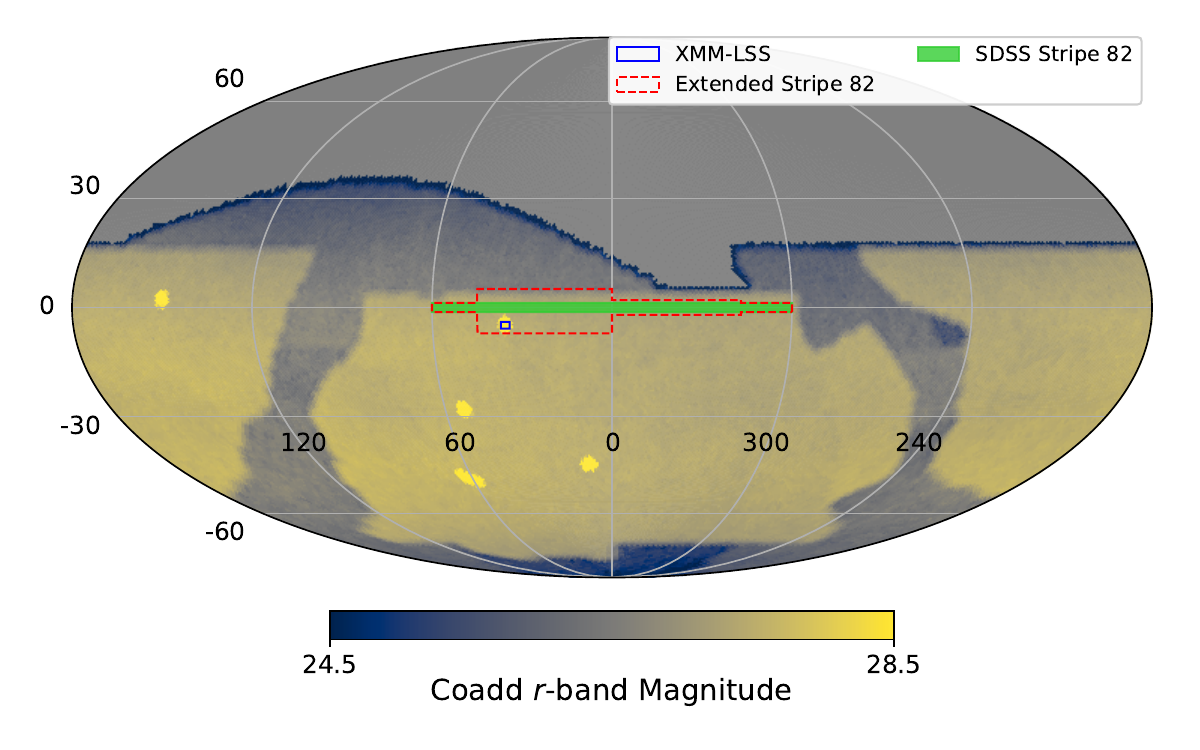}
    \caption{SDSS Stripe 82 (filled green), extended Stripe 82 (dashed red) and XMM-LSS (solid blue) areas projected on the LSST observable sky. Color map indicates the final depth of the coadds in the $r$-band. XMM-LSS region coincides completely with one of the LSST deep drilling fields (bright-yellow regions). The map was generated using the LSST simulation project \texttt{OpSim} \citep{2014SPIE.9150E..15D}.}
    \label{fig:dc_footprint}
\end{figure*}

The astronomical objects included in the release dataset are drawn from three main sources: spectroscopically identified objects in an extended Sloan Digital Sky Survey \citep [SDSS\footnote{\url{https://www.sdss.org/}};][]{2000AJ....120.1579Y} Stripe 82 area with the spectroscopy collected from the 16th data release of SDSS \citep[DR16;][]{2020ApJS..249....3A}, X-ray detected and classified objects in the the XMM-LSS\footnote{\url{https://www.cosmos.esa.int/web/xmm-newton}} \citep{2007MNRAS.382..279P} region, and unidentified variable objects in the original SDSS Stripe 82 area\footnote{\url{http://faculty.washington.edu/ivezic/sdss/catalogs/S82variables.html}} \citep{2007AJ....134..973I}. Fig.\,\ref{fig:dc_footprint} illustrates the source survey region footprint on the LSST sky for the baseline observing strategy. The XMM-LSS area is encompassed by one of the LSST deep drilling fields. 

\begin{figure*}[httb]
    \centering
    \includegraphics[width=0.95\hsize]{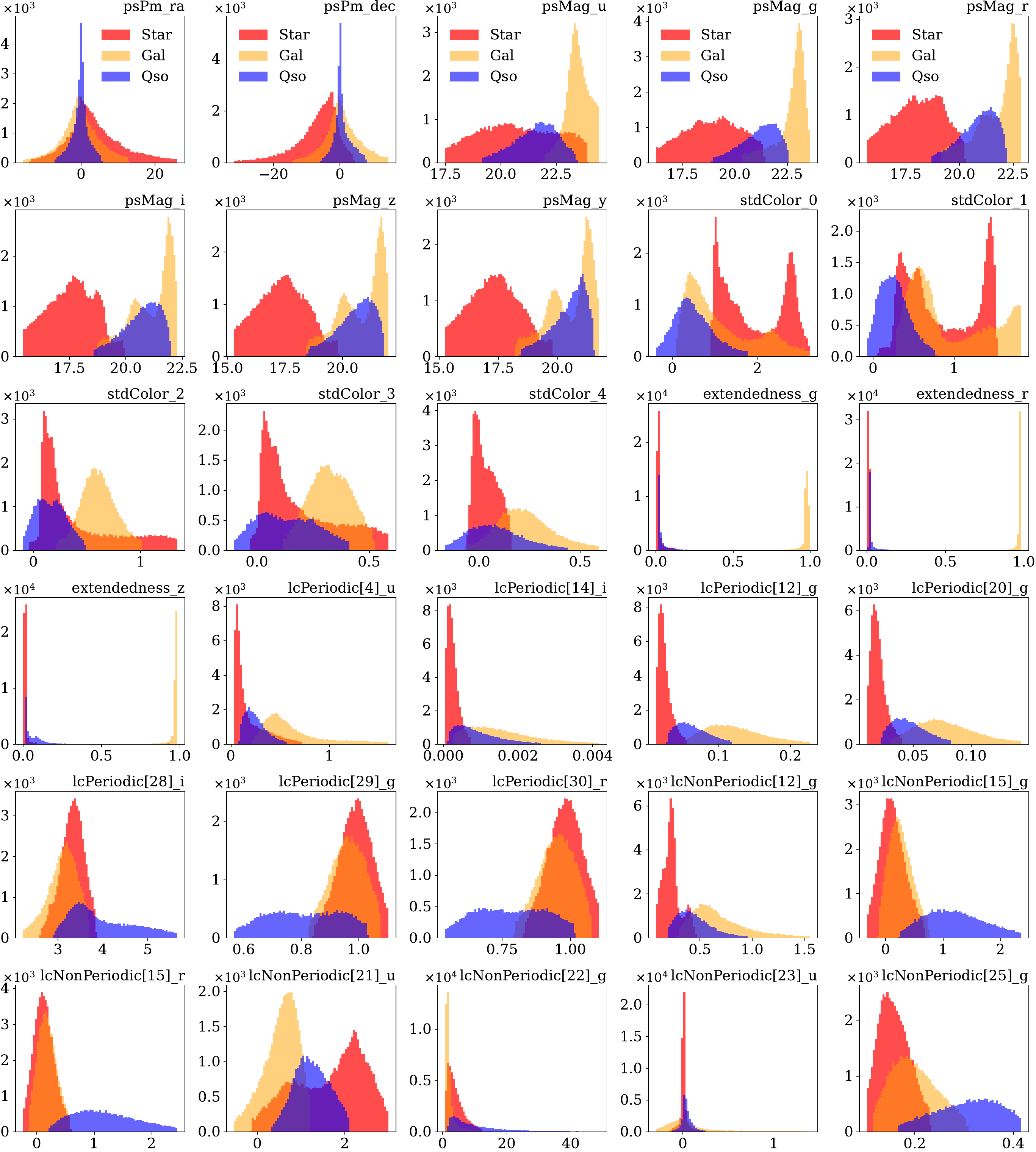}
    \caption{Distributions of 30 features drawn from astrometry(proper motion \texttt{psPm\_ra} and \texttt{psPm\_dec}); photometry (point-source magnitudes \texttt{psMag\_u/g/r/i/z/y} and colors \texttt{stdColor\_0/1/2/3/4}); morphology (extendedness \texttt{estendedness\_g/r/z}); LC features (\texttt{lcPeriodic[12/20/29]\_g}, \texttt{lcPeriodic[14/28]\_i}, \texttt{lcPeriodic[30]\_r}, \texttt{lcNonPeriodic[12/15/22/25]\_g}, \texttt{lcNonPeriodic[4/21/23]\_u}, \texttt{lcNonPeriodic[15]\_r}). Color-coded per class: star (red), galaxy (orange), quasar (blue).}
    \label{fig:attribs}
\end{figure*}

The total number of objects (both combined) in the \texttt{Object} table is $\sim$\SI{440000}{}, after removing $\sim$\SI{5000}{} duplicates found in more than one sources from above.
The total number of epochs in the \texttt{ForcedSource} table is $\sim$5 million. The total number of features (parameters) in the object table is 374. Features are divided into main categories with number of features in each category indicated in the parentheses:
\begin{enumerate}
    \item \texttt{Astrometry(5)}: ra, dec, proper motion and parallax.
    \item \texttt{Photometry(48)}: point and extended source photometry, in both AB magnitdues and fluxes (\SI{}{\nano\Jy}).
    \item \texttt{Color(10)}: derived from the flux ratio between different photometric bands.
    \item \texttt{Morphology(6)}: a real-value quantity between 0 and 1. Values closer to 1 for extended sources while values closer to 0 indicate point-like sources.
    \item \texttt{Light Curve Features(302)}: extracted on the SDSS light curves if available.
    \item \texttt{Redshift(2)}: both spectroscopic and photometric, wherever available.
    \item \texttt{Class Labels(1)}: Star/Gal/Qso (Agn, highZQso), wherever available.
\end{enumerate}
Distributions of a subset of 30 features are shown in Fig.\,\ref{fig:attribs}. It is notable that stars, galaxies and quasars overlap in the feature space, but they may be separated by combining a selection of features containing the most information about interclass difference. For example, galaxies occupy different parts for morphology features when compared to stars and quasars, since the latter two are observed as point sources (Fig.\,\ref{fig:attribs}, 4th row middle and right panels); and similarly for quasars when compared to stars and galaxies for variability features (Fig.\,\ref{fig:attribs}, bottom row from left to right). 

Astrometry measurements were obtained by matching the main catalogs (SDSS Stripe 82 and XMM-LSS) with Gaia EDR3\footnote{\url{https://www.cosmos.esa.int/web/gaia/early-data-release-3}} \citep[Early Data Release 3][]{2016A&A...595A...1G,2021A&A...649A...1G} and the NOIRLab Source Catalog (NSC) data release 2 \citep[DR2\footnote{\url{https://datalab.noirlab.edu/nscdr2/index.php}};][] {2021AJ....161..192N}. Sources in NSC are extracted from reprocessed public images drawn from the NOIRLab Astro Data Archive\footnote{\url{https://astroarchive.noirlab.edu}}. The astrometry for NSC DR2 is calibrated using Gaia DR2~\citep{2018A&A...616A...1G}; its proper motion measurement achieves an accuracy of ~0.2~$\mathrm{mas~yr^{-1}}$ and a precision of ~2.5~$\mathrm{mas~yr^{-1}}$ relative to Gaia DR2~\citep{2021AJ....161..192N}.
For objects with astrometry measurements found in both catalogs, we used the values from Gaia.

The photometry were assembled following a mix-and-match approach. 
In the extended Stripe 82 region, we cross-matched our sources against the Dark Energy Survey~\citep[DES;][]{DES2016} Data Release 2~\citep[DR2;][]{descollaboration2021} photometry catalog, the SDSS Stripe 82 coadded photometry catalog~\citep{annis2014}, and the SDSS DR16 single-epoch photometry catalog~\citep{ahumada2020}. 
DES provides photometry in $grizY$ bands and SDSS provides photometry in $ugriz$ bands. 
When a source is matched with photometry in the same band from catalogs, we choose the photometry following the precedence of: DES DR2 > SDSS Stripe 82 coadd > SDSS DR16. In the XMM-LSS region, the $griz$ photometry were collected from the HSC-VISTA joint catalog (see Section~\ref{subsec:hsc_vista}) and the $u$-band photometry comes from the Canada-France-Hawaii Telescope Legacy Survey (CFHTLS) catalog~\citep{gwyn2012}.

The broad-band colors were first derived using the flux ratios of two adjacent bands and then converted into magnitudes. The error on the color was computed following the standard uncertainty propagation method. 
For the photometry taken from SDSS, the morphology (i.e., \texttt{extendedness} in LSST's nomenclature) is defined as 1 - \texttt{probPSF}. For DES photometry, we define \texttt{extendedness} as 1 - \texttt{class\_star}. Both \texttt{probPSF} and \texttt{class\_star} describe how close the photometry is to a true point source. For the photometry taken from the HSC-VISTA catalog, the LSST pipeline directly outputs the \texttt{extendedness} column.

\begin{figure}[httb]
    \centering
    \includegraphics[width=0.99\hsize]{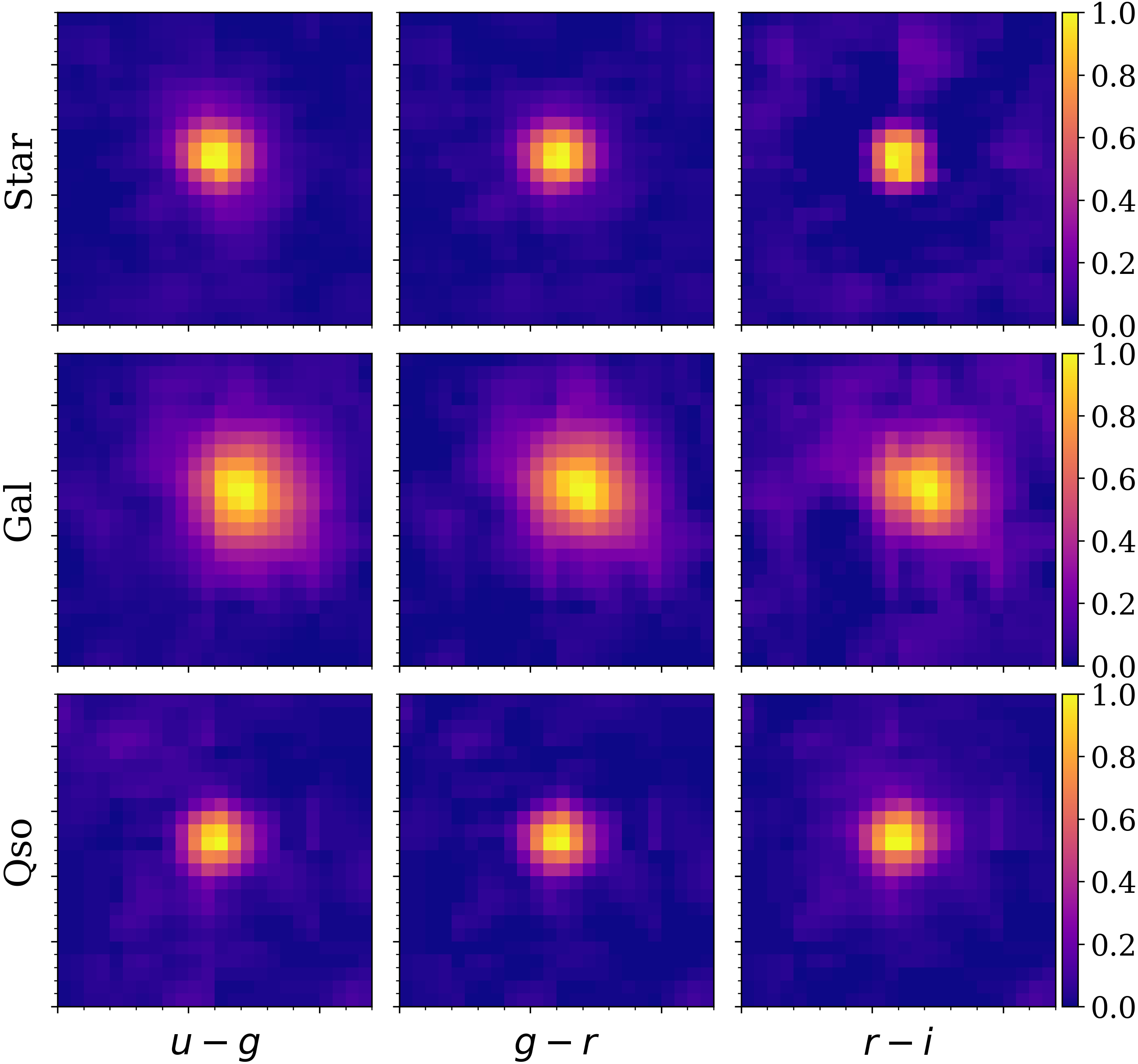}
    \caption{Image cutouts for a randomly selected star (top), galaxy (middle) and quasar (bottom). From left to right are colors $u-g$, $g-r$ and $r-i$. Images are normalized to unity for each color.}
    \label{fig:sgq_inner}
\end{figure}

All objects, except for $\sim$130 of them, in the \texttt{Object} table have pre-generated image thumbnails/cutouts of size 64x64 $\mathrm{pixel}$ from SDSS DR16. The objects without image cutouts were not removed from the main sample. An example of image cutouts for a randomly selected star, galaxy and quasar is shown in Fig.\,\ref{fig:sgq_inner}.

We list the percentage per class label and the assigning method:
\begin{itemize}
    \item \texttt{Star}(\SI{24.5}{\percent}): Spectroscopically confirmed stars (both variable and non-variable)
    \item \texttt{Gal}(\SI{55.9}{\percent}): Spectroscopically confirmed galaxies
    \item \texttt{Qso}(\SI{19.0}{\percent}): Spectroscopically confirmed AGNs or quasars
    \item \texttt{highZQso}(\SI{0.3}{\percent}): A separate catalog of high redshift (z > 4.5) quasars
    \item \texttt{Agn}(\SI{0.3}{\percent}): X-ray classified AGNs in the XMM-LSS region and spectroscopically classified galaxies having emission properties consistent with being a Seyfert or LINER in the extended Stripe 82 region.
\end{itemize}

In addition to the described datasets that are publicly available, a distinct \textit{blinded} dataset, that is $\sim$\SI{10}{\percent} of the training dataset, was constructed and set aside in order to evaluate the performance of the ML and non-ML methods submitted by the participants of the DC. The total object count for a blinded dataset is $\sim$\SI{45100}{}, among which, $\sim$\SI{1000}{} come from the XMM-LSS region, $\sim$\SI{44000}{} come from the Stripe 82 region, and \SI{100}{} come from the separate high-z quasar catalog. Class labels and spectroscopic redshift were removed for those objects in the blinded dataset. Around \SI{21000}{} objects have pre-computed variability features.

We stress important caveats:
\begin{enumerate}
    \item{About $\sim${13,000} objects have no labels, and they come from the SDSS Stripe 82 unidentified variables source catalog~\citep{2007AJ....134..973I}. About $\sim${2500} objects from the XMM-LSS are classified using X-ray data and infrared photometry~\citep{2018MNRAS.478.2132C}, which also have no spectroscopic redshift available. }
    \item Approximately $\sim$\SI{1}{\percent} of the objects do not have optical counterpart i.e., they are bright in X-ray, but are too faint to be detected in the optical band.
    \item CARMA(1,0) and CARMA(2,1) fits are presented as is. 
    Potential bad fits (e.g., perhaps due to limited temporal sampling and/or poor S/N of the photometry) were not removed\footnote{A robust goodness-of-fit metric for CARMA models is not available. The definition for a bad fit can also change given the problem at hand.}
\end{enumerate}

In the following two subsections, we describe how multi-wavelength datasets; high-redshift quasar catalog and how the light curve features are computed.

\subsection{Multi-wavelength data}
Multi-wavelength tables are provided along with the \texttt{Object} table. The multi-wavelength data are obtained by performing positional cross-match between our source positions and other catalogs. A summary of class label distribution per observing mission is listed in Table \ref{tab:stripe82} and \ref{tab:xmm}.
\begin{table}
    \centering
    \begin{tabular}{l|llll}
    Class label              & \texttt{Agn}/\texttt{Qso}   & \texttt{highZQso}   & \texttt{Gal}       &  \texttt{Star}      \\
    \hline
    Stripe 82     & \SI{73000}{}      & \SI{90}{}        & \SI{213000}{}     & \SI{106000}{}      \\
    GALEX         & \SI{36000}{}      & \SI{10}{}        & \SI{45000}{}      & \SI{38000}{}       \\
    UKIDSS        & \SI{36000}{}      & \SI{30}{}        & \SI{87000}{}      & \SI{92000}{}       \\
    Spitzer       & \SI{12000}{}      & \SI{20}{}        & \SI{27000}{}      & \SI{30000}{}       \\
    Herschel      & \SI{2500}{}     &   \SI{10}{}         & \SI{39000}{}      & \SI{12000}{}       \\
    FIRST         & \SI{2000}{}       & \SI{50}{}        & \SI{43000}{}      & \SI{250000}{}       \\
    \end{tabular}
    \caption{The distribution of class labels per catalog. From top to bottom: Stripe 82; GALEX; UKIDSS; Spitzer; Herschel and FIRST.}\label{tab:stripe82}
\end{table}

\begin{table}
    \begin{tabular}{l|lll}
    Class label   & \texttt{Agn}/\texttt{Qso}   & \texttt{Gal}       &  \texttt{Star}      \\
    \hline
    XMM-Newton    & \SI{4000}{}       & N/A     & N/A          \\
    GALEX         & \SI{400}{}     & \SI{300}{}     & \SI{100}{}         \\
    VISTA/VIDEO   & \SI{3500}{}     & \SI{3500}{}     & \SI{370}{}        \\
    Spitzer       & \SI{3500}{}     & \SI{3500}{}     & \SI{350}{}        \\
    Herschel      & \SI{1500}{}     & \SI{1500}{}     & \SI{50}{}        \\
    \hline
    \end{tabular}
    \caption{The distribution of class labels per catalog for objects found in XMM-LSS dataset.
    }
    \label{tab:xmm}
\end{table}

\subsubsection{XMM-LSS}
In the X-ray band of the XMM-LSS field, a number of XMM-Newton surveys of different sensitivities have been collected \citep[e.g.\ Fig.\,3 of][also Table 2 of \citealp{2018MNRAS.478.2132C}]{2015A&ARv..23....1B}. The X-ray source catalog from \citet{2018MNRAS.478.2132C} is adopted for our dataset, which makes use of XMM-Newton observations taken from 2000 to 2017 in the XMM-LSS\footnote{\url{https://personal.psu.edu/wnb3/xmmservs/xmmservs.html}} field, including the \SI{1.3}{\mega\second} new endeavour from the XMM-SERVS survey \citep{2018MNRAS.478.2132C,2021ApJS..256...21N}. With the XMM-SERVS survey, a flux limit of \SI{6.5d-15}{\erg\centi\meter\tothe{-2}\per\second} over \SI{90}{\percent} of the XMM-LSS area is achieved in the \SI{0.5}{}--\SI{10}{\kilo\ev} band. We only include X-ray sources that are classified as AGNs in the dataset \citep[Section 6 of]{2018MNRAS.478.2132C} for the source classification details). The X-ray AGNs are matched to other optical/IR catalogs with likelihood-ratio matching methods as described in Section 4 of \citet{2018MNRAS.478.2132C}.

\subsubsection{HSC and VISTA joint catalogue}~\label{subsec:hsc_vista}
We have provided a jointly processed optical and near-infrared dataset from the HSC\footnote{\url{https://www.naoj.org/Projects/HSC/}} (Hyper Suprime-Cam) Public Data Release 2~\citep{HSC_DR2} deep and ultradeep regions, and the VISTA\footnote{\url{https://www.eso.org/public/teles-instr/paranal-observatory/surveytelescopes/vista/}} (Visible and Infrared Survey Telescope for Astronomy) VIDEO \citep{Jarvis_2012} surveys. The dataset was produced using the LSST Science Pipelines as described in \cite{2018PASJ...70S...5B}. An object detected in any one of the ten bands across these two surveys is measured in every band ensuring that each object will have a measurement in each band. 
This dataset is a prototype developed in preparation for the upcoming LSST data.

\subsubsection{UKIDSS}
In the near-infrared \textit{YJHK} bands, we include 2\,arcsec diameter aperture magnitudes (\texttt{AperMag3}) where available from the UKIDSS DR11plus \citep{2006MNRAS.367..454H, Lawrence07, 2008MNRAS.384..637H, 2009MNRAS.394..675H}. 
SDSS Stripe 82 is partially covered by the UKIDSS Large Area Survey (LAS) in the \textit{YJHK} bands to approximate depths of 20.2, 19.6, 18.8, and 18.2 respectively. The UKIDSS-LAS covers the original Stripe 82 footprint (shown in green in Fig.~\ref{fig:dc_footprint}) but not the full extended area used in this challenge.
We remove duplicate detections from overlapping tiles using \texttt{PriOrSec}, and remove noise and saturated detections using \texttt{mergedClass}.
\subsubsection{Herschel forced photometry}
Far infrared measurements from the \emph{Herschel}\footnote{\url{https://www.herschel.caltech.edu/}} Space Observatory come from HELP: The \emph{Herschel} Extragalactic Legacy Project \citep{10.1093/mnras/stab1526}. This dataset was produced by taking a prior list of Spitzer IRAC detections and providing full Bayesian probability posterior distributions on the object fluxes to account for blending in the low resolution far infrared maps at 250, 350, and 500 microns. 

\subsection{High-redshift quasars}
The catalog of high-redshift known quasars is constructed by collecting all quasars at $z \ge 4.5$ known before October 2020. These quasars are mainly selected using the optical/near-infrared colors, based on the wide-field optical and infrared photometric surveys, e.g., SDSS \citep[][]{York00}, the Pan-STARRS1 survey \citep[PS1,][]{Chambers16}, the DESI Legacy Imaging Surveys \citep{Dey19}, the Hyper Suprime-Cam Subaru Strategic Program survey \citep{Aihara18}, the UKIRT Hemisphere Survey \citep{Dye18}, the UKIDSS-LAS \citep[][]{Lawrence07}, the VISTA Hemisphere Survey \citep[][]{Mcmahon13}, and the Wide-field Infrared Survey Explorer \citep[{\it WISE},][]{Wright10}.

About half of the $z<6$ quasars are from SDSS quasar catalogs and the rest are from several major quasar surveys \citep[e.g.,][]{McGreer13, Banados16, Wang16, Yang19a}. The $z>6$ quasars were mostly collected from quasar surveys like the SDSS high-redshift quasar survey \citep[e.g.,][]{Fan06, Jiang16}, the Canada–France High-$z$ Quasar Survey \citep[e.g.,][]{Willott10}, the PS1 distant quasar survey \citep[e.g.,][]{Banados16, Venemans15, Mazzucchelli17}, the Subaru High-$z$ Exploration of Low-Luminosity Quasars project \citep[e.g.,][]{Matsuoka18}, the DES quasar survey \citep[e.g.,][]{Reed17}, and the reionization-era quasar survey \citep[e.g.,][]{Wang19, Yang19b}. Quasars included in this catalog are all identified through spectroscopic observations. Their redshifts are mainly from the quasar broad emission lines in the rest-frame UV (e.g., Ly$\alpha$, Si\,{\sc iv}, C\,{\sc iv}, and Mg\,{\sc ii}), which result in a redshift uncertainty up to $\sim 0.05$. A small number of them have [C\,{\sc ii}]-based redshifts, with a typical uncertainty of $<$0.001.

\subsection{Light curve features}
\begin{figure}[httb]
  \centering
  \begin{tabular}{c}
    \includegraphics[width=0.99\hsize]{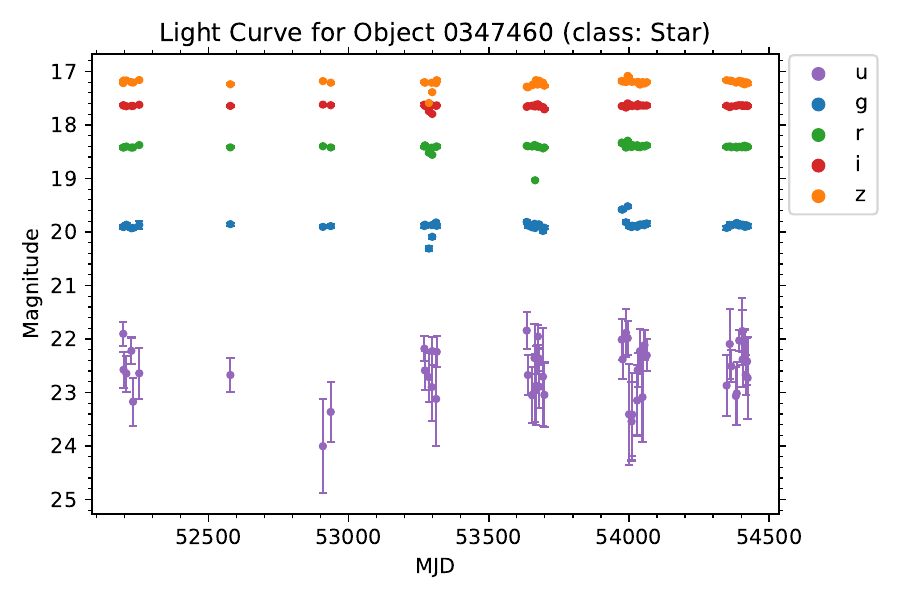} \\ \includegraphics[width=0.99\hsize]{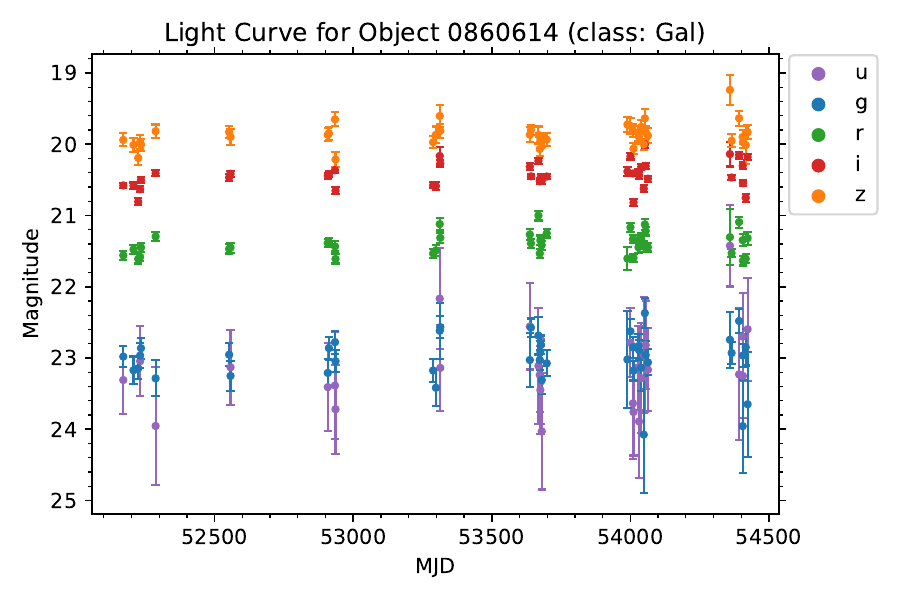} \\
    \includegraphics[width=0.99\hsize]{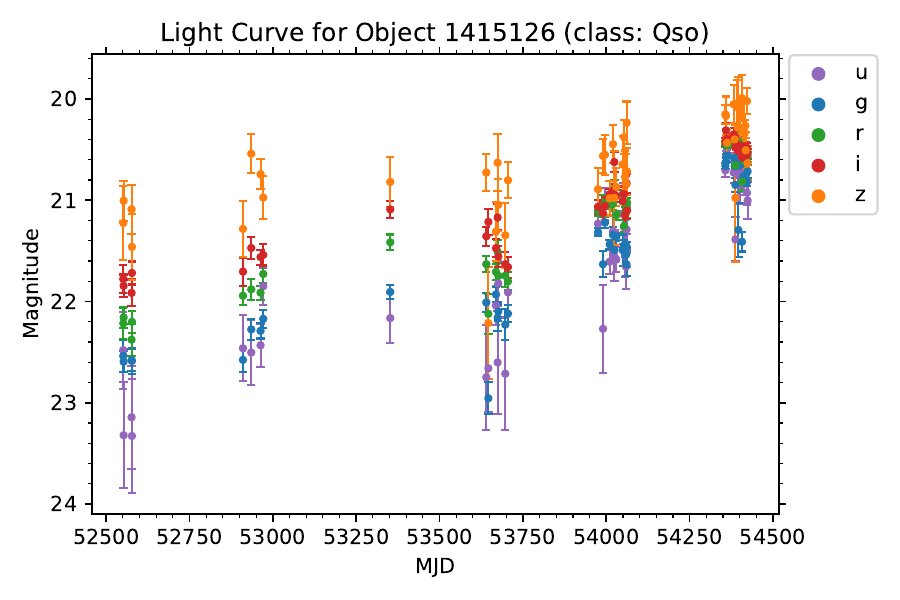} \\
  \end{tabular}
  \caption{Example LCs for the SDSS $ugriz$-bands: $u$-band (purple), $g$-band (blue), $r$-band (green), $i$-band (red) and $z$-band (orange) for a star (top panel), galaxy (middle panel) and QSO (bottom panel). Time units are given in modified Julian date (MJD).}
  \label{fig:ts} 
\end{figure}

The light curve (LC) features are computed for the sources that have corresponding time-domain data from SDSS. These features populate the \texttt{lcPeriodic} and \texttt{lcNonPeriodic} columns. The majority of the features computed are described in \citet{2011ApJ...733...10R}, hereafter \citetalias{2011ApJ...733...10R}. Some additional features computed include those introduced by the Feature Analysis for Time Series (\textsc{FATS}) project\footnote{\url{https://isadoranun.github.io/tsfeat/FeaturesDocumentation.html}} and best-fit CARMA(1,0) (continuous-time auto-regressive moving average model, otherwise known as a damped random walk; DRW) and CARMA(2,1) (otherwise known as a damped harmonic oscillabor; DHO) parameters \citep{2009ApJ...698..895K, kasliwal2017, moreno2019, Yu2022b} obtained using \textsc{EzTao}\footnote{\url{https://github.com/ywx649999311/EzTao}}\citep{Yu2022a}. Both the \citetalias{2011ApJ...733...10R} and FATS features are computed using the \textsc{cesium}\footnote{\url{https://cesium-ml.org/}} software package. An example of LC for a random star, galaxy and QSO for the SDSS $ugriz$-bands are shown in Fig.\,{\ref{fig:ts}}.

\begin{figure}[httb]
    \centering
    \includegraphics[width=0.99\hsize]{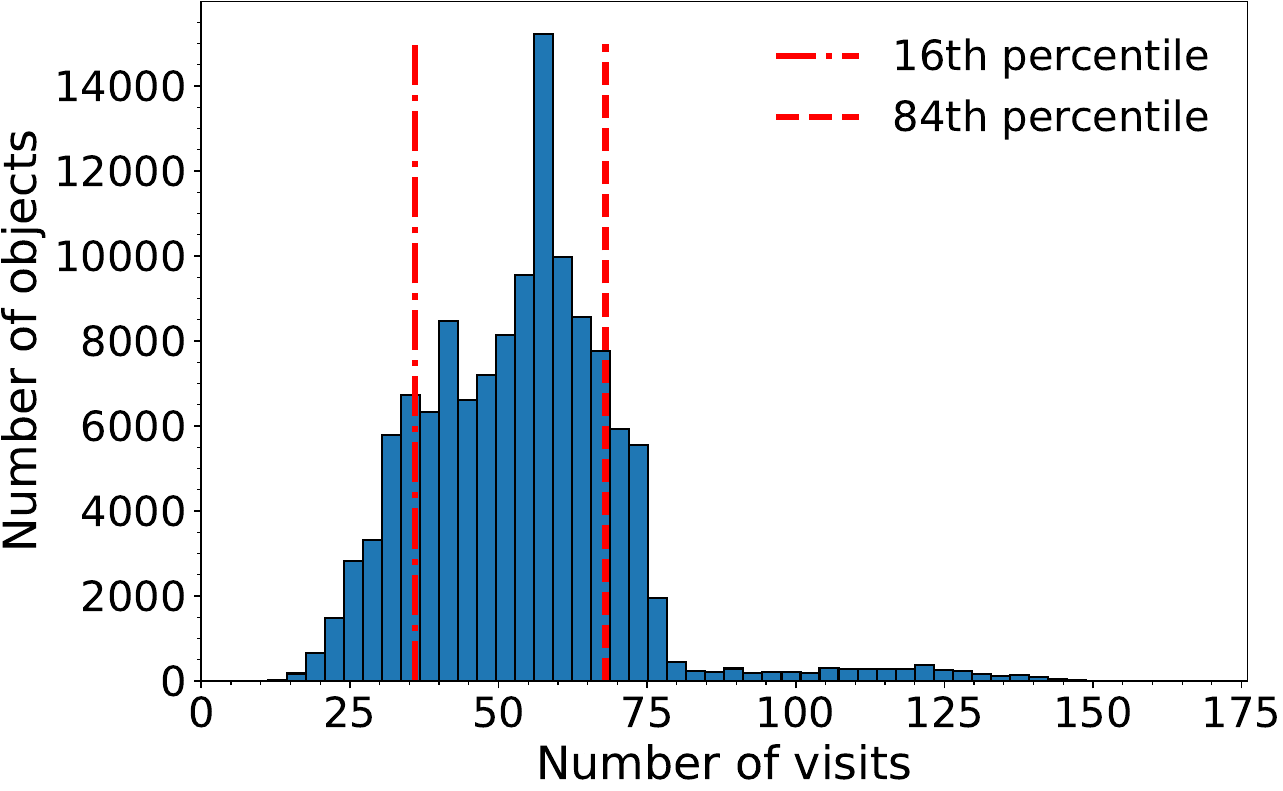}
    \caption{Distribution of the number of visits in the \texttt{ForcedSource} table for objects with known class label. Vertical red lines mark the 16th and 84th percentile values.}
    \label{fig:nvisits}
\end{figure}

We note that a 5-$\sigma$ clipping (in magnitude) was applied before the variability metrics were computed, this decision is mostly driven by the `spurious dimming' of SDSS light curves as discussed by \citet{2010ApJ...714.1194S}. The total count of time domain objects in Stripe 82 is $\sim$\SI{210000}{}. Each SDSS filter, for which LC features have been computed, has the same number of visits and the same cadence. These numbers vary from object to object. The distribution of the number of visits per object is shown in Fig.\,\ref{fig:nvisits}. The bulk of objects have the number of visits between 30 and 70. For comparison, LSST sources will have a larger number of observations \citep{2022ApJS..258....1B,2022arXiv220806203K,2022ApJS..258....3R,2023MNRAS.522.2002P, 2023arXiv230108975C}.

Table~\ref{tab:lc} summarizes the LC features and notation. A full description of the LC features is also publicly available\footnote{\url{https://github.com/RichardsGroup/AGN_DataChallenge/blob/main/docs/04_LC_features.ipynb}}.
\begin{table*}[h]
\centering
\begin{tabular}{lp{13cm}} 
 \toprule
 \toprule
 LC feature & Description \\ 
 \midrule
 \midrule
 \texttt{lcPeriodic 0-3} &  Four best-fit \texttt{CARMA(2,1)/DHO} parameters, fitted in flux
    using a 3-point median filter with a 5-$\sigma$ clipping for outliers removing.
    Fitted data are for $g$-, $r$- and $i$-band light curves only. Fitted light curves have more 30 epochs. \\
 \texttt{lcPeriodic 4-32} & Generalized Lomb Scargle fit/parameters as described in the \citetalias{2011ApJ...733...10R} paper. The exact matching from index to features is given below. Note that first relative phase (\texttt{rel\textunderscore phase\textunderscore0}) are not included since it is negligible. \\
 \bottomrule
 \texttt{lcNonPeriodic 0-20} & Non-periodic features introduced in the \citetalias{2011ApJ...733...10R} paper.\\ 
 \texttt{lcNonPeriodic 21} & Variance divided by the median. \\
 \texttt{lcNonPeriodic 22} & Reduced $\chi^2$ for a constant model with a given degrees of freedom:
 
 \vbox{\begin{equation*} \chi^2/\mathrm{d.o.f.} = \frac{1}{N-1}\sum_{i=1}^{N}\left(\frac{m_i - \bar{m}}{\sigma_i}\right)^2, \end{equation*}} where $\bar{m}$ is the inverse variance weighted average. \\
 \texttt{lcNonPeriodic 23} & Excess variance defined by \citet{2018PASP..130e4501B}:\hfill
 
        \hfill\vbox{\begin{align*} 
        \sigma^2_\text{sys} & \equiv \left\langle\Delta m_{i}^{2}-\sigma_{\text {stat }, i}^{2}\right\rangle \\
            \Delta m_{i} & \equiv \frac{m_{i}-\bar{m}}{\sqrt{1-w_{i} / \sum w_{j}}}, \\
                   w_{i} & \equiv \sigma_{\text{stat}, i}^{-2}, \\
                 \bar{m} & \equiv \frac{\sum w_{j} m_{j}}{\sum w_{j}},
        \end{align*}} 
        where $\sigma_{\text{sys}}$ is the excess variance and  $\sigma_{\text{stat}, i}$ is the photometric uncertainty.\\
\texttt{lcNonPeriodic 24} & Normalized excess variance \citep{2013ApJ...771....9A}:\hfill

            \vbox{\begin{equation*}
              \sigma_{\text{sys, norm }}^{2} \equiv \frac{\sigma_{\text{sys}}^{2}}{N \bar{m}^{2}}.
                  \end{equation*}}
            \\
\texttt{lcNonPeriodic 25} & Range of a cumulative sum \citep{2011ApJ...735...68K}:\hfill

            \vbox{\begin{align*}
              R_{\mathrm{CS}} &=\max (S)-\min (S), \\
                        S_{l} &=\frac{1}{N \sigma} \sum_{i=1}^{l}\left(m_{i}-\bar{m}\right).
                  \end{align*}}
            \\
\texttt{lcNonPeriodic 26} &  The von Neumann ratio:\hfill

            \vbox{\begin{equation*}
                  \eta=\frac{1}{(N-1) \sigma^{2}} \sum_{i=1}^{N-1}\left(m_{i+1}-m_{i}\right)^{2}.
                      \end{equation*}}
            \\
\texttt{lcNonPeriodic 27-28} & The best fit CARMA(1,0)/DRW parameters. Light curves with less than 10 epochs were not fitted. 
            \\
\texttt{lcNonPeriodic 27} & The driving amplitude $\sigma$, or $\beta_{0}$ in the CARMA notation \citep{2009ApJ...698..895K}. 
            \\
\texttt{lcNonPeriodic 28} & The characteristic timescale described by the DRW model, or $1/\alpha_{1}$ in the CARMA notation.
            \\
\bottomrule
\end{tabular}
\caption{LC features divided in two groups consisting of 33 periodic and 29 non-periodic features. Feature numeration starts with 0.}
\label{tab:lc}
\end{table*}

\section{Machine learning methods}\label{sec:ML}
The classification of sources from wide-field surveys is one of the most fundamental problems in astronomy. Efficient classification will be difficult if using classical techniques for manual inspection and it is expected that ML applications will be helpful in automating the process. When trained on big astronomical data, ML methods tend to outperform traditional methods based on explicit programming \citep[e.g.,][]{2010MNRAS.406..342B,2016ApJS..225...31L,2019arXiv190407248B}.

Broadly speaking, ML methods can be divided into supervised and unsupervised methods \citep{10.5555/3379017}. Supervised Learning is a ML paradigm that relies on labeled data for acquiring the input-output relationship information for classification and regression problems. The supervised ML methods we used are: support vector machine \citep[SVM,][]{cortes1995support}, random forest \citep[RF,][]{ho1995random}, extreme gradient boosting \citep[XGB,][]{2016arXiv160302754C} and artificial neural networks \citep[ANN,][]{Cybenko1989ApproximationBS}. These methods have been widely applied to numerous classification problems in astronomy \citep[e.g.][]{2019arXiv190407248B} and for AGN classification \citep{2008MNRAS.391..369C,2014MNRAS.437..968C,2014ApJ...782...41D,2021MNRAS.501.3951C,2021A&A...645A.103D,2021A&A...651A.108P,2021ApJ...920...68C}. For more details on each of supervised ML method we used, we refer to \citet{10.5555/3379017}.

Unsupervised methods discover hidden patterns in the data without the need for human intervention and are mostly used for clustering and dimensionality reduction. Commonly used are: $k$-means \citep{1056489}, Gaussian mixture of models \citep{Reynolds2009} for clustering; and principal component analysis \citep[PCA,][]{1986pca..book.....J}, autoencoders \citep{Kramer1991NonlinearPC}, t-distributed stochastic neighbor embedding \citep[t-SNE,][]{vanDerMaaten2008} for dimensionality reduction. 

Deep learning (DL) is another broader family of ML methods that relies exclusively on the use of ANNs with a large number of hidden layers, hence \textit{deep}. The abundance of imaging data in astronomy has naturally led to the application of deep convolutional neural networks \citep[CNNs,][]{2015Natur.521..436L}. One advantage of DL is that it allows for simultaneous training of models with multiple inputs (e.g., catalog features + images) and for multiple outputs \citep[e.g., regression and classification,][]{chollet2015keras}. An important property of DL is the application of transfer learning \citep{2018arXiv180801974T} i.e., a model trained for one task is re-purposed on a second related task \citep{2022A&A...666A.171D}. For the application of DL in astronomy, we refer to a review by \citet{2022arXiv221103796S}.

Many ML methods are developed as a combination of the above mentioned methods. Therefore, it is common to train a dimensionality reduction model and then train supervised or unsupervised models using the latent features. We used one unsupervised model built in this manner: deep embedding clustering \citep[DEC;][]{2015arXiv151106335X,guo2017improved}, which consists of an ANN based on autoencoder for dimensionality reduction and the preservation of the latent space, followed by clustering using latent features. Additionally, latent features are used to for visualizing complex multidimensional data space in 2 or 3 dimensions \citep{2020A&A...639A..84C,2021POBeo.100..241J}, which is often the very beginning of the ML experiment setup.



\section{Results}\label{sec:results}
In this section, we summarize four out of the five ``solutions'' submitted to the DC (since one using CNNs and transfer learning has been already been presented by \citealt{2022A&A...666A.171D}). One submission used a non-ML method extending the traditional approach based on color-color diagrams by adding magnitude standard deviation and the coefficients of correlation between light curves in two wavebands. A second submission trained supervised models: XGB, RF, SVM, ANN and one unsupervised model (DEC) for the Star/Gal/QSO classification and clustering respectively by adding LC features. The third submission trained CNNs that use projected time series data onto 2D images for Star/Gal/QSO classification. A final submission trained a RF on a subsample consisting of stars and quasars only, when all data features, except for flags and redshift, are taken into account. A summary of the contributions, ML methods, train and test sample sizes and dimensionality {\ bf(number of features)} and model performance is listed in Table \ref{tab:summary}, indicating which of the co-authors submitted the solution. For a measure of performance, we report accuracy and completeness. 

In the following subsections, we elaborate in detail each of the submitted solution while following the main workflow: feature selection; preprocessing (if needed); model training and evaluating on test and blinded datasets. The uncertainties on the performance were estimated with k-fold cross-validation \citep{stone1974}. We split the training dataset into $k=10$ subsumples. One of the groups is used for testing while the rest are used for training. This process is repeated $k$ times, with each group being used once for testing. The evaluation results are then averaged to give an overall train performance. During the cross-validation process, after every training iteration, we also evaluate the performance on the blinded dataset that is always kept aside.

\begin{table*}[httb]
\centering
\resizebox{\linewidth}{!}{%
\begin{tabular}{lrcllllcccc} 
 \toprule
 \toprule
 \multirow{3}{*}{Contribution} & \multicolumn{2}{r}{\multirow{3}{*}{ML method}} & \multirow{3}{*}{Data type} & \multicolumn{2}{c}{\multirow{2}{*}{Sample size}} & \multicolumn{5}{c}{\hspace{20pt} Performance} \\
 & & & & & & & \multicolumn{2}{c}{Accuracy} & \multicolumn{2}{c}{Completeness} \\
 & & & & Train & Blinded & Dim. & Train & Blinded & Train & Blinded\\
 \midrule
 \midrule
 V.P. and M.P. & \multicolumn{2}{r}{WEDGE+EXT+COL} & tabular & $\SI{10000}{}$ & $\SI{6000}{}$ & \SI{10}{} & \SI{0.932(1)}{} & \SI{0.950(2)}{} & \SI{0.818(3)}{} & \SI{0.831(4)}{}\\
 \midrule
 
 \multicolumn{2}{l}{\multirow{7}{*}{Đ.S., I.J. and SER-SAG}}
 & XGB & tabular & $\SI{380000}{}$ & $\SI{44000}{}$ & \SI{64}{} & \SI{0.980(1)}{} & \SI{0.970(1)}{} & \SI{0.894(2)}{} & \SI{0.834(1)}{}\\
 & & XGB & tabular & $\SI{128000}{}$ & $\SI{15000}{}$ & \SI{64}{} & \SI{0.983(4)}{} & \SI{0.978(3)}{} & \SI{0.929(3)}{} & \SI{0.882(3)}{}\\
 & & RF  & tabular & $\SI{128000}{}$ & $\SI{15000}{}$ & \SI{64}{}  & \SI{0.982(4)}{} & \SI{0.976(2)}{} & \SI{0.920(4)}{} & \SI{0.866(2)}{}\\
 & & SVM & tabular & $\SI{128000}{}$ & $\SI{15000}{}$ & \SI{64}{}  & \SI{0.982(5)}{} & \SI{0.976(3)}{} & \SI{0.919(5)}{} & \SI{0.870(2)}{}\\
 & & ANN & tabular & $\SI{128000}{}$ & $\SI{15000}{}$ & \SI{64}{}  & \SI{0.982(4)}{} & \SI{0.975(4)}{} & \SI{0.914(5)}{} & \SI{0.858(6)}{}\\
 & & ANN & tabular+im. & $\SI{128000}{}$ & $\SI{15000}{}$ & \SI{64}{}  & \SI{0.982(5)}{} & \SI{0.974(4)}{} & \SI{0.913(5)}{} & \SI{0.852(6)}{}\\
 & & DEC & tabular & $\SI{128000}{}$ &  $\SI{15000}{}$ & \SI{64}{}  & \SI{0.973(8)}{} & \SI{0.959(6)}{} & \SI{0.867(4)}{} & \SI{0.787(5)}{}\\
 \midrule
 W.Y. & & CNN & tabular+im. & $\SI{152000}{}$ & $\SI{17000}{}$ & \SI{1070}{} & \SI{0.975(3)}{} & \SI{0.975(3)}{} & \SI{0.900(5)}{} & \SI{0.860(4)}{}\\
 \midrule
 G.T.R. & & RF & tabular & $\SI{61000}{}$ & $\SI{3000}{}$ & \SI{380}{} & \SI{0.995(1)}{} & \SI{0.994(1)}{} & \SI{0.946(5)}{} & \SI{0.924(5)}{}\\
 \midrule
 L.D. & & CNN & images & $\SI{350000}{}$ & $\SI{25000}{}$ & \SI{50176}{} & N/A & N/A & N/A & N/A\\ 
 \bottomrule
\end{tabular}
}

\caption{Columns from left to right: contributions, listed by participants; ML methods used; data type used for model training (tabular, images or tabular+images); training and test sample sizes; dimensionality or the number of features used; and the performance (purity and completeness) for the train and blinded datasets. The sample sizes are rounded down to the nearest thousand. The performance of the CNN model by L.D.\,(bottom row) is described by \citet{2022A&A...666A.171D}, and the model is not applicable (N/A) to images in our datasets due to low resolution.} 
\label{tab:summary}
\end{table*}

\subsection{A preliminary classical approach -- V.P. and M.P.}
As already mentioned, variability is an intrinsic property of AGN and a promising selection tool, as both the time scales of the variations and the overall trends are different from other, mainly stellar, sources. Power spectra of AGN are characterized by a typical red noise behavior \citep{2009ApJ...698..895K,2010ApJ...708..927K,2013ApJ...765..106Z}, with most of the variation arising on long time scales \citep{2002MNRAS.332..231U}. This timescale behavior means that the longer the observed baseline, the better the selection through variability \citep[see e.g.][]{2019A&A...627A..33D}. The LSST, with its dense and long coverage, will be the best survey to exploit this selection technique. Moreover, the implementation of DIA on the entire dataset will also enable the selection of low-luminosity AGN dominated by their host galaxy.

Unfortunately, the AGN DC does not contain difference images and the population of confirmed AGN in the region where optical light curves are available (SDSS Stripe82) is biased towards bright quasars. In spite of the limitation of the archival data, which prevented us from testing the capabilities of DIA on AGN science, we examined some of the light curve features not included in the original dataset, before using ML with all the available LSST-like data products.

Intensive X-ray, UV and optical monitoring campaigns of AGN show that, whatever the intrinsic physical mechanism (thermal propagating fluctuations, hydrodynamical instabilities, reprocessing of high energy coronal photons  by the accretion disk; \citealt{2018MNRAS.480.2881M}) we can expect correlation between adjacent regions of the electromagnetic spectrum, such as the LSST bluer and redder bands. As SDSS images in the $ugriz$-bands were taken  close in time (at 71.7 second intervals; \citealt{1998AJ....116.3040G}), we calculated the Pearson correlation coefficient between pairs of light curves in different bands, along with the average magnitude and the standard deviation per each band. We restricted our analysis to the $gri$-bands as they have the highest signal-to-noise ratio. A preliminary analysis that include the $uz$-bands have shown to bring no improvement.

\begin{figure}[h]
    \centering
    \includegraphics[width=0.99\hsize]{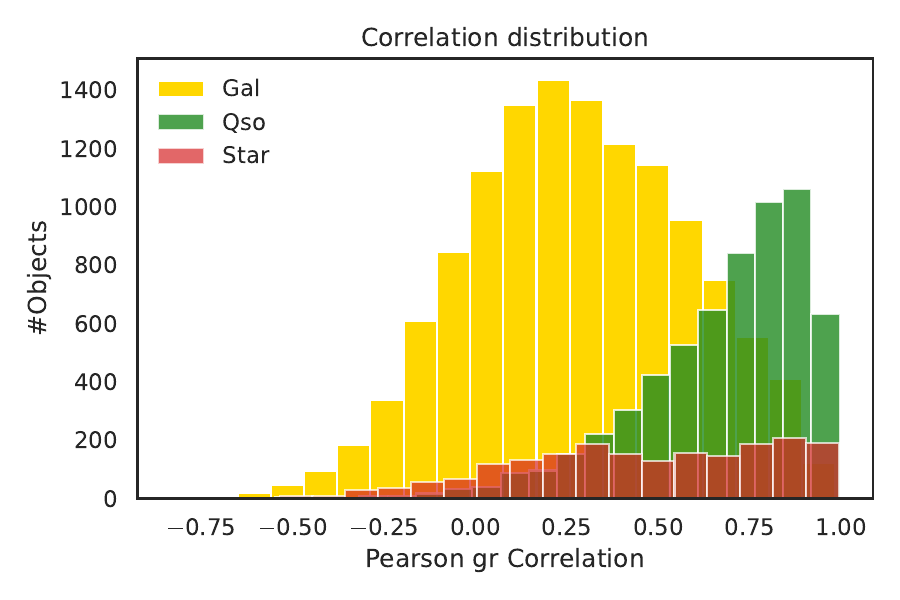}
    \caption{Distribution of the $gr$-bands Pearson correlation coefficients for galaxies (blue), QSOs (green) and stars (red).}
    \label{fig:hist_vp}
\end{figure}

\begin{figure}[httb]
    \centering
    \includegraphics[width=0.99\hsize]{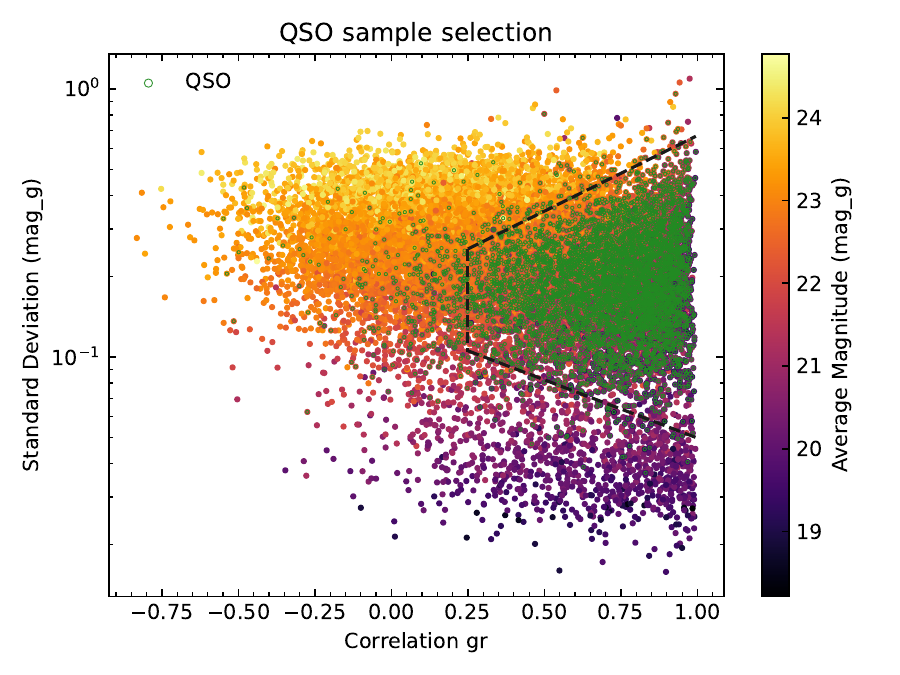}
    \caption{Standard deviation of the light curves vs. $g-r$ band correlation for a random sample of sources. Points are color-coded according to the g-band average magnitude. The black dashed lines define the $wedge$ where QSOs (green points) tend to group.}
    \label{fig:wedge_vp}
\end{figure}

The distributions of the correlation coefficients for sources labeled as QSO, Star or Galaxy clearly show that they belong to three different populations (Fig.\,\ref{fig:hist_vp}). We selected random samples of 10,000 objects and looked at how they distributed on a plot with $g$-band magnitude standard deviation vs. $gr$-bands correlation coefficient. The majority of quasars (more than $90\%$) tend to group in a well defined $wedge$ of the space\footnote{Comprised in $x > 0.25$, $y < e^{1.3x - 1.7}$, $y > e^{1x - 2}$.}, as shown in  Fig.\,\ref{fig:wedge_vp}. Thus, we  identified the region of interest for our sources and tested the $wedge$ selection on the $blinded$ $dataset$ which was provided at the end of the AGN DC. We find that the light curve variance and correlation among bands alone, allow us to produce samples of AGN with a completeness\footnote{The ratio of QSOs selected and total number of QSOs in the sample.} of \SI{90.9}{\percent}, albeit with low purity\footnote{The ratio of the selected QSOs and the total number of selected sources.} \SI{52.0}{\percent}. The low purity is expected since these two features alone tend to identify intrinsic correlated variability above a certain variance threshold, but do not include a full characterization of AGN properties allowing one to disentangle them from, e.g., stars.

\begin{figure}[httb]
    \centering
    \includegraphics[width=0.99\hsize]{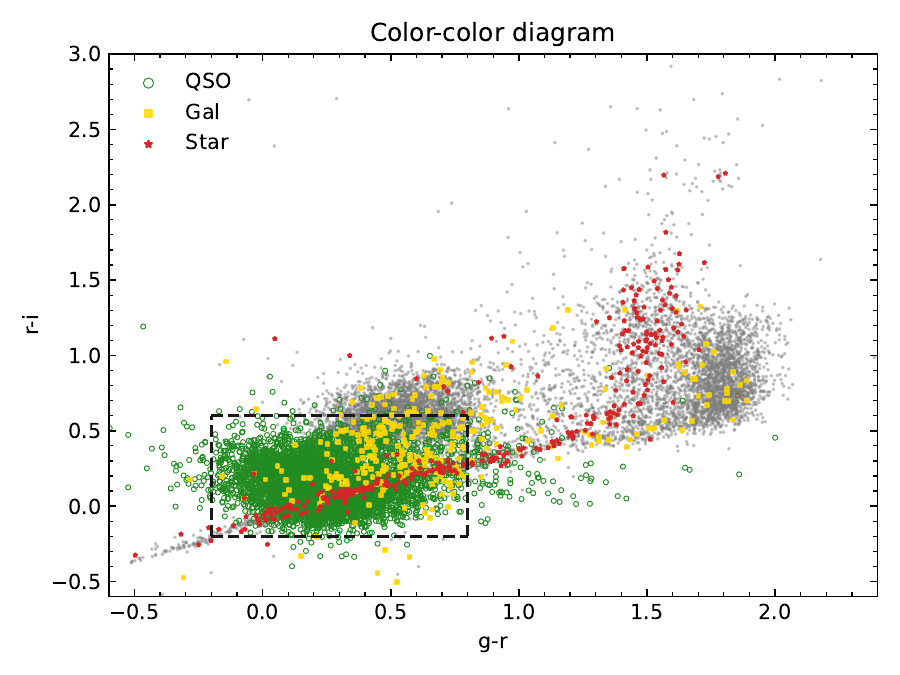}
    \caption{Color-color diagram showing the distribution of Galaxies, QSOs and Stars selected by the \textit{$wedge$}, with a cut in LSST extendedness > 0.95. Gray dots represent the total sample, while the black dashed lines highlight the $box$ where QSOs tend to group.}
    \label{fig:color_vp}
\end{figure}

It is possible to remove most of the contaminants and reach a purity of \SI{95}{\percent} (and decreasing the completeness by less than \SI{10}{\percent}) by adding the extendedness and color information, which is particularly useful in the selection of candidate AGN \citep{2002AJ....123.2945R}. As our AGNs were mainly bright quasars, we made a cut requiring the LSST extendedness parameter to be greater than 0.95. Then, we used a $r-i$ vs $g-r$ color-color diagram to select candidate quasars among the sources deriving from the wedge+extendedness criteria by defining a box in which they tend to group\footnote{Comprised in $-0.2 < g-r < 0.8$ and $-0.2 < r-i < 0.6$.} (see Fig.\,\ref{fig:color_vp}).

In order to evaluate the contribution of these three selection criteria, we tested them both singularly and combined together (see Table \ref{tab:summary_vp}). The best result overall is obtained by the combination of all the selection criteria, with a purity of \SI{95.0}{\percent} (Tables \ref{tab:summary_vp}, \ref{tab:summary}). All the methods alone, strongly suffer from contamination and the extendedness seems to show the best performance. However, this is  mainly due to the bias towards point-like quasars in the AGN DC sample and does not reflect the true diversity of the AGN population. For this reason, it is also worth noting that the pair wedge+color returns completeness and purity of \SI{83.0}{\percent} and \SI{85.7}{\percent} respectively, without making any assumption on the morphology  of the source (Table \ref{tab:summary_vp}). This rate of success is extremely promising in the LSST perspective of applying the selection directly to sources detected on difference images, where sources will be point-like and contamination will only be due to transients, variable stars or bogus events. 

We point out that the lower performance with respect to ML approaches presented in the following subsections has to be expected, since we did not use any advanced light curve feature. However, in  spite of the lower performance, we demonstrate that correlation analysis among different bands enables a very fast and cheap first order selection of candidate QSOs (and possibly less luminous AGN).  Furthermore, in the case of LSST where the low-luminosity AGN population will be detectable through DIA, correlation will help to probe the low S/N regime disentangling intrinsic variability from spurious uncorrelated noise.
\begin{table*}[httb]
\centering
\begin{tabular}{c c c c c c c} 
 \toprule
 \toprule
 {} & {Galaxy} & {QSO} & {Star} & {Total} & {Completeness (QSO)} & {Purity (QSO)} \\
 \midrule
 \midrule
 {Sample} & {13066} & {6177} & {1955} & {21238} & {} & {} \\
 {Wedge} & {4767} & {5614} & {404} & {10789} & {\SI{90.9}{\percent}} & {\SI{52.0}{\percent}} \\
 {Extendedness} & {1385} & {6029} & {1796} & {9211} & {\SI{97.6}{\percent}} & {\SI{65.4}{\percent}} \\
 {Color} & {2763} & {5538} & {956} & {9257} & {\SI{90.3}{\percent}} & {\SI{59.8}{\percent}} \\
 {Wedge+Ext} & {325} & {5492} & {331} & {6148} & {\SI{89.5}{\percent}} & {\SI{89.3}{\percent}} \\
 {Col+Ext} & {612} & {5492} & {950} & {7054} & {\SI{89.5}{\percent}} & {\SI{77.8}{\percent}} \\
 {Wedge+Col} & {716} & {5093} & {130} & {5989} & {\SI{83.0}{\percent}} & {\SI{85.7}{\percent}} \\
 {Wedge+Ext+Col} & {138} & {5055} & {125} & {5318} & {\SI{82.4}{\percent}} & {\SI{95.0}{\percent}} \\
 \bottomrule
\end{tabular}
\caption{Results of the selection of QSOs on the blinded dataset, using a classical approach and combinations of different selection criteria.}
\label{tab:summary_vp}
\end{table*}

\subsection{AGN, galaxy, star classification -- Đ.S., I.J. and SER-SAG}\label{ss:exp1}
\begin{figure}[httb]
    \centering
    \includegraphics[width=0.99\hsize]{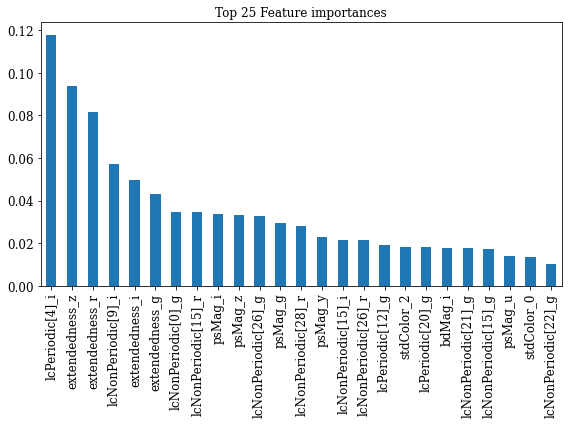}
    \caption{Top 25 features ranked by importance for XGB classifier run on a full dataset.}
    \label{fig:XGBtop25}
\end{figure}

Initial feature selection was done through trial and error. Bearing in mind that XGB supports missing (null) values by default, we first train a few XGB classifiers on a total set of $\sim$\SI{380000}{} containing all objects with known labels using a smaller subset of features. We further examined feature histograms with good visual separation between at least two classes (especially star-QSO and galaxy-QSO). After numerous tests, we find optimal 64 features for which we report accuracy of $\SI{98.0\pm0.1}{\percent}$ and $\SI{97.0\pm0.1}{\percent}$ on a test and blinded datasets. A large fraction of the dataset contains missing values for many of the features. We’ve performed a few methods of data imputation: median, hot deck and KNN imputation, however, the preliminary results are poor. Therefore, we keep the same 64 features and filter out objects with missing values for which we obtain a subset of $\sim$\SI{128000}{} objects divided into $\sim$\SI{55000}{} stars, $\sim$\SI{45000}{} galaxies and $\sim$\SI{28000}{} quasars. With such setup, we proceed with ML steps before training separate XGB, RF, SVM and NN models. The most important 20 features in the dataset from the initial XGB analysis are given in Fig.\,\ref{fig:XGBtop25}. More than a half of the top 25 features are LC features, both periodic and non-periodic.

\begin{figure*}
  \centering
  \begin{tabular}{cccc}
    \includegraphics[width=0.24\hsize]{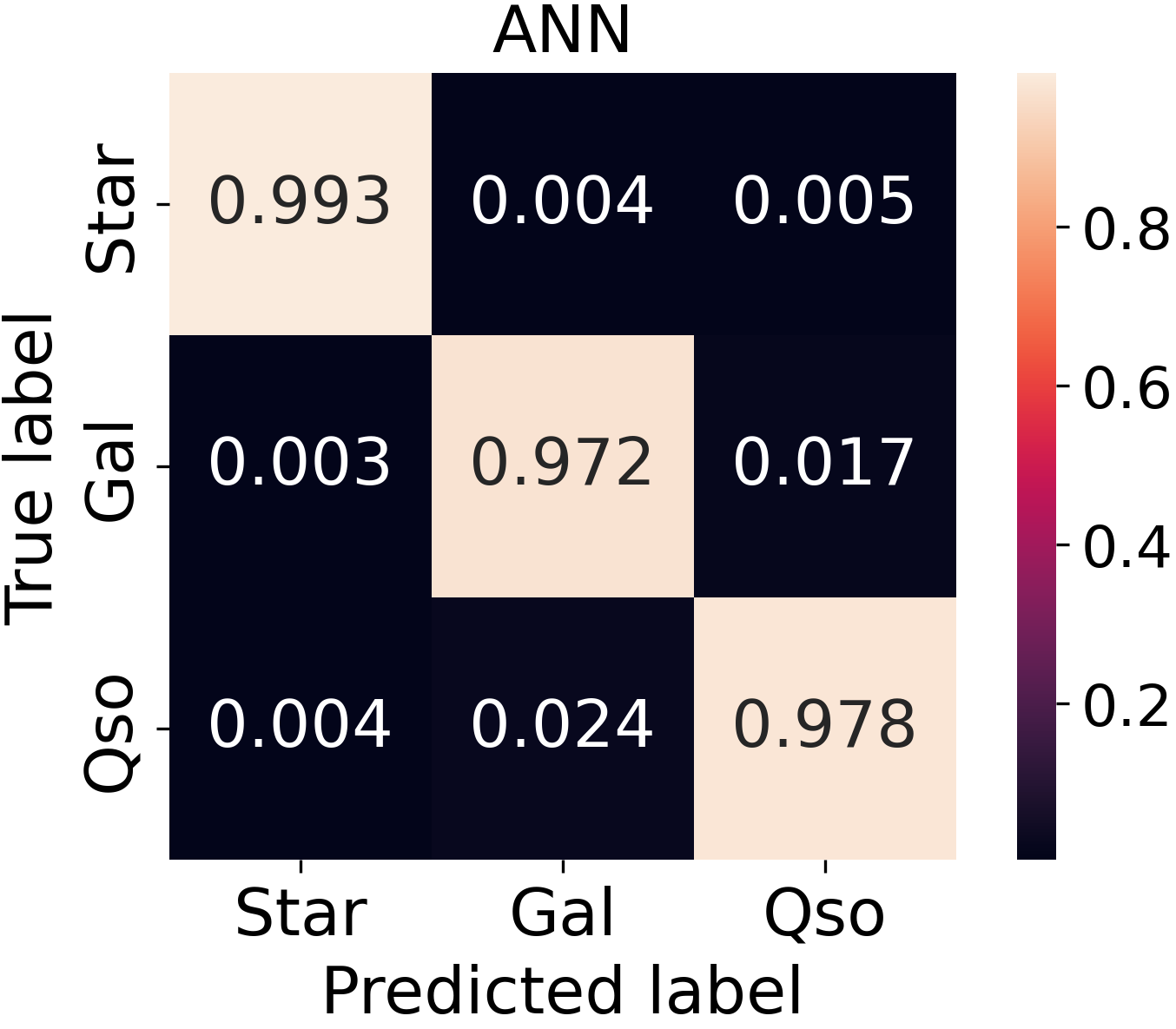} & \includegraphics[width=0.24\hsize]{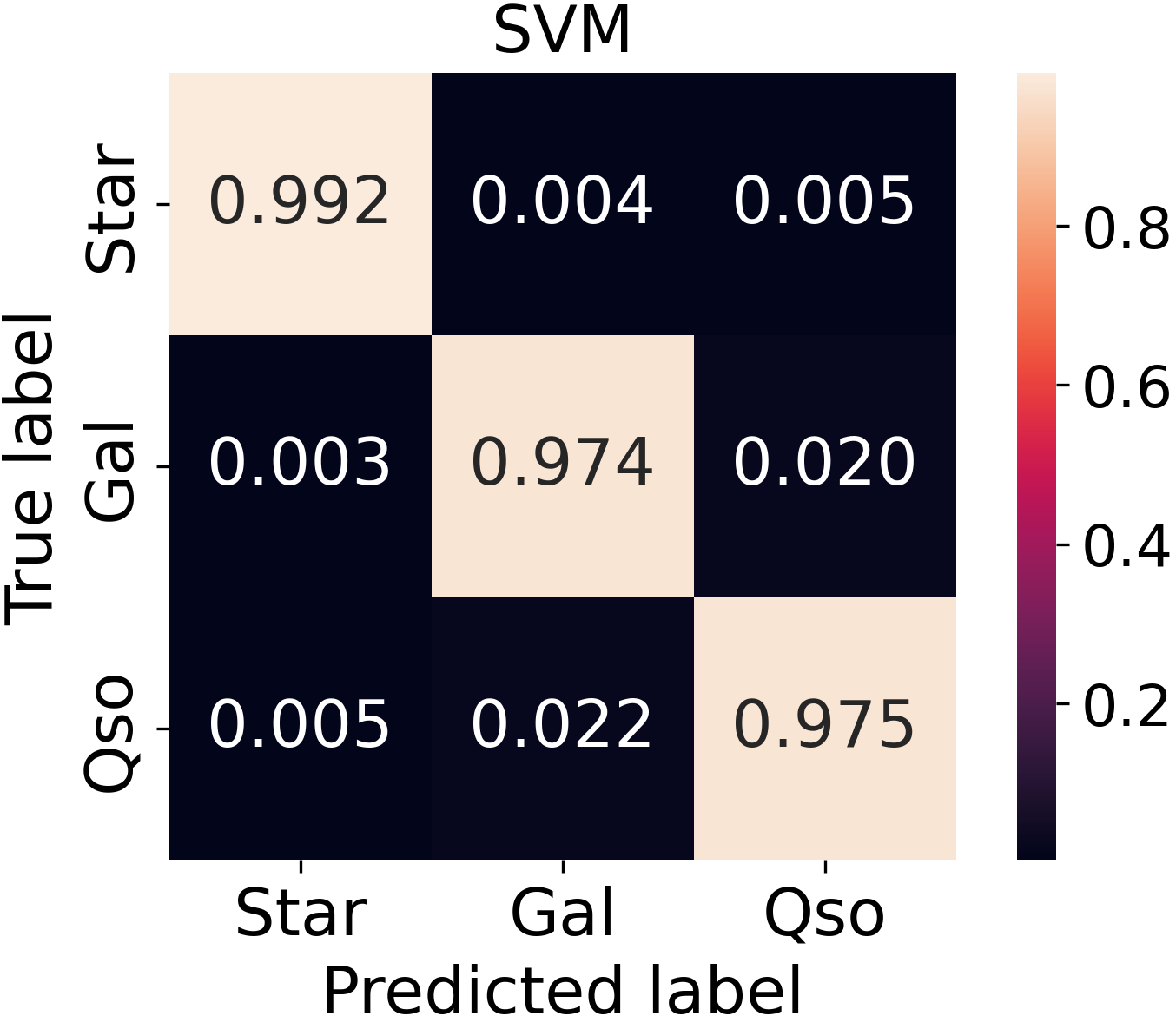} &
    \includegraphics[width=0.24\hsize]{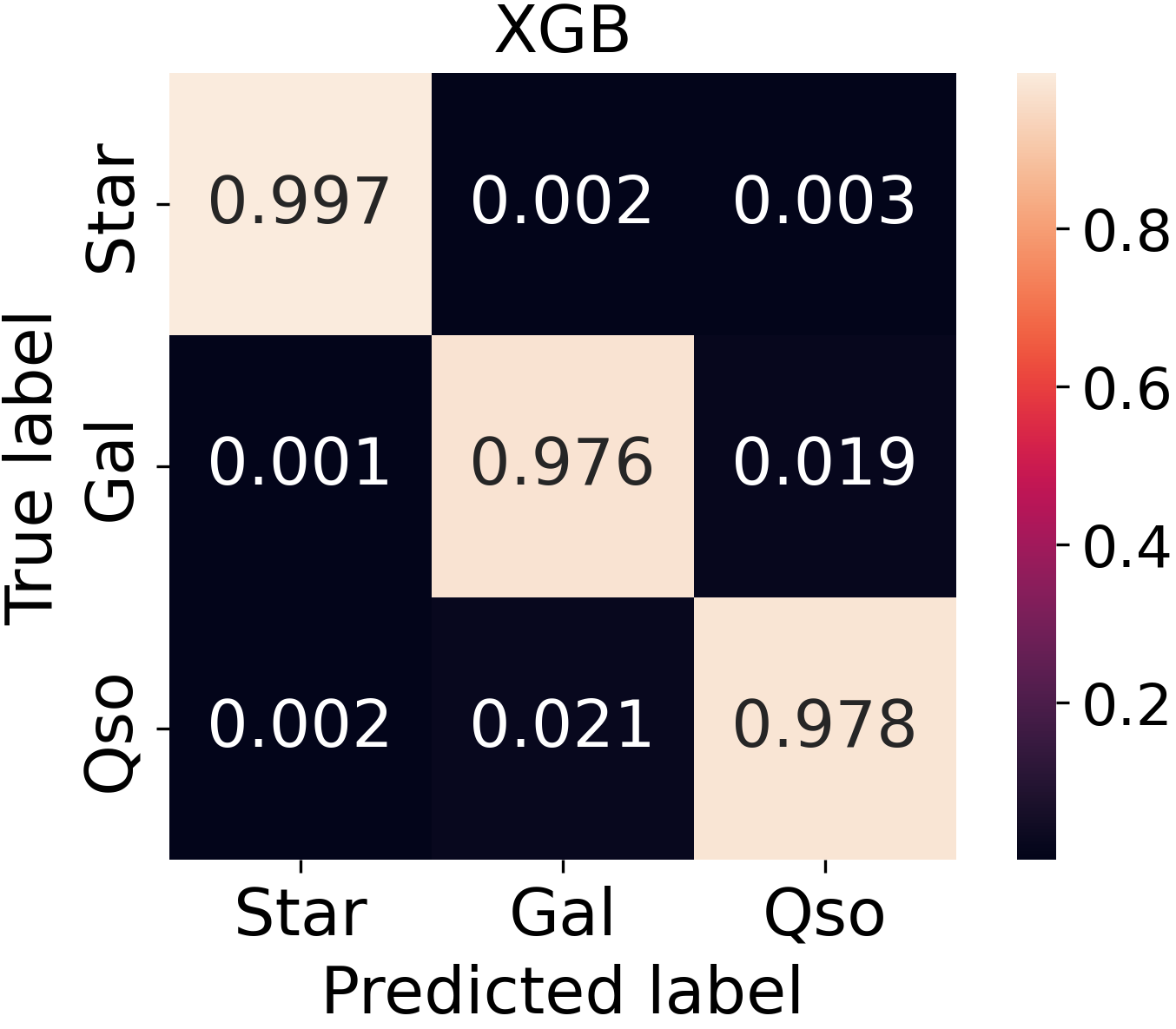} &
    \includegraphics[width=0.24\hsize]{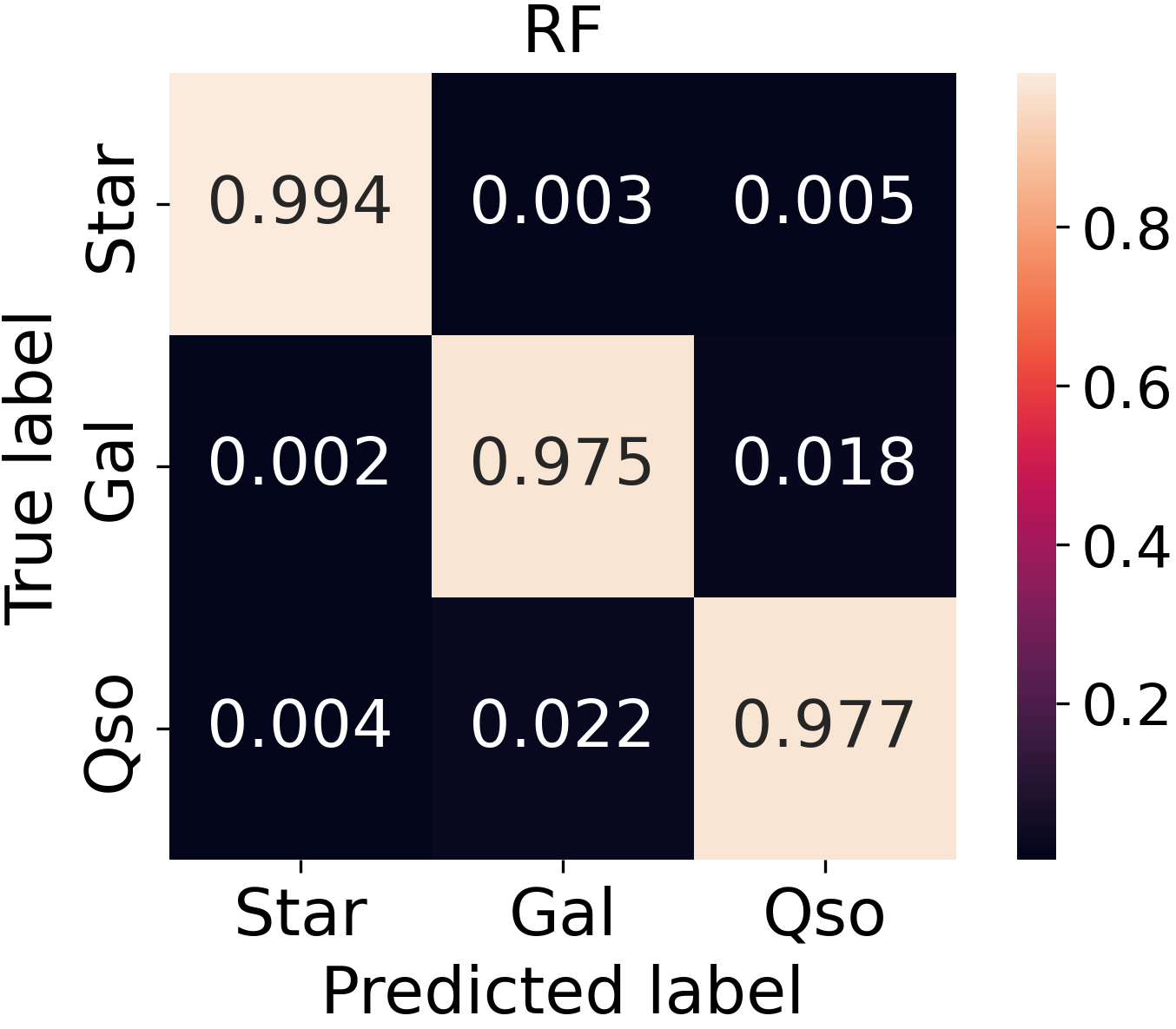}
    \\
    \includegraphics[width=0.24\hsize]{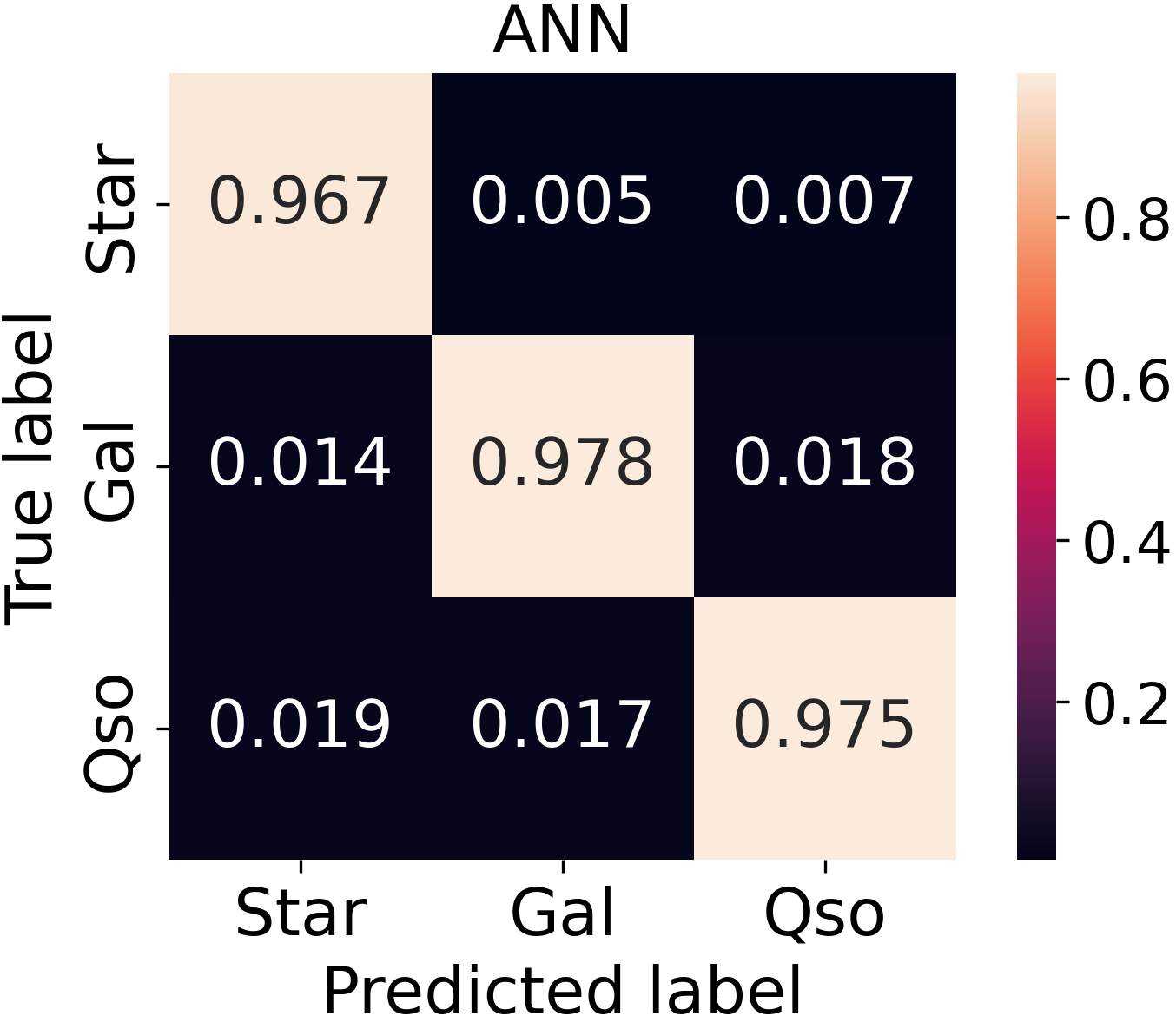} &
    \includegraphics[width=0.24\hsize]{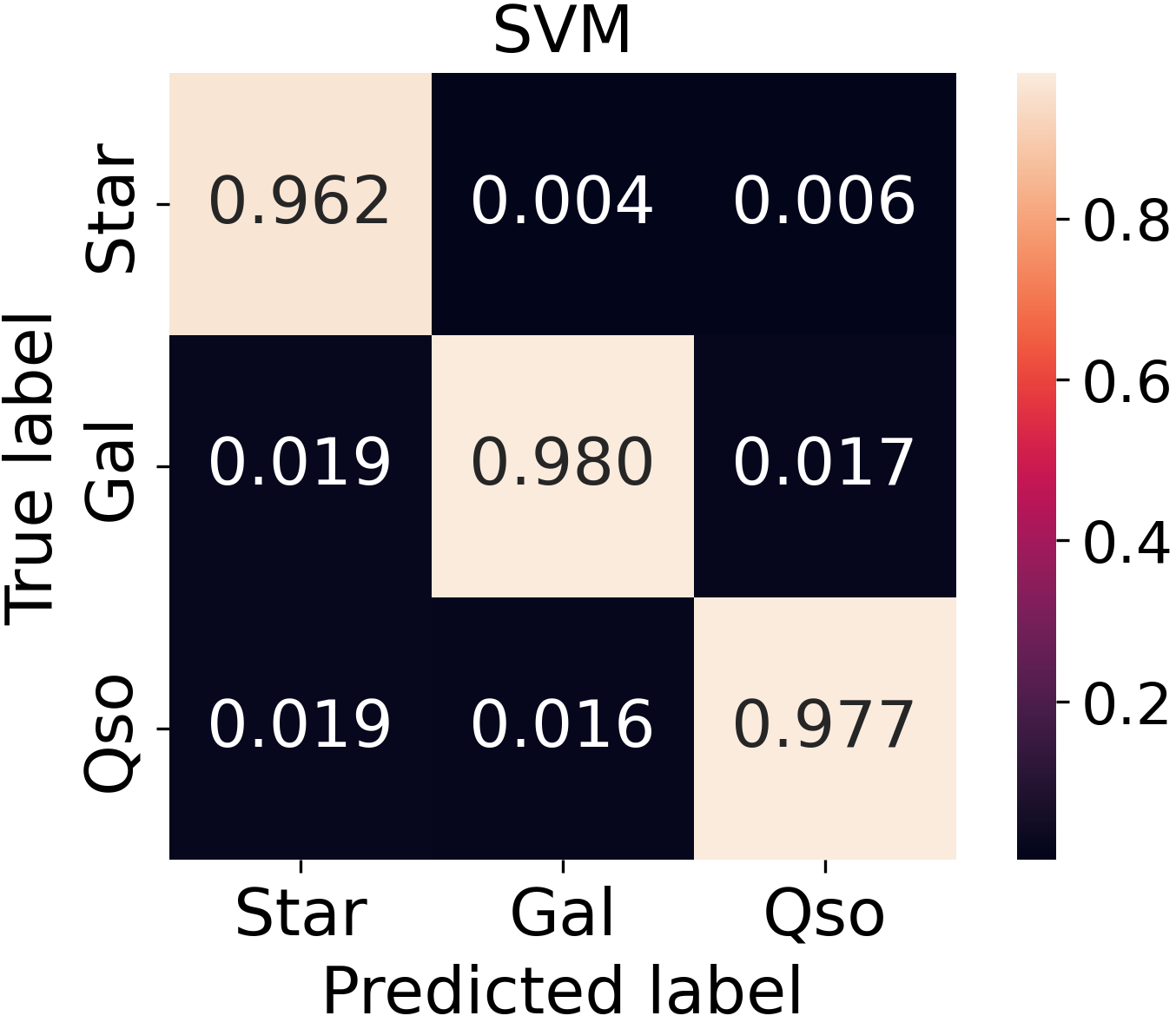} &
    \includegraphics[width=0.24\hsize]{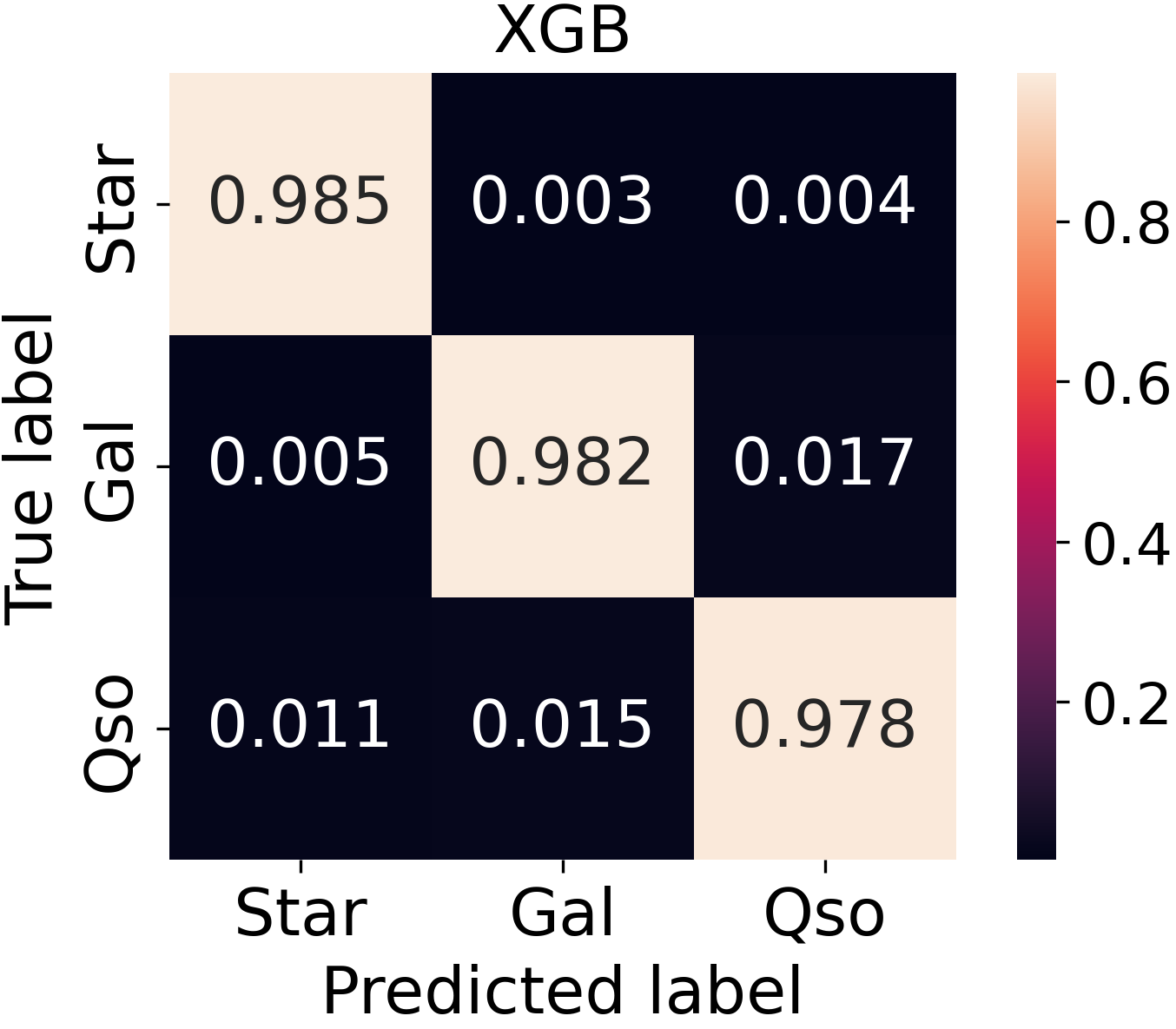} &
    \includegraphics[width=0.24\hsize]{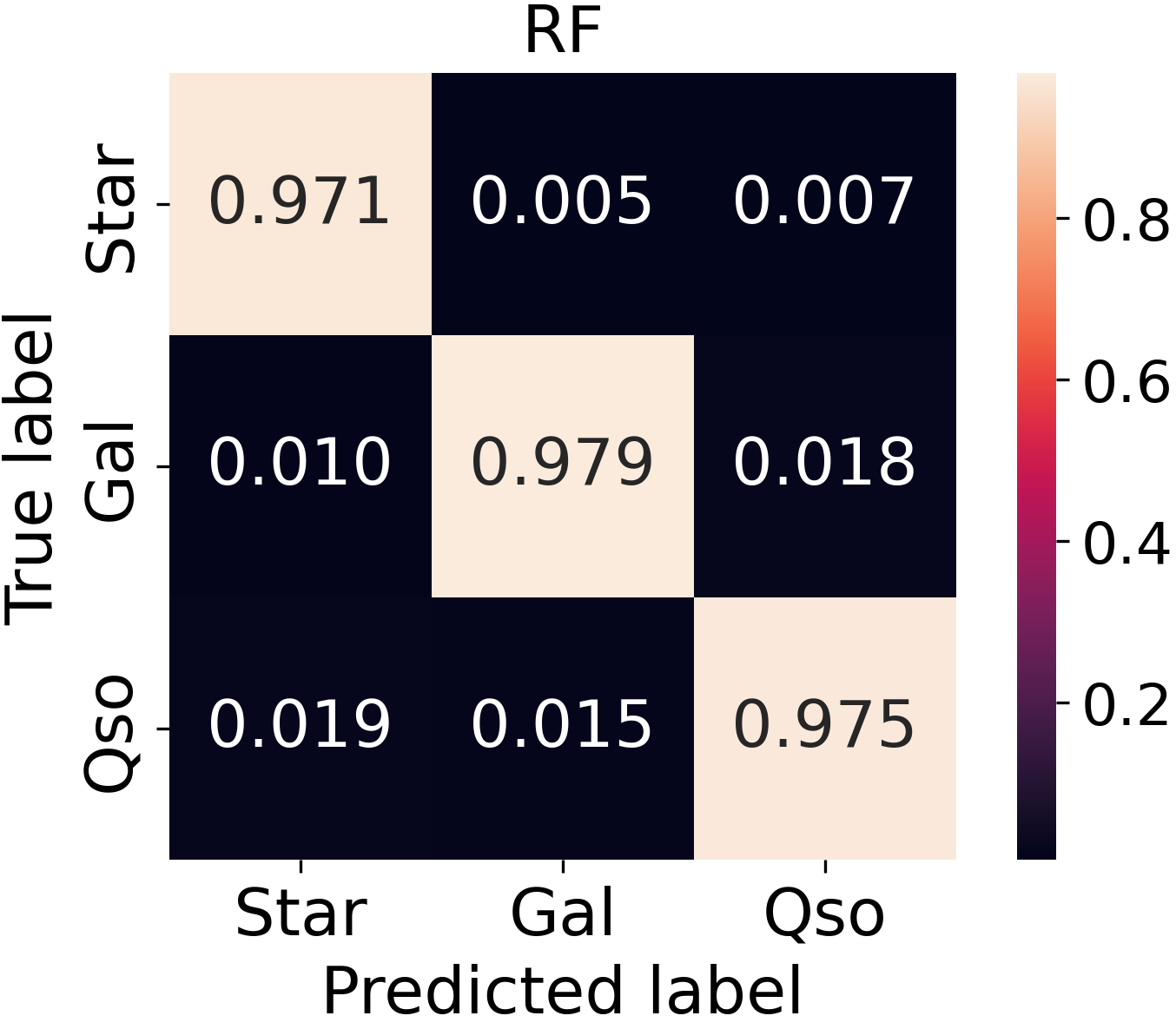}
    \\
  \end{tabular}
  \caption{Confusion matrices normalized by purity for tabular data. From left to right: ANN, SVM, XGB, RF. Upper panels were computed on a test set, while lower panels were computed for a blinded set. True labels are placed on vertical axis, while the predicted labels are on the horizontal axis.}
  \label{fig:cf_tab} 
\end{figure*}

We divide the subset dataset(s) into training, validation and test sets, which account for 70, 20 and \SI{10}{\percent} of the subset object count respectively. Before applying any of the ML methods, data pre-processing is required. We standardize the training data to zero mean and standard deviation of unity. Using mean and standard deviations obtained for each train feature, we normalize validation and test sets.
We used both supervised and unsupervised ML methods with the goal of comparing the performance of each method in order to establish a solid foundation towards building more advanced models. For supervised learning, we were able to achieve a high accuracy for each method (e.g., XGB accuracy of \SI{98.3\pm0.4}{\percent} and \SI{97.8\pm0.3}{\percent} for the training and blinded datasets respectively) when the light curve features are taken into account. However, the dataset is dominated by bright quasars. The XGB and RF perform the best overall on this subset of data, which is very often the case for ML applications on tabular data. The performance rating measured by classification accuracy is XGB>RF>SVM>NN (Table \ref{tab:summary}). Confusion matrices are illustrated in Fig.\,\ref{fig:cf_tab}.
Feature importance for the latter four methods follow the similar trend as for the initial XGB analysis. Applying the same methods on the same set when the LC features are ignored, reduces the classification accuracy to \SI{95}{\percent}.

In addition to tabular data, we investigated whether the addition of pixel information of image cutouts could contribute to achieving accuracy>\SI{99}{\percent}. We investigated deep ANNs with joint image and tabular data as inputs, as well as using an autoencoder with a simple bottleneck architecture for image dimensionality reduction to $\sim$10 latent features that are concatenated to the tabular data. No improvement has been observed in comparison with the models trained using tabular data only. The pixel resolution is simply too low for any morphology traits to be learned by the model.

Unsupervised methods are usually statistically weaker and, are in general outperformed by supervised methods on similar problems. Essentially, clustering is grouping the instances based on their similarity without the help of class labels. In this ML paradigm, the model learns the similarities (i.e., the distance between instances in multidimensional parameter space), not the mapping function between input features and class label as is the case with supervised methods. The class naturally emerges in the representation of the data clusters. In this way, we can expect that latent features of stars will differ at measurable level from the latent features of galaxies and QSOs. However, the number of clusters is in principle unknown. We set the number of clusters to be equal to 3, which is the number of ground-truth categories (star/galaxy/QSO). When the objects are passed to the model, it outputs soft assignments (the probabilities that given objects belong to given clusters). Each object is assigned a label of the most probable cluster. The DEC model performance is evaluated by unsupervised clustering accuracy \citep[][Eq.\,10]{2015arXiv151106335X} computed in the following way. The optimal one-to-one matching between the set of cluster labels and the set of class labels is found, so as to maximise the agreement between the mapped cluster labels and true class labels on the given set of objects. Such mapped cluster labels constitute the predictions of the clustering model. Then the accuracy of the predictions given the true classes is computed.

\begin{figure}[httb]
  \centering
  \begin{tabular}{cc}
    \includegraphics[width=0.46\hsize]{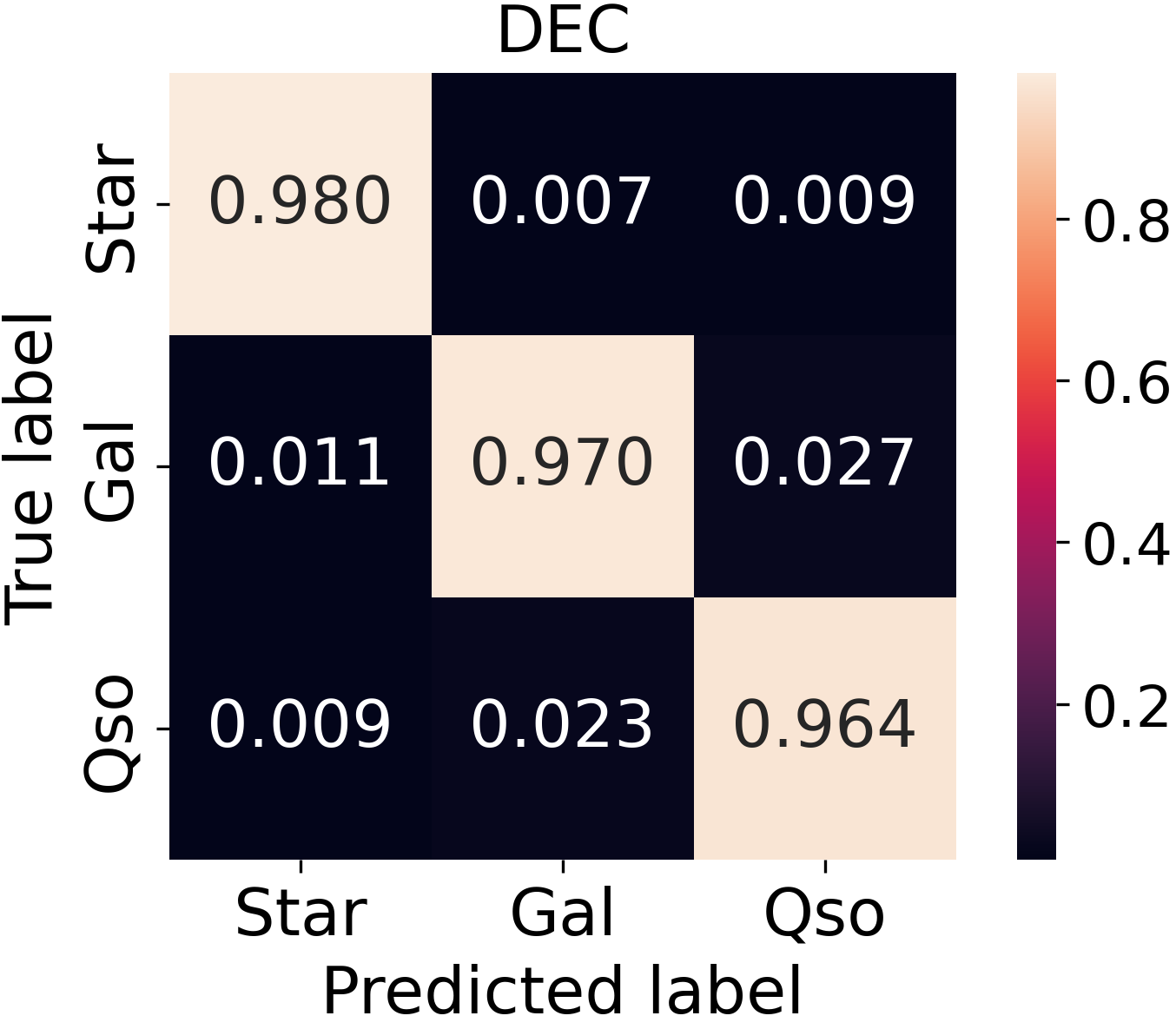} & \includegraphics[width=0.46\hsize]{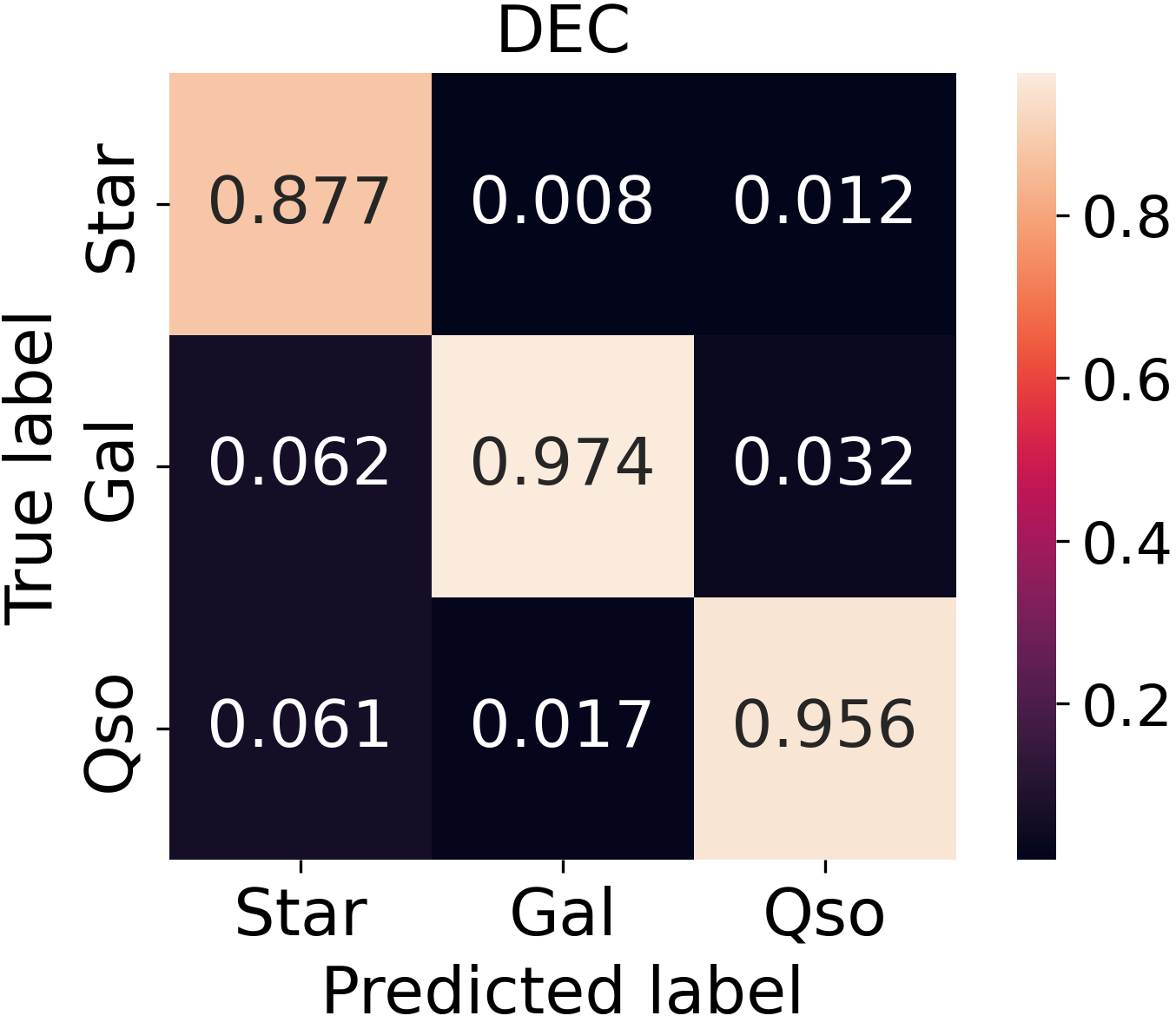}
  \end{tabular}
  \caption{Same as Figure~\ref{fig:cf_tab}, for DEC model. The performance is shown for the test set (left) and the blinded set (right).}
  \label{fig:cf_idec} 
\end{figure}

We obtained \SI{97.3\pm0.8}{\percent} and \SI{95.9\pm0.6}{\percent} (Table \ref{tab:summary}) clustering accuracy on the same test and blinded datasets respectively as for the supervised methods. When applying $k$-means or Gaussian mixture of models on the latent space, a clustering accuracy of $\sim$\SI{94}{\percent} is obtained, however, these numbers are heavily influenced by the initial weights of the ML model. Confusion matrices for DEC are shown in Fig.\,\ref{fig:cf_idec}. 

For the datasets used in our analysis, we expect a similar ranking of methods with respect to measured performance within a possible framework developed by \citet{2022AJ....164....6S}, that attempts to incorporate measurement error in astronomical classification problems. We expect a drop in performance if such experiments are repeated.

\subsection{Density maps -- W.Y.}\label{subsec:wy}
\begin{figure}[httb]
    \centering
    \includegraphics[width=0.99\hsize]{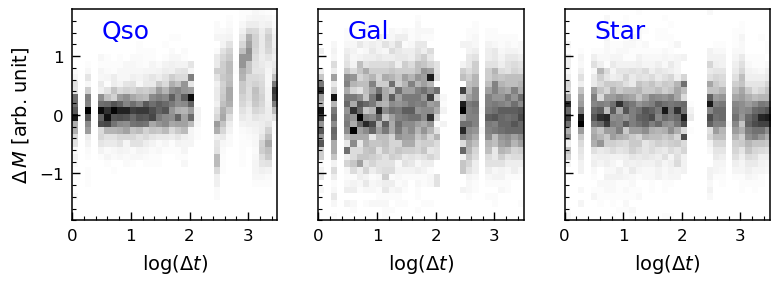}
    \caption{Computed density map in the $r$-band for one randomly selected quasar (left), galaxy (middle) and star (right).}
    \label{fig:DM_WY}
\end{figure}

We also explored the approach of first projecting time-series data onto 2D images (i.e., density maps) and then performing classification using the constructed density maps \citep[e.g.,][]{2017arXiv170906257M}.
Density maps are essentially a 2D distribution of variability power in the timescale and magnitude space; example density maps and the distinguishing power brought by this method are shown in Figure~\ref{fig:DM_WY}. The actual algorithm used to generate those density maps is a modified version of the one used in \citet{2017arXiv170906257M}. 
Projecting time-series data onto fixed size 2D images allows us to exploit the power of CNNs~\citep{2015Natur.521..436L}---the gold standard for ML-based image classification.

The total size of the density map sub-sample is $\sim$\SI{152000}, divided into $\sim$\SI{32000} stars, $\sim$\SI{52400}{} galaxies and $\sim$\SI{67600} QSOs.
We removed objects with less than 5 epochs in their $r$-band light curves and dropped objects having missing values (e.g., NaN) for the features utilized to train the model (see next paragraph). Objects with missing values typically fall into one of two categories: 1) too faint to have been detected in all "ugrizY" bands; 2) too extended (and/or faint) to have a reliably determined proper motion from Gaia EDR3~\citep{2021A&A...649A...1G} or NSC DR2~\citep{2021AJ....161..192N}.
Two versions of the above dataset are created to test our ML model: one only containing objects with low temporal sampling (number of epochs is less than 30 in the $r$-band) and one containing \SI{20}{\percent} of all objects randomly selected from the parent sample (without any additional cut regarding temporal sampling).  

\begin{figure}[httb]
  \centering
  \begin{tabular}{cc}
    \includegraphics[width=0.46\hsize]{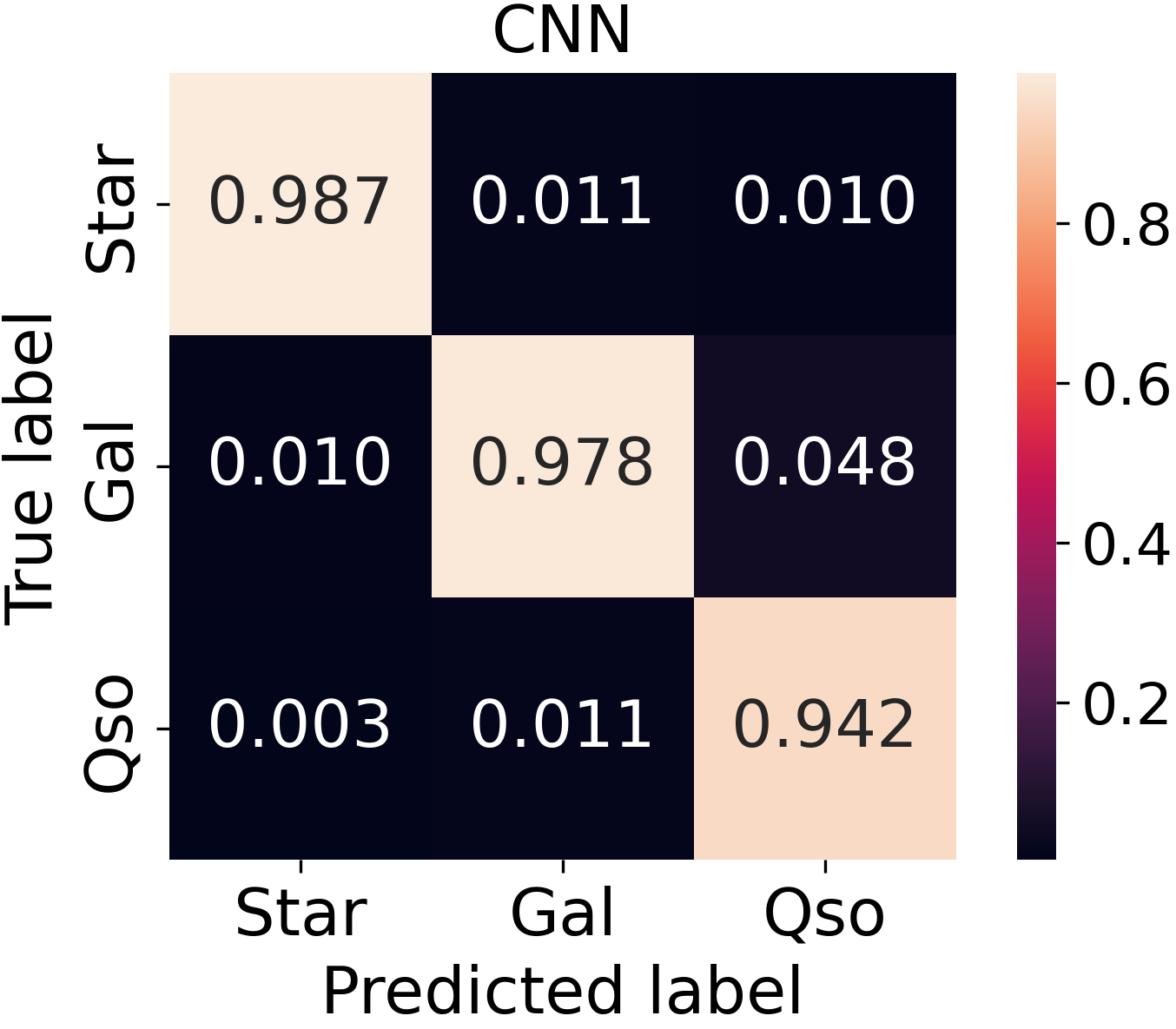} & \includegraphics[width=0.46\hsize]{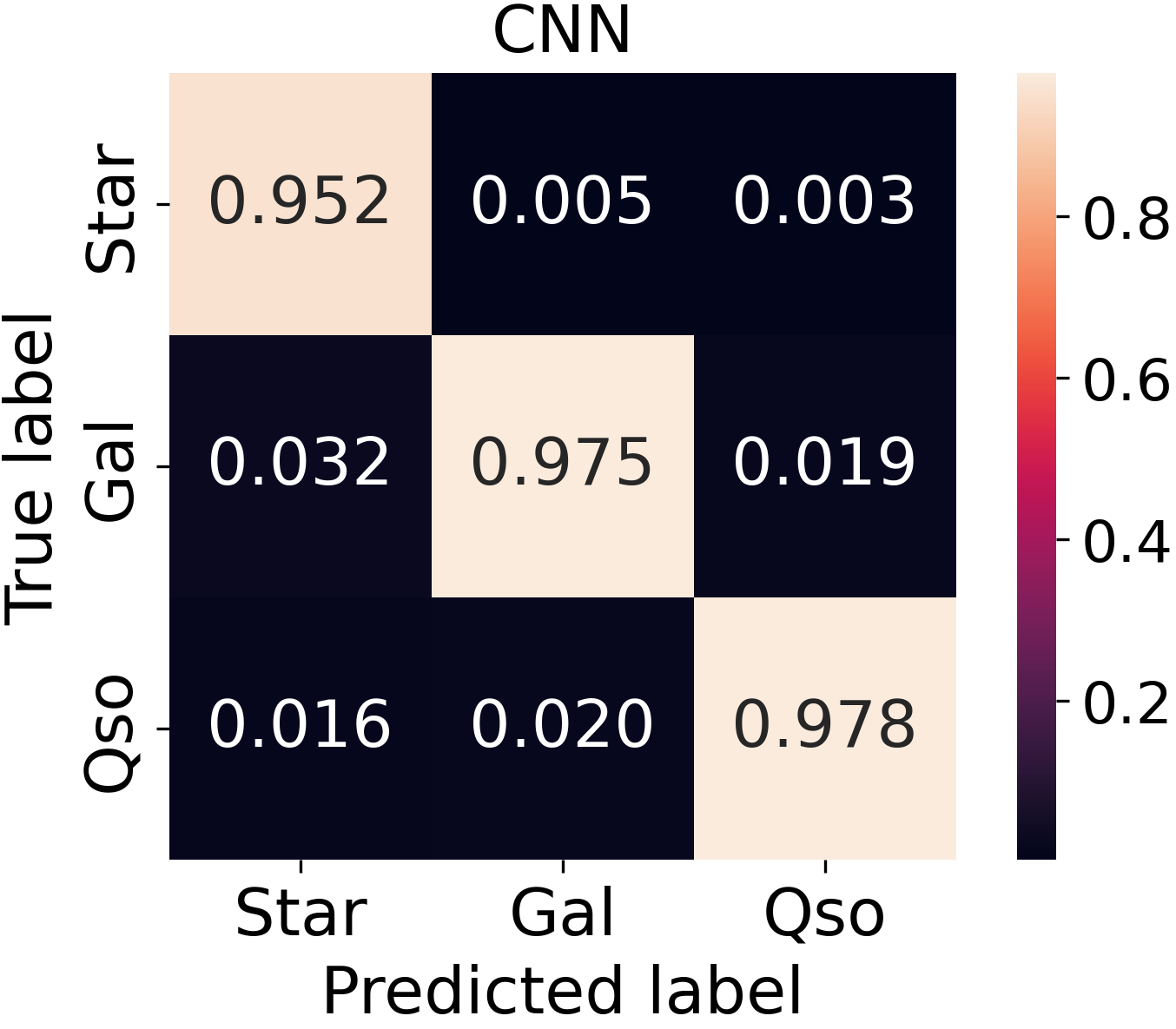}
  \end{tabular}
  \caption{Confusion matrices for a CNN model with density maps included. The performance is shown for the test set (left) and the blinded set (right).}
  \label{fig:cf_dm} 
\end{figure}

Our final classification model starts with extracting latent features from the constructed density maps using a convolutional neural network and then concatenate the latent features with a selection of features/columns from the \texttt{Object} table to form the final input for ANN that is used for classification. The columns picked from the \texttt{Object} table are the five optical colors (i.e., \texttt{stdColor[0-4]}), proper motion, and eight time-series features (i.e., \texttt{lcNonPeriodic[9, 13, 15, 19, 20, 22, 25, 26]}). The time-series features were cherry-picked from the feature importance rank of a random-forest classifier trained/tested on the full catalog of 374 features.
A very high accuracy of \SI{97.5}{\percent} (Table \ref{tab:summary}) were achieved for both low-cadence and high-cadence sub-samples, which demonstrates the robustness of the classification trained on density maps which are computationally inexpensive. Confusion matrices are shown in Figure~\ref{fig:cf_dm}.

\subsection{Separating quasars from stars -- G.T.R}
\begin{figure}[httb]
  \centering
  \begin{tabular}{cc}
    \includegraphics[width=0.45\hsize]{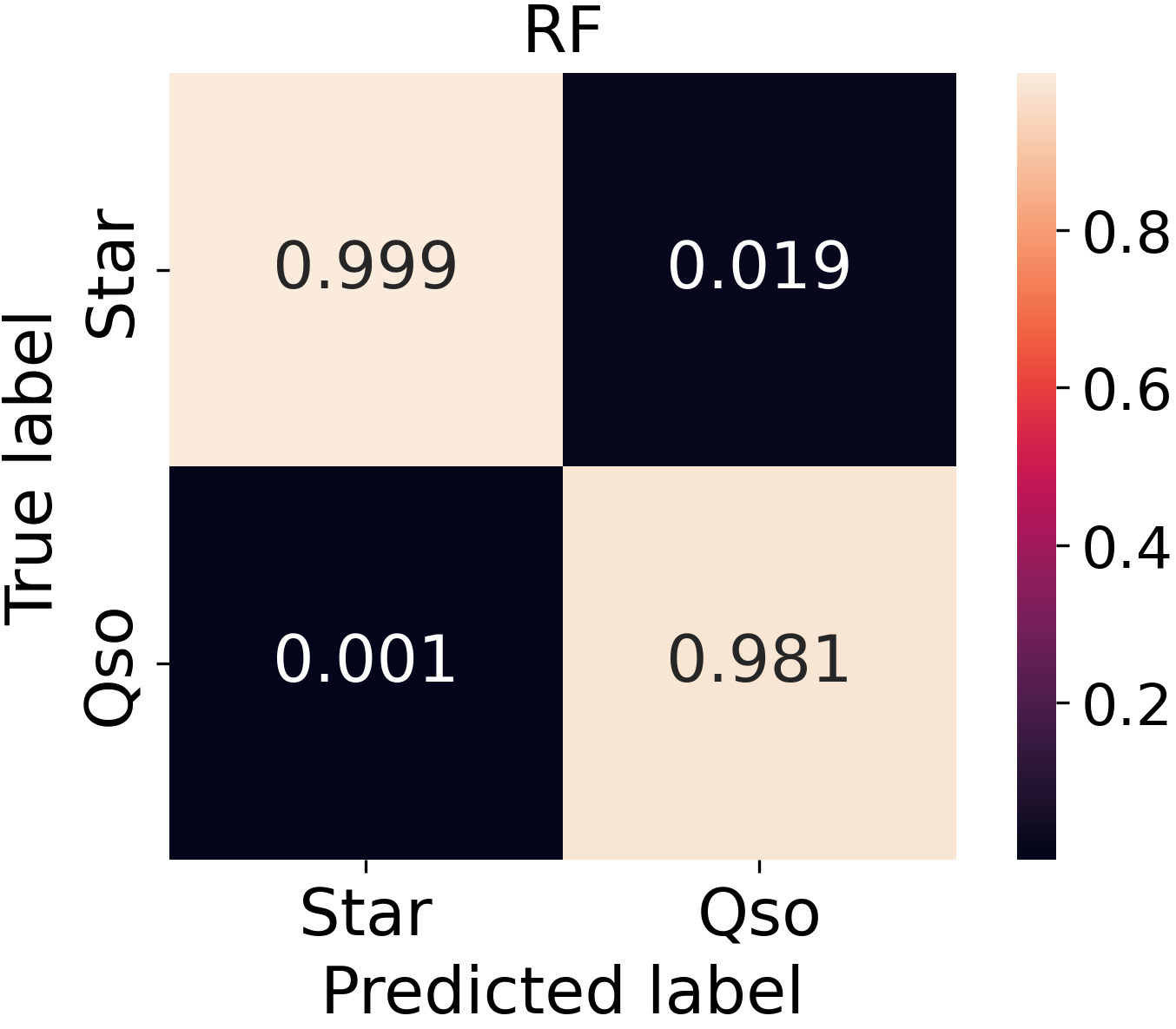} & \includegraphics[width=0.45\hsize]{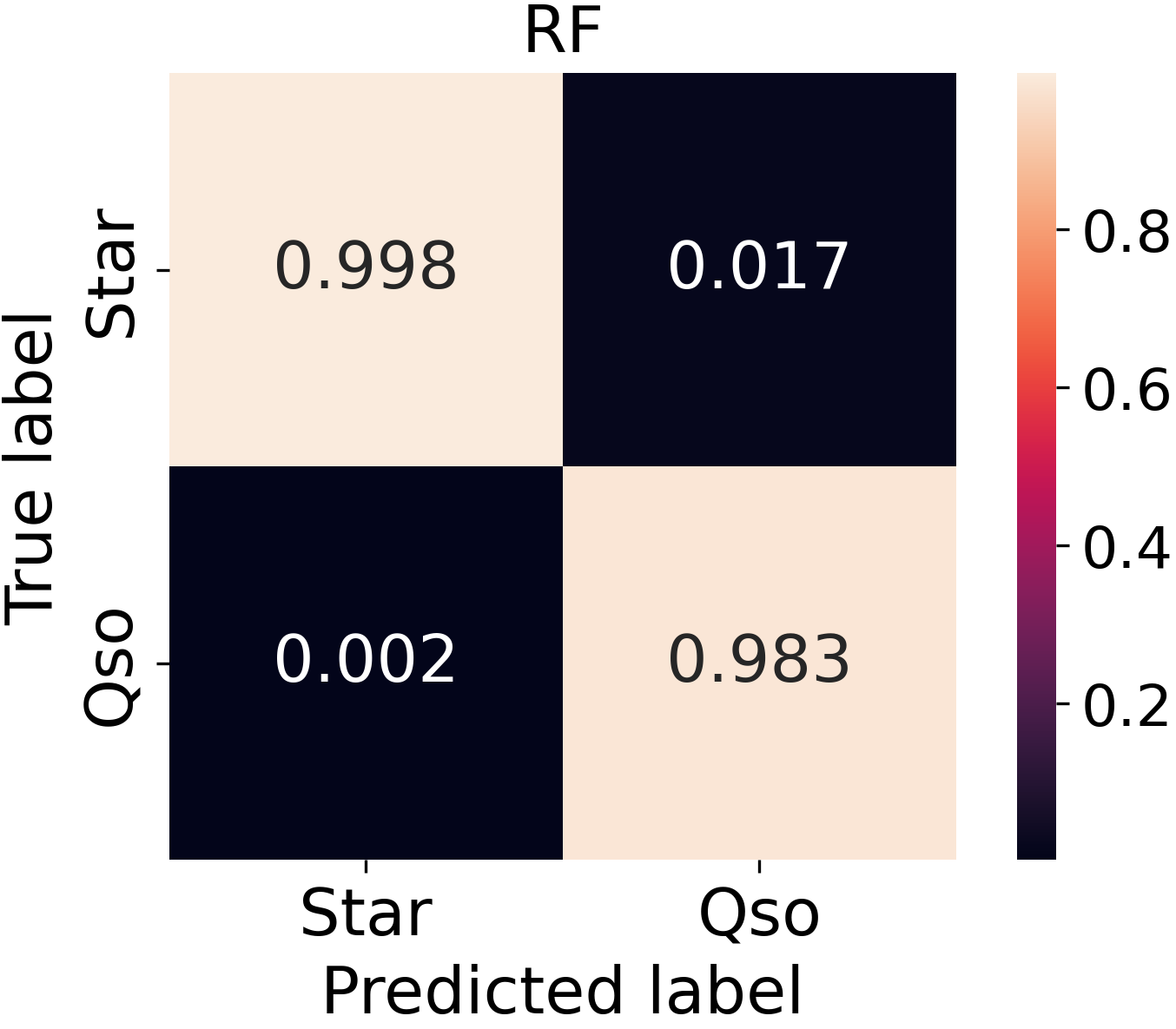}
  \end{tabular}
  \caption{Confusion matrices for a star/quasar RF classification model. The performance is shown for the test set (left) and the blinded set (right).}
  \label{fig:cf_gtr} 
\end{figure}

In this experiment, we use all data features except for flags and the redshift, which results in a subsample $\sim$\SI{10000} quasars and $\sim$\SI{50000} stars after excluding all objects with any missing values.
The excluded objects either fall into the same two categories of objects containing missing values listed in Section~\ref{subsec:wy} or have less than 30 epochs in their $r$-band light curves. The strict cut on temporal sampling is given by the fact that only light curves with more than 30 epochs are fitted with CARMA(2,1), whose parameters are \texttt{lcPeriodic[0-3]}.
The main difference between this experiment and the one in Section \ref{ss:exp1} is that the training dataset contains more data features while at the same time fewer objects due to filtering out those with missing and/or \texttt{NaN} values. We used RF for separating quasars from stars with an incredibly high accuracy of \SI{99.6}{\percent} (Table \ref{tab:summary}) and a similar performance on the blinded dataset. The white dwarfs and M-type stars can be seen to not be confused for quasars, which is a common source of error. Confusion matrices are shown in Fig.\,\ref{fig:cf_gtr}.

\section{Summary and Discussion}\label{sec:summary}
The LSST AGNSC-DC was designed to fulfill the growing need for efficient ML-based AGN selection methods. While we focus on simulated LSST data,  these tools should benefit a much wider AGN community. We addressed the problem of star, galaxy, and quasar classification from both a classical and a ML perspective while mimicking the future LSST catalog data. We used both supervised and unsupervised ML approaches. We followed the standard procedure of dividing the dataset into training, validation, and test datasets. The blinded dataset was revealed much later, after the submission of proposed solutions was finished. The performance of each method, sample size and the dimensionality is listed in the Table \ref{tab:summary}. We obtain high performance for supervised models and slightly lower for the unsupervised models for train and test datasets. We obtain slightly lower performance on the blinded dataset. The addition of LC features engineered from the $gri$-bands significantly improves the classification accuracy by a few percents, however, the dataset is skewed toward bright quasars. Fainter AGNs where the host galaxy contamination is strong, as well as sources in the regions where colors overlap will strongly benefit from variability \citep{2019A&A...627A..33D}. Further improvements of the ML methods that rely on LC features will be the inclusion of LC features computed using the missing $uzy$-bands. We are limited by the number of visits per object that is between 30 and 70 (Fig.\,\ref{fig:nvisits}), and uneven temporal sampling (Fig.\,\ref{fig:ts}), which are the two main obstacles for direct application of DL, but at the same time, the main reason for adopting LC feature engineering. The number of visits, higher by at least an order of magnitude, will allow us to train DL models directly on the light curves. 

In addition, we find that 64x64 $\mathrm{pixel}$ image cutouts are not sufficient for extracting pixel-level information. However, in the parallel effort done on low-redshift/low-luminosity AGNs \citep{2022A&A...666A.171D}, it was found that images of resolution 224x224 $\mathrm{pixel}$ can be used to directly extract meaningful information that can help in disentangling AGN vs non-AGN hosts. Moreover, the quality of images in all 6 bands that will be provided by LSST will allow us to develop DL models for separating AGNs from galaxies in a highly efficient manner at the cost of additional computing power. 

In addition to LSST data, we can also make use of the data from other surveys -- not just in the optical, but also in the X-ray, ultraviolet, infrared and radio wavebands. Such data allow us to make use of differences in object SEDs across a more extended wavelength range. For example, \citet{2015ApJS..221...27M} have shown that adding just a single infrared data point to SDSS \textit{ugriz} optical bands significantly improves the classification probability for AGNs vs stars. However, these  other surveys do not cover the same area of sky and adding multi-wavelength data will have the effect of breaking our largely monolithic LSST survey area into considerably smaller regions. Surveys at different wavelengths most often have different spatial resolutions, leading to situations where a single source in one survey is associated with multiple sources in another waveband. Forced photometry or probabilistic cross-matching are techniques to mitigate this problem \citep{2008ApJ...679..301B,2016ascl.soft04008L, 2017ApJS..230....9N, 2021ascl.soft02014B, 2022A&A...661A...3S}. Simple recognition of the problem may enable it to be avoided in many cases; however, the depth of LSST means that there will be few sources that are truly isolated. 

We conclude that LSST-like data enable the development of highly accurate star/galaxy/quasar classifiers, mainly using only $gri$-bands and without spectral information. Promising results have been achieved in spite of limitations in resolution, survey area, cadence, and baseline of the SDSS data (compared to what LSST will be capable of). Thus the new survey will allow us not only to improve the performances of the algorithms successfully tested within the DC, but also to develop more sophisticated techniques e.g., exploitation of pixel-level information on source cutouts, analysis of the population of weak AGNs dominated by their host galaxy trough DIA and identifying new types of AGNs. The AGN DC collection of multi-wavelength data is an important legacy for future research beyond AGNs science that will be included in the follow up work. 

\section*{Data Statement}
The DC was hosted on the SciServer\footnote{\url{https://www.sciserver.org/}} platform. Each participant of the DC had a verified account on the platform. The datasets released in the DC are publicly available on Zenodo\footnote{\url{https://doi.org/10.5281/zenodo.6878414}}~\citep{weixiang_yu_2022_6878414} under a Creative Commons Attribution 4.0 International Public License. The notebooks which reproduce the results of this work are designed as end-to-end and follow the flowchart of this work. The notebooks are publicly available on GitHub\footnote{\url{https://github.com/RichardsGroup/AGN_DataChallenge}}. The notebooks have been successfully executed on different servers with different operating systems which can also support GPU acceleration, including SciServer.

\section*{Author Contribution Statement}
Đ.S. and I.J. submitted the winning solution to the data challenge and drafted the manuscript; W.Y. and G.T.R. curated the data challenge; M.J.T., Q.N. and R.S. supplied the multi-wavelength data and associated paper text; Đ.S., I.J, W.Y., V.P., A.K., M.N., D.I., L.P., M.P., A.C. and G.T.R. contributed to solutions to the data challenge. All authors contributed to the writing and editing of the draft.

\section*{Acknowledgements}
Prizes for participating in data challenge were funded by the LSST Corporation's Enabling Science Program. Đ.S.\, acknowledges the support by the F.R.S. FNRS under grant PDR T.0116.21. Đ.S.\,and L.Č.P acknowledge support by the Astronomical Observatory (the contract \textnumero451-03-68/2022-14/200002), through the grants by the Ministry of Education, Science, and Technological Development of the Republic of Serbia. Đ.S.\, acknowledges support by the Science Fund of the Republic of Serbia, PROMIS \textnumero6060916, BOWIE. D.I., A.B.K, and L.Č.P. acknowledge funding provided by the University of Belgrade - Faculty of Mathematics (the contract \textnumero451-03-68/2022-14/200104) through the grants by the Ministry of Education, Science, and Technological Development of the Republic of Serbia. D.I. acknowledges the support of the Alexander von Humboldt Foundation. A.B.K. and L.Č.P thank the support by Chinese Academy of Sciences President's International Fellowship Initiative (PIFI) for visiting scientist.  M.J.T. acknowledges support from ANID (Fondecyt Proyecto 3220516). S.P. acknowledges financial support from the Conselho Nacional de Desenvolvimento Científico e Tecnológico (CNPq) Fellowship (\textnumero164753/2020-6) and the Polish Funding Agency National Science Centre, project \textnumero2017/26/A/ST9/00756 (MAESTRO 9). A.Ć. acknowledges support from the Fermi Research Alliance, LLC under Contract {\textnumero}DE-AC02-07CH11359 with the U.S.\ Department of Energy (DOE), Office of Science, Office of High Energy Physics.

The authors thank Feige Wang and Jinyi Yang for constructing and providing the \texttt{highZQso} catalog.

This research makes use of the SciServer science platform (www.sciserver.org). SciServer is a collaborative research environment for large-scale data-driven science. It is being developed at, and administered by, the Institute for Data Intensive Engineering and Science at Johns Hopkins University. SciServer is funded by the National Science Foundation through the Data Infrastructure Building Blocks (DIBBs) program and others, as well as by the Alfred P. Sloan Foundation and the Gordon and Betty Moore Foundation.

\textit{Software:} \textsc{python} \citep{van1995python}, \textsc{jupyter} \citep{soton403913}.

\textit{ML packages:} \textsc{numpy} and \textsc{scipy} \citep{2011CSE....13b..22V}, \textsc{pandas} \citep{mckinney2010data}, \textsc{scikit-learn} \citep{pedregosa2011scikit}, \textsc{keras} \citep{chollet2015keras}, \textsc{tensorflow} \citep{2016arXiv160508695A}.

\textit{Data visualization:} \textsc{matplotlib} \citep{4160265}, \textsc{seaborn} \citep{michael_waskom_2017_883859}.

\bibliography{bibliography}

\begin{thebibliography}{}
\expandafter\ifx\csname natexlab\endcsname\relax\def\natexlab#1{#1}\fi
\providecommand{\url}[1]{\href{#1}{#1}}
\providecommand{\dodoi}[1]{doi:~\href{http://doi.org/#1}{\nolinkurl{#1}}}
\providecommand{\doeprint}[1]{\href{http://ascl.net/#1}{\nolinkurl{http://ascl.net/#1}}}
\providecommand{\doarXiv}[1]{\href{https://arxiv.org/abs/#1}{\nolinkurl{https://arxiv.org/abs/#1}}}

\bibitem[{{Abadi} {et~al.}(2016){Abadi}, {Barham}, {Chen}, {Chen}, {Davis},
  {Dean}, {Devin}, {Ghemawat}, {Irving}, {Isard}, {Kudlur}, {Levenberg},
  {Monga}, {Moore}, {Murray}, {Steiner}, {Tucker}, {Vasudevan}, {Warden},
  {Wicke}, {Yu}, \& {Zheng}}]{2016arXiv160508695A}
{Abadi}, M., {Barham}, P., {Chen}, J., {et~al.} 2016, arXiv e-prints,
  arXiv:1605.08695.
\newblock \doarXiv{1605.08695}

\bibitem[{{Ahumada} {et~al.}(2020){Ahumada}, {Prieto}, {Almeida}, {Anders},
  {Anderson}, {Andrews}, {Anguiano}, {Arcodia}, {Armengaud}, {Aubert}, {Avila},
  {Avila-Reese}, {Badenes}, {Balland}, {Barger}, {Barrera-Ballesteros}, {Basu},
  {Bautista}, {Beaton}, {Beers}, {Benavides}, {Bender}, {Bernardi}, {Bershady},
  {Beutler}, {Bidin}, {Bird}, {Bizyaev}, {Blanc}, {Blanton}, {Boquien},
  {Borissova}, {Bovy}, {Brandt}, {Brinkmann}, {Brownstein}, {Bundy}, {Bureau},
  {Burgasser}, {Burtin}, {Cano-D{\'\i}az}, {Capasso}, {Cappellari}, {Carrera},
  {Chabanier}, {Chaplin}, {Chapman}, {Cherinka}, {Chiappini}, {Doohyun Choi},
  {Chojnowski}, {Chung}, {Clerc}, {Coffey}, {Comerford}, {Comparat}, {da
  Costa}, {Cousinou}, {Covey}, {Crane}, {Cunha}, {Ilha}, {Dai}, {Damsted},
  {Darling}, {Davidson}, {Davies}, {Dawson}, {De}, {de la Macorra}, {De Lee},
  {Queiroz}, {Deconto Machado}, {de la Torre}, {Dell'Agli}, {du Mas des
  Bourboux}, {Diamond-Stanic}, {Dillon}, {Donor}, {Drory}, {Duckworth},
  {Dwelly}, {Ebelke}, {Eftekharzadeh}, {Davis Eigenbrot}, {Elsworth},
  {Eracleous}, {Erfanianfar}, {Escoffier}, {Fan}, {Farr},
  {Fern{\'a}ndez-Trincado}, {Feuillet}, {Finoguenov}, {Fofie},
  {Fraser-McKelvie}, {Frinchaboy}, {Fromenteau}, {Fu}, {Galbany}, {Garcia},
  {Garc{\'\i}a-Hern{\'a}ndez}, {Oehmichen}, {Ge}, {Maia}, {Geisler}, {Gelfand},
  {Goddy}, {Gonzalez-Perez}, {Grabowski}, {Green}, {Grier}, {Guo}, {Guy},
  {Harding}, {Hasselquist}, {Hawken}, {Hayes}, {Hearty}, {Hekker}, {Hogg},
  {Holtzman}, {Horta}, {Hou}, {Hsieh}, {Huber}, {Hunt}, {Chitham}, {Imig},
  {Jaber}, {Angel}, {Johnson}, {Jones}, {J{\"o}nsson}, {Jullo}, {Kim},
  {Kinemuchi}, {Kirkpatrick}, {Kite}, {Klaene}, {Kneib}, {Kollmeier}, {Kong},
  {Kounkel}, {Krishnarao}, {Lacerna}, {Lan}, {Lane}, {Law}, {Le Goff}, {Leung},
  {Lewis}, {Li}, {Lian}, {Lin}, {Long}, {Longa-Pe{\~n}a}, {Lundgren}, {Lyke},
  {Ted Mackereth}, {MacLeod}, {Majewski}, {Manchado}, {Maraston}, {Martini},
  {Masseron}, {Masters}, {Mathur}, {McDermid}, {Merloni}, {Merrifield},
  {M{\'e}sz{\'a}ros}, {Miglio}, {Minniti}, {Minsley}, {Miyaji}, {Mohammad},
  {Mosser}, {Mueller}, {Muna}, {Mu{\~n}oz-Guti{\'e}rrez}, {Myers}, {Nadathur},
  {Nair}, {Nandra}, {do Nascimento}, {Nevin}, {Newman}, {Nidever}, {Nitschelm},
  {Noterdaeme}, {O'Connell}, {Olmstead}, {Oravetz}, {Oravetz}, {Osorio},
  {Pace}, {Padilla}, {Palanque-Delabrouille}, {Palicio}, {Pan}, {Pan},
  {Parker}, {Paviot}, {Peirani}, {Ram{\'r}ez}, {Penny}, {Percival},
  {Perez-Fournon}, {P{\'e}rez-R{\`a}fols}, {Petitjean}, {Pieri},
  {Pinsonneault}, {Poovelil}, {Povick}, {Prakash}, {Price-Whelan}, {Raddick},
  {Raichoor}, {Ray}, {Rembold}, {Rezaie}, {Riffel}, {Riffel}, {Rix}, {Robin},
  {Roman-Lopes}, {Rom{\'a}n-Z{\'u}{\~n}iga}, {Rose}, {Ross}, {Rossi},
  {Rowlands}, {Rubin}, {Salvato}, {S{\'a}nchez}, {S{\'a}nchez-Menguiano},
  {S{\'a}nchez-Gallego}, {Sayres}, {Schaefer}, {Schiavon}, {Schimoia},
  {Schlafly}, {Schlegel}, {Schneider}, {Schultheis}, {Schwope}, {Seo},
  {Serenelli}, {Shafieloo}, {Shamsi}, {Shao}, {Shen}, {Shetrone}, {Shirley},
  {Aguirre}, {Simon}, {Skrutskie}, {Slosar}, {Smethurst}, {Sobeck}, {Sodi},
  {Souto}, {Stark}, {Stassun}, {Steinmetz}, {Stello}, {Stermer},
  {Storchi-Bergmann}, {Streblyanska}, {Stringfellow}, {Stutz}, {Su{\'a}rez},
  {Sun}, {Taghizadeh-Popp}, {Talbot}, {Tayar}, {Thakar}, {Theriault}, {Thomas},
  {Thomas}, {Tinker}, {Tojeiro}, {Toledo}, {Tremonti}, {Troup}, {Tuttle},
  {Unda-Sanzana}, {Valentini}, {Vargas-Gonz{\'a}lez}, {Vargas-Maga{\~n}a},
  {V{\'a}zquez-Mata}, {Vivek}, {Wake}, {Wang}, {Weaver}, {Weijmans}, {Wild},
  {Wilson}, {Wilson}, {Wolthuis}, {Wood-Vasey}, {Yan}, {Yang}, {Y{\`e}che},
  {Zamora}, {Zarrouk}, {Zasowski}, {Zhang}, {Zhao}, {Zhao}, {Zheng}, {Zheng},
  {Zhu}, \& {Zou}}]{2020ApJS..249....3A}
{Ahumada}, R., {Prieto}, C.~A., {Almeida}, A., {et~al.} 2020, \apjs, 249, 3,
  \dodoi{10.3847/1538-4365/ab929e}

\bibitem[{Ahumada {et~al.}(2020)Ahumada, Prieto, Almeida, Anders, Anderson,
  Andrews, Anguiano, Arcodia, Armengaud, Aubert, Avila, {Avila-Reese}, Badenes,
  Balland, Barger, {Barrera-Ballesteros}, Basu, Bautista, Beaton, Beers,
  Benavides, Bender, Bernardi, Bershady, Beutler, Bidin, Bird, Bizyaev, Blanc,
  Blanton, Boquien, Borissova, Bovy, Brandt, Brinkmann, Brownstein, Bundy,
  Bureau, Burgasser, Burtin, {Cano-D{\'i}az}, Capasso, Cappellari, Carrera,
  Chabanier, Chaplin, Chapman, Cherinka, Chiappini, Doohyun~Choi, Chojnowski,
  Chung, Clerc, Coffey, Comerford, Comparat, {da Costa}, Cousinou, Covey,
  Crane, Cunha, Ilha, Dai, Damsted, Darling, Davidson, Davies, Dawson, De, {de
  la Macorra}, De~Lee, Queiroz, Deconto~Machado, {de la Torre}, Dell'Agli, {du
  Mas des Bourboux}, {Diamond-Stanic}, Dillon, Donor, Drory, Duckworth, Dwelly,
  Ebelke, Eftekharzadeh, Davis~Eigenbrot, Elsworth, Eracleous, Erfanianfar,
  Escoffier, Fan, Farr, {Fern{\'a}ndez-Trincado}, Feuillet, Finoguenov, Fofie,
  {Fraser-McKelvie}, Frinchaboy, Fromenteau, Fu, Galbany, Garcia,
  {Garc{\'i}a-Hern{\'a}ndez}, Oehmichen, Ge, Maia, Geisler, Gelfand, Goddy,
  {Gonzalez-Perez}, Grabowski, Green, Grier, Guo, Guy, Harding, Hasselquist,
  Hawken, Hayes, Hearty, Hekker, Hogg, Holtzman, Horta, Hou, Hsieh, Huber,
  Hunt, Chitham, Imig, Jaber, Angel, Johnson, Jones, J{\"o}nsson, Jullo, Kim,
  Kinemuchi, Kirkpatrick, Kite, Klaene, Kneib, Kollmeier, Kong, Kounkel,
  Krishnarao, Lacerna, Lan, Lane, Law, Le~Goff, Leung, Lewis, Li, Lian, Lin,
  Long, {Longa-Pe{\~n}a}, Lundgren, Lyke, Ted~Mackereth, MacLeod, Majewski,
  Manchado, Maraston, Martini, Masseron, Masters, Mathur, McDermid, Merloni,
  Merrifield, M{\'e}sz{\'a}ros, Miglio, Minniti, Minsley, Miyaji, Mohammad,
  Mosser, Mueller, Muna, {Mu{\~n}oz-Guti{\'e}rrez}, Myers, Nadathur, Nair,
  Nandra, {do Nascimento}, Nevin, Newman, Nidever, Nitschelm, Noterdaeme,
  O'Connell, Olmstead, Oravetz, Oravetz, Osorio, Pace, Padilla,
  {Palanque-Delabrouille}, Palicio, Pan, Pan, Parker, Paviot, Peirani,
  Ram{\'r}ez, Penny, Percival, {Perez-Fournon}, {P{\'e}rez-R{\`a}fols},
  Petitjean, Pieri, Pinsonneault, Poovelil, Povick, Prakash, {Price-Whelan},
  Raddick, Raichoor, Ray, Rembold, Rezaie, Riffel, Riffel, Rix, Robin,
  {Roman-Lopes}, {Rom{\'a}n-Z{\'u}{\~n}iga}, Rose, Ross, Rossi, Rowlands,
  Rubin, Salvato, S{\'a}nchez, {S{\'a}nchez-Menguiano}, {S{\'a}nchez-Gallego},
  Sayres, Schaefer, Schiavon, Schimoia, Schlafly, Schlegel, Schneider,
  Schultheis, Schwope, Seo, Serenelli, Shafieloo, Shamsi, Shao, Shen, Shetrone,
  Shirley, Aguirre, Simon, Skrutskie, Slosar, Smethurst, Sobeck, Sodi, Souto,
  Stark, Stassun, Steinmetz, Stello, Stermer, {Storchi-Bergmann}, Streblyanska,
  Stringfellow, Stutz, Su{\'a}rez, Sun, {Taghizadeh-Popp}, Talbot, Tayar,
  Thakar, Theriault, Thomas, Thomas, Tinker, Tojeiro, Toledo, Tremonti, Troup,
  Tuttle, {Unda-Sanzana}, Valentini, {Vargas-Gonz{\'a}lez},
  {Vargas-Maga{\~n}a}, {V{\'a}zquez-Mata}, Vivek, Wake, Wang, Weaver, Weijmans,
  Wild, Wilson, Wilson, Wolthuis, {Wood-Vasey}, Yan, Yang, Y{\`e}che, Zamora,
  Zarrouk, Zasowski, Zhang, Zhao, Zhao, Zheng, Zheng, Zhu, \&
  Zou}]{ahumada2020}
Ahumada, R., Prieto, C.~A., Almeida, A., {et~al.} 2020, The Astrophysical
  Journal Supplement Series, 249, 3, \dodoi{10.3847/1538-4365/ab929e}

\bibitem[{{Aihara} {et~al.}(2018){Aihara}, {Arimoto}, {Armstrong}, {Arnouts},
  {Bahcall}, {Bickerton}, {Bosch}, {Bundy}, {Capak}, {Chan}, {Chiba}, {Coupon},
  {Egami}, {Enoki}, {Finet}, {Fujimori}, {Fujimoto}, {Furusawa}, {Furusawa},
  {Goto}, {Goulding}, {Greco}, {Greene}, {Gunn}, {Hamana}, {Harikane},
  {Hashimoto}, {Hattori}, {Hayashi}, {Hayashi}, {He{\l}miniak}, {Higuchi},
  {Hikage}, {Ho}, {Hsieh}, {Huang}, {Huang}, {Ikeda}, {Imanishi}, {Inoue},
  {Iwasawa}, {Iwata}, {Jaelani}, {Jian}, {Kamata}, {Karoji}, {Kashikawa},
  {Katayama}, {Kawanomoto}, {Kayo}, {Koda}, {Koike}, {Kojima}, {Komiyama},
  {Konno}, {Koshida}, {Koyama}, {Kusakabe}, {Leauthaud}, {Lee}, {Lin}, {Lin},
  {Lupton}, {Mandelbaum}, {Matsuoka}, {Medezinski}, {Mineo}, {Miyama},
  {Miyatake}, {Miyazaki}, {Momose}, {More}, {More}, {Moritani}, {Moriya},
  {Morokuma}, {Mukae}, {Murata}, {Murayama}, {Nagao}, {Nakata}, {Niida},
  {Niikura}, {Nishizawa}, {Obuchi}, {Oguri}, {Oishi}, {Okabe}, {Okamoto},
  {Okura}, {Ono}, {Onodera}, {Onoue}, {Osato}, {Ouchi}, {Price}, {Pyo}, {Sako},
  {Sawicki}, {Shibuya}, {Shimasaku}, {Shimono}, {Shirasaki}, {Silverman},
  {Simet}, {Speagle}, {Spergel}, {Strauss}, {Sugahara}, {Sugiyama}, {Suto},
  {Suyu}, {Suzuki}, {Tait}, {Takada}, {Takata}, {Tamura}, {Tanaka}, {Tanaka},
  {Tanaka}, {Tanaka}, {Terai}, {Terashima}, {Toba}, {Tominaga}, {Toshikawa},
  {Turner}, {Uchida}, {Uchiyama}, {Umetsu}, {Uraguchi}, {Urata}, {Usuda},
  {Utsumi}, {Wang}, {Wang}, {Wong}, {Yabe}, {Yamada}, {Yamanoi}, {Yasuda},
  {Yeh}, {Yonehara}, \& {Yuma}}]{Aihara18}
{Aihara}, H., {Arimoto}, N., {Armstrong}, R., {et~al.} 2018, \pasj, 70, S4,
  \dodoi{10.1093/pasj/psx066}

\bibitem[{Aihara {et~al.}(2019)Aihara, AlSayyad, Ando, Armstrong, Bosch, Egami,
  Furusawa, Furusawa, Goulding, Harikane, Hikage, Ho, Hsieh, Huang, Ikeda,
  Imanishi, Ito, Iwata, Jaelani, Kakuma, Kawana, Kikuta, Kobayashi, Koike,
  Komiyama, Li, Liang, Lin, Luo, Lupton, Lust, MacArthur, Matsuoka, Mineo,
  Miyatake, Miyazaki, More, Murata, Namiki, Nishizawa, Oguri, Okabe, Okamoto,
  Okura, Ono, Onodera, Onoue, Osato, Ouchi, Shibuya, Strauss, Sugiyama, Suto,
  Takada, Takagi, Takata, Takita, Tanaka, Terai, Toba, Uchiyama, Utsumi, Wang,
  Wang, \& Yamada}]{HSC_DR2}
Aihara, H., AlSayyad, Y., Ando, M., {et~al.} 2019, arXiv:1905.12221 [astro-ph],
  \dodoi{10.1093/pasj/psz103}

\bibitem[{{Allevato} {et~al.}(2013){Allevato}, {Paolillo}, {Papadakis}, \&
  {Pinto}}]{2013ApJ...771....9A}
{Allevato}, V., {Paolillo}, M., {Papadakis}, I., \& {Pinto}, C. 2013, \apj,
  771, 9, \dodoi{10.1088/0004-637X/771/1/9}

\bibitem[{Annis {et~al.}(2014)Annis, {Soares-Santos}, Strauss, Becker,
  Dodelson, Fan, Gunn, Hao, Ivezi{\'c}, Jester, Jiang, Johnston, Kubo,
  Lampeitl, Lin, Lupton, Miknaitis, Seo, Simet, \& Yanny}]{annis2014}
Annis, J., {Soares-Santos}, M., Strauss, M.~A., {et~al.} 2014, ApJ, 794, 120,
  \dodoi{10.1088/0004-637X/794/2/120}

\bibitem[{{Antonucci}(1993)}]{1993ARA&A..31..473A}
{Antonucci}, R. 1993, \araa, 31, 473,
  \dodoi{10.1146/annurev.aa.31.090193.002353}

\bibitem[{{Ba{\~n}ados} {et~al.}(2016){Ba{\~n}ados}, {Venemans}, {Decarli},
  {Farina}, {Mazzucchelli}, {Walter}, {Fan}, {Stern}, {Schlafly}, {Chambers},
  {Rix}, {Jiang}, {McGreer}, {Simcoe}, {Wang}, {Yang}, {Morganson}, {De Rosa},
  {Greiner}, {Balokovi{\'c}}, {Burgett}, {Cooper}, {Draper}, {Flewelling},
  {Hodapp}, {Jun}, {Kaiser}, {Kudritzki}, {Magnier}, {Metcalfe}, {Miller},
  {Schindler}, {Tonry}, {Wainscoat}, {Waters}, \& {Yang}}]{Banados16}
{Ba{\~n}ados}, E., {Venemans}, B.~P., {Decarli}, R., {et~al.} 2016, \apjs, 227,
  11, \dodoi{10.3847/0067-0049/227/1/11}

\bibitem[{{Banerji} {et~al.}(2010){Banerji}, {Lahav}, {Lintott}, {Abdalla},
  {Schawinski}, {Bamford}, {Andreescu}, {Murray}, {Raddick}, {Slosar},
  {Szalay}, {Thomas}, \& {Vandenberg}}]{2010MNRAS.406..342B}
{Banerji}, M., {Lahav}, O., {Lintott}, C.~J., {et~al.} 2010, \mnras, 406, 342,
  \dodoi{10.1111/j.1365-2966.2010.16713.x}

\bibitem[{{Baron}(2019)}]{2019arXiv190407248B}
{Baron}, D. 2019, arXiv e-prints, arXiv:1904.07248.
\newblock \doarXiv{1904.07248}

\bibitem[{{Bellm}(2014)}]{2014htu..conf...27B}
{Bellm}, E. 2014, in The Third Hot-wiring the Transient Universe Workshop, ed.
  P.~R. {Wozniak}, M.~J. {Graham}, A.~A. {Mahabal}, \& R.~{Seaman}, 27--33.
\newblock \doarXiv{1410.8185}

\bibitem[{{Bernstein} {et~al.}(2018){Bernstein}, {Abbott}, {Armstrong},
  {Burke}, {Diehl}, {Gruendl}, {Johnson}, {Li}, {Rykoff}, {Walker}, {Wester},
  \& {Yanny}}]{2018PASP..130e4501B}
{Bernstein}, G.~M., {Abbott}, T.~M.~C., {Armstrong}, R., {et~al.} 2018, \pasp,
  130, 054501, \dodoi{10.1088/1538-3873/aaa753}

\bibitem[{Berry {et~al.}(2019)Berry, Mohamed, \& Yap}]{10.5555/3379017}
Berry, M.~W., Mohamed, A., \& Yap, B.~W. 2019, Supervised and Unsupervised
  Learning for Data Science, 1st edn. (Springer Publishing Company,
  Incorporated)

\bibitem[{{Bianco} {et~al.}(2022){Bianco}, {Ivezi{\'c}}, {Jones}, {Graham},
  {Marshall}, {Saha}, {Strauss}, {Yoachim}, {Ribeiro}, {Anguita}, {Bauer},
  {Bauer}, {Bellm}, {Blum}, {Brandt}, {Brough}, {Catelan}, {Clarkson},
  {Connolly}, {Gawiser}, {Gizis}, {Hlo{\v{z}}ek}, {Kaviraj}, {Liu}, {Lochner},
  {Mahabal}, {Mandelbaum}, {McGehee}, {Neilsen}, {Olsen}, {Peiris}, {Rhodes},
  {Richards}, {Ridgway}, {Schwamb}, {Scolnic}, {Shemmer}, {Slater}, {Slosar},
  {Smartt}, {Strader}, {Street}, {Trilling}, {Verma}, {Vivas}, {Wechsler}, \&
  {Willman}}]{2022ApJS..258....1B}
{Bianco}, F.~B., {Ivezi{\'c}}, {\v{Z}}., {Jones}, R.~L., {et~al.} 2022, \apjs,
  258, 1, \dodoi{10.3847/1538-4365/ac3e72}

\bibitem[{{Bonoli} {et~al.}(1979){Bonoli}, {Braccesi}, {Federici}, {Zitelli},
  \& {Formiggini}}]{1979A&AS...35..391B}
{Bonoli}, F., {Braccesi}, A., {Federici}, L., {Zitelli}, V., \& {Formiggini},
  L. 1979, \aaps, 35, 391

\bibitem[{{Bosch} {et~al.}(2018){Bosch}, {Armstrong}, {Bickerton}, {Furusawa},
  {Ikeda}, {Koike}, {Lupton}, {Mineo}, {Price}, {Takata}, {Tanaka}, {Yasuda},
  {AlSayyad}, {Becker}, {Coulton}, {Coupon}, {Garmilla}, {Huang}, {Krughoff},
  {Lang}, {Leauthaud}, {Lim}, {Lust}, {MacArthur}, {Mandelbaum}, {Miyatake},
  {Miyazaki}, {Murata}, {More}, {Okura}, {Owen}, {Swinbank}, {Strauss},
  {Yamada}, \& {Yamanoi}}]{2018PASJ...70S...5B}
{Bosch}, J., {Armstrong}, R., {Bickerton}, S., {et~al.} 2018, \pasj, 70, S5,
  \dodoi{10.1093/pasj/psx080}

\bibitem[{{Bramich}(2008)}]{2008MNRAS.386L..77B}
{Bramich}, D.~M. 2008, \mnras, 386, L77,
  \dodoi{10.1111/j.1745-3933.2008.00464.x}

\bibitem[{{Brandt} \& {Alexander}(2015)}]{2015A&ARv..23....1B}
{Brandt}, W.~N., \& {Alexander}, D.~M. 2015, \aapr, 23, 1,
  \dodoi{10.1007/s00159-014-0081-z}

\bibitem[{{Buchner} {et~al.}(2021){Buchner}, {Salvato}, {Budav{\'a}ri}, \&
  {Fotopoulou}}]{2021ascl.soft02014B}
{Buchner}, J., {Salvato}, M., {Budav{\'a}ri}, T., \& {Fotopoulou}, S. 2021,
  {nway: Bayesian cross-matching of astronomical catalogs}, Astrophysics Source
  Code Library, record ascl:2102.014.
\newblock \doeprint{2102.014}

\bibitem[{{Budav{\'a}ri} \& {Szalay}(2008)}]{2008ApJ...679..301B}
{Budav{\'a}ri}, T., \& {Szalay}, A.~S. 2008, \apj, 679, 301,
  \dodoi{10.1086/587156}

\bibitem[{{Butler} \& {Bloom}(2011)}]{2011AJ....141...93B}
{Butler}, N.~R., \& {Bloom}, J.~S. 2011, \aj, 141, 93,
  \dodoi{10.1088/0004-6256/141/3/93}

\bibitem[{{Carballo} {et~al.}(2008){Carballo}, {Gonz{\'a}lez-Serrano}, {Benn},
  \& {Jim{\'e}nez-Luj{\'a}n}}]{2008MNRAS.391..369C}
{Carballo}, R., {Gonz{\'a}lez-Serrano}, J.~I., {Benn}, C.~R., \&
  {Jim{\'e}nez-Luj{\'a}n}, F. 2008, \mnras, 391, 369,
  \dodoi{10.1111/j.1365-2966.2008.13896.x}

\bibitem[{{Cavuoti} {et~al.}(2014){Cavuoti}, {Brescia}, {D'Abrusco}, {Longo},
  \& {Paolillo}}]{2014MNRAS.437..968C}
{Cavuoti}, S., {Brescia}, M., {D'Abrusco}, R., {Longo}, G., \& {Paolillo}, M.
  2014, \mnras, 437, 968, \dodoi{10.1093/mnras/stt1961}

\bibitem[{{Chambers} {et~al.}(2016){Chambers}, {Magnier}, {Metcalfe},
  {Flewelling}, {Huber}, {Waters}, {Denneau}, {Draper}, {Farrow}, {Finkbeiner},
  {Holmberg}, {Koppenhoefer}, {Price}, {Rest}, {Saglia}, {Schlafly}, {Smartt},
  {Sweeney}, {Wainscoat}, {Burgett}, {Chastel}, {Grav}, {Heasley}, {Hodapp},
  {Jedicke}, {Kaiser}, {Kudritzki}, {Luppino}, {Lupton}, {Monet}, {Morgan},
  {Onaka}, {Shiao}, {Stubbs}, {Tonry}, {White}, {Ba{\~n}ados}, {Bell},
  {Bender}, {Bernard}, {Boegner}, {Boffi}, {Botticella}, {Calamida},
  {Casertano}, {Chen}, {Chen}, {Cole}, {Deacon}, {Frenk}, {Fitzsimmons},
  {Gezari}, {Gibbs}, {Goessl}, {Goggia}, {Gourgue}, {Goldman}, {Grant},
  {Grebel}, {Hambly}, {Hasinger}, {Heavens}, {Heckman}, {Henderson}, {Henning},
  {Holman}, {Hopp}, {Ip}, {Isani}, {Jackson}, {Keyes}, {Koekemoer}, {Kotak},
  {Le}, {Liska}, {Long}, {Lucey}, {Liu}, {Martin}, {Masci}, {McLean}, {Mindel},
  {Misra}, {Morganson}, {Murphy}, {Obaika}, {Narayan}, {Nieto-Santisteban},
  {Norberg}, {Peacock}, {Pier}, {Postman}, {Primak}, {Rae}, {Rai}, {Riess},
  {Riffeser}, {Rix}, {R{\"o}ser}, {Russel}, {Rutz}, {Schilbach}, {Schultz},
  {Scolnic}, {Strolger}, {Szalay}, {Seitz}, {Small}, {Smith}, {Soderblom},
  {Taylor}, {Thomson}, {Taylor}, {Thakar}, {Thiel}, {Thilker}, {Unger},
  {Urata}, {Valenti}, {Wagner}, {Walder}, {Walter}, {Watters}, {Werner},
  {Wood-Vasey}, \& {Wyse}}]{Chambers16}
{Chambers}, K.~C., {Magnier}, E.~A., {Metcalfe}, N., {et~al.} 2016, arXiv
  e-prints, arXiv:1612.05560.
\newblock \doarXiv{1612.05560}

\bibitem[{{Chang} {et~al.}(2021){Chang}, {Hsieh}, {Wang}, {Lin}, {Lim}, {Toba},
  {Zhong}, \& {Chang}}]{2021ApJ...920...68C}
{Chang}, Y.-Y., {Hsieh}, B.-C., {Wang}, W.-H., {et~al.} 2021, \apj, 920, 68,
  \dodoi{10.3847/1538-4357/ac167c}

\bibitem[{{Chen} {et~al.}(2021){Chen}, {Goto}, {Kim}, {Wang}, {Santos}, {Ho},
  {Hashimoto}, {Poliszczuk}, {Pollo}, {Trippe}, {Miyaji}, {Toba}, {Malkan},
  {Serjeant}, {Pearson}, {Hwang}, {Kim}, {Shim}, {Lu}, {Hsiao}, {Huang},
  {Herrera-Endoqui}, {Bravo-Navarro}, \& {Matsuhara}}]{2021MNRAS.501.3951C}
{Chen}, B.~H., {Goto}, T., {Kim}, S.~J., {et~al.} 2021, \mnras, 501, 3951,
  \dodoi{10.1093/mnras/staa3865}

\bibitem[{{Chen} {et~al.}(2018){Chen}, {Brandt}, {Luo}, {Ranalli}, {Yang},
  {Alexander}, {Bauer}, {Kelson}, {Lacy}, {Nyland}, {Tozzi}, {Vito},
  {Cirasuolo}, {Gilli}, {Jarvis}, {Lehmer}, {Paolillo}, {Schneider}, {Shemmer},
  {Smail}, {Sun}, {Tanaka}, {Vaccari}, {Vignali}, {Xue}, {Banerji}, {Chow},
  {H{\"a}u{\ss}ler}, {Norris}, {Silverman}, \& {Trump}}]{2018MNRAS.478.2132C}
{Chen}, C. T.~J., {Brandt}, W.~N., {Luo}, B., {et~al.} 2018, \mnras, 478, 2132,
  \dodoi{10.1093/mnras/sty1036}

\bibitem[{{Chen} \& {Guestrin}(2016)}]{2016arXiv160302754C}
{Chen}, T., \& {Guestrin}, C. 2016, arXiv e-prints, arXiv:1603.02754.
\newblock \doarXiv{1603.02754}

\bibitem[{Chollet {et~al.}(2015)}]{chollet2015keras}
Chollet, F., {et~al.} 2015, Keras,  GitHub.
\newblock \url{https://github.com/fchollet/keras}

\bibitem[{{Clarke} {et~al.}(2020){Clarke}, {Scaife}, {Greenhalgh}, \&
  {Griguta}}]{2020A&A...639A..84C}
{Clarke}, A.~O., {Scaife}, A.~M.~M., {Greenhalgh}, R., \& {Griguta}, V. 2020,
  \aap, 639, A84, \dodoi{10.1051/0004-6361/201936770}

\bibitem[{Cortes \& Vapnik(1995)}]{cortes1995support}
Cortes, C., \& Vapnik, V. 1995, Machine learning, 20, 273

\bibitem[{Cybenko(1989)}]{Cybenko1989ApproximationBS}
Cybenko, G.~V. 1989, Mathematics of Control, Signals and Systems, 2, 303

\bibitem[{{Czerny} {et~al.}(2022){Czerny}, {Cao}, {Jaiswal}, {Karas}, {Khadka},
  {Mart{\'\i}nez-Aldama}, {Naddaf}, {Panda}, {Pozo Nu{\~n}ez}, {Prince},
  {Ratra}, {Sniegowska}, {Yu}, \& {Zaja{\v{c}}ek}}]{2022arXiv220906563C}
{Czerny}, B., {Cao}, S., {Jaiswal}, V.~K., {et~al.} 2022, arXiv e-prints,
  arXiv:2209.06563.
\newblock \doarXiv{2209.06563}

\bibitem[{{Czerny} {et~al.}(2023){Czerny}, {Panda}, {Prince}, {Jaiswal},
  {Zajacek}, {Martinez Aldama}, {Kozlowski}, {Kovacevic}, {Ilic}, {Popovic},
  {Pozo Nunez}, {Hoenig}, \& {Brandt}}]{2023arXiv230108975C}
{Czerny}, B., {Panda}, S., {Prince}, R., {et~al.} 2023, arXiv e-prints,
  arXiv:2301.08975, \dodoi{10.48550/arXiv.2301.08975}

\bibitem[{{Dark Energy Survey Collaboration} {et~al.}(2016){Dark Energy Survey
  Collaboration}, Abbott, Abdalla, Aleksi{\'c}, Allam, Amara, Bacon, Balbinot,
  Banerji, Bechtol, {Benoit-L{\'e}vy}, Bernstein, Bertin, Blazek, Bonnett,
  Bridle, Brooks, Brunner, {Buckley-Geer}, Burke, Caminha, Capozzi, Carlsen,
  {Carnero-Rosell}, Carollo, {Carrasco-Kind}, Carretero, Castander, Clerkin,
  Collett, Conselice, Crocce, Cunha, D'Andrea, {da Costa}, Davis, Desai, Diehl,
  Dietrich, Dodelson, Doel, {Drlica-Wagner}, Estrada, Etherington, Evrard,
  Fabbri, Finley, Flaugher, Foley, Fosalba, Frieman, {Garc{\'i}a-Bellido},
  Gaztanaga, Gerdes, Giannantonio, Goldstein, Gruen, Gruendl, Guarnieri,
  Gutierrez, Hartley, Honscheid, Jain, James, Jeltema, Jouvel, Kessler, King,
  Kirk, Kron, Kuehn, Kuropatkin, Lahav, Li, Lima, Lin, Maia, Makler, Manera,
  Maraston, Marshall, Martini, McMahon, Melchior, Merson, Miller, Miquel, Mohr,
  {Morice-Atkinson}, Naidoo, Neilsen, Nichol, Nord, Ogando, Ostrovski, Palmese,
  Papadopoulos, Peiris, Peoples, Percival, Plazas, Reed, Refregier, Romer,
  Roodman, Ross, Rozo, Rykoff, Sadeh, Sako, S{\'a}nchez, Sanchez, Santiago,
  Scarpine, Schubnell, {Sevilla-Noarbe}, Sheldon, Smith, Smith,
  {Soares-Santos}, Sobreira, Soumagnac, Suchyta, Sullivan, Swanson, Tarle,
  Thaler, Thomas, Thomas, Tucker, Vieira, Vikram, Walker, Wechsler, Weller,
  Wester, Whiteway, Wilcox, Yanny, Zhang, \& Zuntz}]{DES2016}
{Dark Energy Survey Collaboration}, Abbott, T., Abdalla, F.~B., {et~al.} 2016,
  Mon Not R Astron Soc, 460, 1270, \dodoi{10.1093/mnras/stw641}

\bibitem[{{De Cicco} {et~al.}(2019){De Cicco}, {Paolillo}, {Falocco},
  {Poulain}, {Brandt}, {Bauer}, {Vagnetti}, {Longo}, {Grado}, {Ragosta},
  {Botticella}, {Pignata}, {Vaccari}, {Radovich}, {Salvato}, {Covone},
  {Napolitano}, {Marchetti}, \& {Schipani}}]{2019A&A...627A..33D}
{De Cicco}, D., {Paolillo}, M., {Falocco}, S., {et~al.} 2019, \aap, 627, A33,
  \dodoi{10.1051/0004-6361/201935659}

\bibitem[{{De Cicco} {et~al.}(2021){De Cicco}, {Bauer}, {Paolillo}, {Cavuoti},
  {S{\'a}nchez-S{\'a}ez}, {Brandt}, {Pignata}, {Vaccari}, \&
  {Radovich}}]{2021A&A...645A.103D}
{De Cicco}, D., {Bauer}, F.~E., {Paolillo}, M., {et~al.} 2021, \aap, 645, A103,
  \dodoi{10.1051/0004-6361/202039193}

\bibitem[{{Delgado} {et~al.}(2014){Delgado}, {Saha}, {Chandrasekharan}, {Cook},
  {Petry}, \& {Ridgway}}]{2014SPIE.9150E..15D}
{Delgado}, F., {Saha}, A., {Chandrasekharan}, S., {et~al.} 2014, in Society of
  Photo-Optical Instrumentation Engineers (SPIE) Conference Series, Vol. 9150,
  Society of Photo-Optical Instrumentation Engineers (SPIE) Conference Series,
  15, \dodoi{10.1117/12.2056898}

\bibitem[{{DES Collaboration} {et~al.}(2021){DES Collaboration}, Abbott,
  Adamow, Aguena, Allam, Amon, Annis, Avila, Bacon, Banerji, Bechtol, Becker,
  Bernstein, Bertin, Bhargava, Bridle, Brooks, Burke, Rosell, Kind, Carretero,
  Castander, Cawthon, Chang, Choi, Conselice, Costanzi, Crocce, {da Costa},
  Davis, De~Vicente, DeRose, Desai, Diehl, Dietrich, {Drlica-Wagner}, Eckert,
  {Elvin-Poole}, Everett, Evrard, Ferrero, Fert{\'e}, Flaugher, Fosalba,
  Friedel, Frieman, {Garc{\'i}a-Bellido}, Gelman, Gerdes, Giannantonio, Gill,
  Gruen, Gruendl, Gschwend, Gutierrez, Hartley, Hinton, Hollowood, Honscheid,
  Huterer, James, Jeltema, Johnson, Kent, Kron, Kuehn, Kuropatkin, Lahav, Li,
  Lidman, Lin, MacCrann, Maia, Manning, March, Marshall, Martini, Melchior,
  Menanteau, Miquel, Morgan, Myles, Neilsen, Ogando, Palmese,
  {Paz-Chinch{\'o}n}, Petravick, Pieres, Plazas, Pond, {Rodriguez-Monroy},
  Romer, Roodman, Rykoff, Sako, Sanchez, Santiago, Serrano, {Sevilla-Noarbe},
  Smith, Smith, {Soares-Santos}, Suchyta, Swanson, Tarle, Thomas, To, Tremblay,
  Troxel, Tucker, Turner, Varga, Walker, Wechsler, Weller, Wester, Wilkinson,
  Yanny, Zhang, Nikutta, Fitzpatrick, Jacques, Scott, Olsen, Huang, Herrera,
  Juneau, Nidever, Weaver, Adean, Correia, {de Freitas}, Freitas, Singulani, \&
  {Vila-Verde}}]{descollaboration2021}
{DES Collaboration}, Abbott, T. M.~C., Adamow, M., {et~al.} 2021,
  arXiv:2101.05765 [astro-ph].
\newblock \doeprint{2101.05765}

\bibitem[{{Dey} {et~al.}(2019){Dey}, {Schlegel}, {Lang}, {Blum}, {Burleigh},
  {Fan}, {Findlay}, {Finkbeiner}, {Herrera}, {Juneau}, {Landriau}, {Levi},
  {McGreer}, {Meisner}, {Myers}, {Moustakas}, {Nugent}, {Patej}, {Schlafly},
  {Walker}, {Valdes}, {Weaver}, {Y{\`e}che}, {Zou}, {Zhou}, {Abareshi},
  {Abbott}, {Abolfathi}, {Aguilera}, {Alam}, {Allen}, {Alvarez}, {Annis},
  {Ansarinejad}, {Aubert}, {Beechert}, {Bell}, {BenZvi}, {Beutler}, {Bielby},
  {Bolton}, {Brice{\~n}o}, {Buckley-Geer}, {Butler}, {Calamida}, {Carlberg},
  {Carter}, {Casas}, {Castander}, {Choi}, {Comparat}, {Cukanovaite}, {Delubac},
  {DeVries}, {Dey}, {Dhungana}, {Dickinson}, {Ding}, {Donaldson}, {Duan},
  {Duckworth}, {Eftekharzadeh}, {Eisenstein}, {Etourneau}, {Fagrelius},
  {Farihi}, {Fitzpatrick}, {Font-Ribera}, {Fulmer}, {G{\"a}nsicke},
  {Gaztanaga}, {George}, {Gerdes}, {Gontcho}, {Gorgoni}, {Green}, {Guy},
  {Harmer}, {Hernandez}, {Honscheid}, {Huang}, {James}, {Jannuzi}, {Jiang},
  {Joyce}, {Karcher}, {Karkar}, {Kehoe}, {Kneib}, {Kueter-Young}, {Lan},
  {Lauer}, {Le Guillou}, {Le Van Suu}, {Lee}, {Lesser}, {Perreault Levasseur},
  {Li}, {Mann}, {Marshall}, {Mart{\'\i}nez-V{\'a}zquez}, {Martini}, {du Mas des
  Bourboux}, {McManus}, {Meier}, {M{\'e}nard}, {Metcalfe},
  {Mu{\~n}oz-Guti{\'e}rrez}, {Najita}, {Napier}, {Narayan}, {Newman}, {Nie},
  {Nord}, {Norman}, {Olsen}, {Paat}, {Palanque-Delabrouille}, {Peng},
  {Poppett}, {Poremba}, {Prakash}, {Rabinowitz}, {Raichoor}, {Rezaie},
  {Robertson}, {Roe}, {Ross}, {Ross}, {Rudnick}, {Safonova}, {Saha},
  {S{\'a}nchez}, {Savary}, {Schweiker}, {Scott}, {Seo}, {Shan}, {Silva},
  {Slepian}, {Soto}, {Sprayberry}, {Staten}, {Stillman}, {Stupak}, {Summers},
  {Sien Tie}, {Tirado}, {Vargas-Maga{\~n}a}, {Vivas}, {Wechsler}, {Williams},
  {Yang}, {Yang}, {Yapici}, {Zaritsky}, {Zenteno}, {Zhang}, {Zhang}, {Zhou}, \&
  {Zhou}}]{Dey19}
{Dey}, A., {Schlegel}, D.~J., {Lang}, D., {et~al.} 2019, \aj, 157, 168,
  \dodoi{10.3847/1538-3881/ab089d}

\bibitem[{{Doert} \& {Errando}(2014)}]{2014ApJ...782...41D}
{Doert}, M., \& {Errando}, M. 2014, \apj, 782, 41,
  \dodoi{10.1088/0004-637X/782/1/41}

\bibitem[{{Doorenbos} {et~al.}(2022){Doorenbos}, {Torbaniuk}, {Cavuoti},
  {Paolillo}, {Longo}, {Brescia}, {Sznitman}, \&
  {M{\'a}rquez-Neila}}]{2022A&A...666A.171D}
{Doorenbos}, L., {Torbaniuk}, O., {Cavuoti}, S., {et~al.} 2022, \aap, 666,
  A171, \dodoi{10.1051/0004-6361/202243900}

\bibitem[{{Dye} {et~al.}(2018){Dye}, {Lawrence}, {Read}, {Fan}, {Kerr},
  {Varricatt}, {Furnell}, {Edge}, {Irwin}, {Hambly}, {Lucas}, {Almaini},
  {Chambers}, {Green}, {Hewett}, {Liu}, {McGreer}, {Best}, {Zhang}, {Sutorius},
  {Froebrich}, {Magnier}, {Hasinger}, {Lederer}, {Bold}, \& {Tedds}}]{Dye18}
{Dye}, S., {Lawrence}, A., {Read}, M.~A., {et~al.} 2018, \mnras, 473, 5113,
  \dodoi{10.1093/mnras/stx2622}

\bibitem[{{Eckert} {et~al.}(2021){Eckert}, {Gaspari}, {Gastaldello}, {Le Brun},
  \& {O'Sullivan}}]{2021Univ....7..142E}
{Eckert}, D., {Gaspari}, M., {Gastaldello}, F., {Le Brun}, A. M.~C., \&
  {O'Sullivan}, E. 2021, Universe, 7, 142, \dodoi{10.3390/universe7050142}

\bibitem[{{Fabian}(2012)}]{2012ARA&A..50..455F}
{Fabian}, A.~C. 2012, \araa, 50, 455,
  \dodoi{10.1146/annurev-astro-081811-125521}

\bibitem[{{Fan} {et~al.}(2006){Fan}, {Strauss}, {Richards}, {Hennawi},
  {Becker}, {White}, {Diamond-Stanic}, {Donley}, {Jiang}, {Kim}, {Vestergaard},
  {Young}, {Gunn}, {Lupton}, {Knapp}, {Schneider}, {Brandt}, {Bahcall},
  {Barentine}, {Brinkmann}, {Brewington}, {Fukugita}, {Harvanek}, {Kleinman},
  {Krzesinski}, {Long}, {Neilsen}, {Nitta}, {Snedden}, \& {Voges}}]{Fan06}
{Fan}, X., {Strauss}, M.~A., {Richards}, G.~T., {et~al.} 2006, \aj, 131, 1203,
  \dodoi{10.1086/500296}

\bibitem[{{Ferrarese} \& {Merritt}(2000)}]{2000ApJ...539L...9F}
{Ferrarese}, L., \& {Merritt}, D. 2000, \apjl, 539, L9, \dodoi{10.1086/312838}

\bibitem[{{Gaia Collaboration} {et~al.}(2016){Gaia Collaboration}, {Prusti},
  {de Bruijne}, {Brown}, {Vallenari}, {Babusiaux}, {Bailer-Jones}, {Bastian},
  {Biermann}, {Evans}, {Eyer}, {Jansen}, {Jordi}, {Klioner}, {Lammers},
  {Lindegren}, {Luri}, {Mignard}, {Milligan}, {Panem}, {Poinsignon},
  {Pourbaix}, {Randich}, {Sarri}, {Sartoretti}, {Siddiqui}, {Soubiran},
  {Valette}, {van Leeuwen}, {Walton}, {Aerts}, {Arenou}, {Cropper}, {Drimmel},
  {H{\o}g}, {Katz}, {Lattanzi}, {O'Mullane}, {Grebel}, {Holland}, {Huc},
  {Passot}, {Bramante}, {Cacciari}, {Casta{\~n}eda}, {Chaoul}, {Cheek}, {De
  Angeli}, {Fabricius}, {Guerra}, {Hern{\'a}ndez}, {Jean-Antoine-Piccolo},
  {Masana}, {Messineo}, {Mowlavi}, {Nienartowicz}, {Ord{\'o}{\~n}ez-Blanco},
  {Panuzzo}, {Portell}, {Richards}, {Riello}, {Seabroke}, {Tanga},
  {Th{\'e}venin}, {Torra}, {Els}, {Gracia-Abril}, {Comoretto},
  {Garcia-Reinaldos}, {Lock}, {Mercier}, {Altmann}, {Andrae}, {Astraatmadja},
  {Bellas-Velidis}, {Benson}, {Berthier}, {Blomme}, {Busso}, {Carry},
  {Cellino}, {Clementini}, {Cowell}, {Creevey}, {Cuypers}, {Davidson}, {De
  Ridder}, {de Torres}, {Delchambre}, {Dell'Oro}, {Ducourant}, {Fr{\'e}mat},
  {Garc{\'\i}a-Torres}, {Gosset}, {Halbwachs}, {Hambly}, {Harrison}, {Hauser},
  {Hestroffer}, {Hodgkin}, {Huckle}, {Hutton}, {Jasniewicz}, {Jordan},
  {Kontizas}, {Korn}, {Lanzafame}, {Manteiga}, {Moitinho}, {Muinonen},
  {Osinde}, {Pancino}, {Pauwels}, {Petit}, {Recio-Blanco}, {Robin}, {Sarro},
  {Siopis}, {Smith}, {Smith}, {Sozzetti}, {Thuillot}, {van Reeven}, {Viala},
  {Abbas}, {Abreu Aramburu}, {Accart}, {Aguado}, {Allan}, {Allasia},
  {Altavilla}, {{\'A}lvarez}, {Alves}, {Anderson}, {Andrei}, {Anglada Varela},
  {Antiche}, {Antoja}, {Ant{\'o}n}, {Arcay}, {Atzei}, {Ayache}, {Bach},
  {Baker}, {Balaguer-N{\'u}{\~n}ez}, {Barache}, {Barata}, {Barbier}, {Barblan},
  {Baroni}, {Barrado y Navascu{\'e}s}, {Barros}, {Barstow}, {Becciani},
  {Bellazzini}, {Bellei}, {Bello Garc{\'\i}a}, {Belokurov}, {Bendjoya},
  {Berihuete}, {Bianchi}, {Bienaym{\'e}}, {Billebaud}, {Blagorodnova},
  {Blanco-Cuaresma}, {Boch}, {Bombrun}, {Borrachero}, {Bouquillon}, {Bourda},
  {Bouy}, {Bragaglia}, {Breddels}, {Brouillet}, {Br{\"u}semeister},
  {Bucciarelli}, {Budnik}, {Burgess}, {Burgon}, {Burlacu}, {Busonero}, {Buzzi},
  {Caffau}, {Cambras}, {Campbell}, {Cancelliere}, {Cantat-Gaudin}, {Carlucci},
  {Carrasco}, {Castellani}, {Charlot}, {Charnas}, {Charvet}, {Chassat},
  {Chiavassa}, {Clotet}, {Cocozza}, {Collins}, {Collins}, {Costigan}, {Crifo},
  {Cross}, {Crosta}, {Crowley}, {Dafonte}, {Damerdji}, {Dapergolas}, {David},
  {David}, {De Cat}, {de Felice}, {de Laverny}, {De Luise}, {De March}, {de
  Martino}, {de Souza}, {Debosscher}, {del Pozo}, {Delbo}, {Delgado},
  {Delgado}, {di Marco}, {Di Matteo}, {Diakite}, {Distefano}, {Dolding}, {Dos
  Anjos}, {Drazinos}, {Dur{\'a}n}, {Dzigan}, {Ecale}, {Edvardsson}, {Enke},
  {Erdmann}, {Escolar}, {Espina}, {Evans}, {Eynard Bontemps}, {Fabre},
  {Fabrizio}, {Faigler}, {Falc{\~a}o}, {Farr{\`a}s Casas}, {Faye}, {Federici},
  {Fedorets}, {Fern{\'a}ndez-Hern{\'a}ndez}, {Fernique}, {Fienga}, {Figueras},
  {Filippi}, {Findeisen}, {Fonti}, {Fouesneau}, {Fraile}, {Fraser}, {Fuchs},
  {Furnell}, {Gai}, {Galleti}, {Galluccio}, {Garabato}, {Garc{\'\i}a-Sedano},
  {Gar{\'e}}, {Garofalo}, {Garralda}, {Gavras}, {Gerssen}, {Geyer}, {Gilmore},
  {Girona}, {Giuffrida}, {Gomes}, {Gonz{\'a}lez-Marcos},
  {Gonz{\'a}lez-N{\'u}{\~n}ez}, {Gonz{\'a}lez-Vidal}, {Granvik}, {Guerrier},
  {Guillout}, {Guiraud}, {G{\'u}rpide}, {Guti{\'e}rrez-S{\'a}nchez}, {Guy},
  {Haigron}, {Hatzidimitriou}, {Haywood}, {Heiter}, {Helmi}, {Hobbs},
  {Hofmann}, {Holl}, {Holland}, {Hunt}, {Hypki}, {Icardi}, {Irwin}, {Jevardat
  de Fombelle}, {Jofr{\'e}}, {Jonker}, {Jorissen}, {Julbe}, {Karampelas},
  {Kochoska}, {Kohley}, {Kolenberg}, {Kontizas}, {Koposov}, {Kordopatis},
  {Koubsky}, {Kowalczyk}, {Krone-Martins}, {Kudryashova}, {Kull}, {Bachchan},
  {Lacoste-Seris}, {Lanza}, {Lavigne}, {Le Poncin-Lafitte}, {Lebreton},
  {Lebzelter}, {Leccia}, {Leclerc}, {Lecoeur-Taibi}, {Lemaitre}, {Lenhardt},
  {Leroux}, {Liao}, {Licata}, {Lindstr{\o}m}, {Lister}, {Livanou}, {Lobel},
  {L{\"o}ffler}, {L{\'o}pez}, {Lopez-Lozano}, {Lorenz}, {Loureiro},
  {MacDonald}, {Magalh{\~a}es Fernandes}, {Managau}, {Mann}, {Mantelet},
  {Marchal}, {Marchant}, {Marconi}, {Marie}, {Marinoni}, {Marrese},
  {Marschalk{\'o}}, {Marshall}, {Mart{\'\i}n-Fleitas}, {Martino}, {Mary},
  {Matijevi{\v{c}}}, {Mazeh}, {McMillan}, {Messina}, {Mestre}, {Michalik},
  {Millar}, {Miranda}, {Molina}, {Molinaro}, {Molinaro}, {Moln{\'a}r},
  {Moniez}, {Montegriffo}, {Monteiro}, {Mor}, {Mora}, {Morbidelli}, {Morel},
  {Morgenthaler}, {Morley}, {Morris}, {Mulone}, {Muraveva}, {Musella},
  {Narbonne}, {Nelemans}, {Nicastro}, {Noval}, {Ord{\'e}novic},
  {Ordieres-Mer{\'e}}, {Osborne}, {Pagani}, {Pagano}, {Pailler}, {Palacin},
  {Palaversa}, {Parsons}, {Paulsen}, {Pecoraro}, {Pedrosa}, {Pentik{\"a}inen},
  {Pereira}, {Pichon}, {Piersimoni}, {Pineau}, {Plachy}, {Plum}, {Poujoulet},
  {Pr{\v{s}}a}, {Pulone}, {Ragaini}, {Rago}, {Rambaux}, {Ramos-Lerate},
  {Ranalli}, {Rauw}, {Read}, {Regibo}, {Renk}, {Reyl{\'e}}, {Ribeiro},
  {Rimoldini}, {Ripepi}, {Riva}, {Rixon}, {Roelens}, {Romero-G{\'o}mez},
  {Rowell}, {Royer}, {Rudolph}, {Ruiz-Dern}, {Sadowski}, {Sagrist{\`a}
  Sell{\'e}s}, {Sahlmann}, {Salgado}, {Salguero}, {Sarasso}, {Savietto},
  {Schnorhk}, {Schultheis}, {Sciacca}, {Segol}, {Segovia}, {Segransan},
  {Serpell}, {Shih}, {Smareglia}, {Smart}, {Smith}, {Solano}, {Solitro},
  {Sordo}, {Soria Nieto}, {Souchay}, {Spagna}, {Spoto}, {Stampa}, {Steele},
  {Steidelm{\"u}ller}, {Stephenson}, {Stoev}, {Suess}, {S{\"u}veges}, {Surdej},
  {Szabados}, {Szegedi-Elek}, {Tapiador}, {Taris}, {Tauran}, {Taylor},
  {Teixeira}, {Terrett}, {Tingley}, {Trager}, {Turon}, {Ulla}, {Utrilla},
  {Valentini}, {van Elteren}, {Van Hemelryck}, {van Leeuwen}, {Varadi},
  {Vecchiato}, {Veljanoski}, {Via}, {Vicente}, {Vogt}, {Voss}, {Votruba},
  {Voutsinas}, {Walmsley}, {Weiler}, {Weingrill}, {Werner}, {Wevers},
  {Whitehead}, {Wyrzykowski}, {Yoldas}, {{\v{Z}}erjal}, {Zucker}, {Zurbach},
  {Zwitter}, {Alecu}, {Allen}, {Allende Prieto}, {Amorim},
  {Anglada-Escud{\'e}}, {Arsenijevic}, {Azaz}, {Balm}, {Beck}, {Bernstein},
  {Bigot}, {Bijaoui}, {Blasco}, {Bonfigli}, {Bono}, {Boudreault}, {Bressan},
  {Brown}, {Brunet}, {Bunclark}, {Buonanno}, {Butkevich}, {Carret}, {Carrion},
  {Chemin}, {Ch{\'e}reau}, {Corcione}, {Darmigny}, {de Boer}, {de Teodoro}, {de
  Zeeuw}, {Delle Luche}, {Domingues}, {Dubath}, {Fodor}, {Fr{\'e}zouls},
  {Fries}, {Fustes}, {Fyfe}, {Gallardo}, {Gallegos}, {Gardiol}, {Gebran},
  {Gomboc}, {G{\'o}mez}, {Grux}, {Gueguen}, {Heyrovsky}, {Hoar}, {Iannicola},
  {Isasi Parache}, {Janotto}, {Joliet}, {Jonckheere}, {Keil}, {Kim},
  {Klagyivik}, {Klar}, {Knude}, {Kochukhov}, {Kolka}, {Kos}, {Kutka}, {Lainey},
  {LeBouquin}, {Liu}, {Loreggia}, {Makarov}, {Marseille}, {Martayan},
  {Martinez-Rubi}, {Massart}, {Meynadier}, {Mignot}, {Munari}, {Nguyen},
  {Nordlander}, {Ocvirk}, {O'Flaherty}, {Olias Sanz}, {Ortiz}, {Osorio},
  {Oszkiewicz}, {Ouzounis}, {Palmer}, {Park}, {Pasquato}, {Peltzer}, {Peralta},
  {P{\'e}turaud}, {Pieniluoma}, {Pigozzi}, {Poels}, {Prat}, {Prod'homme},
  {Raison}, {Rebordao}, {Risquez}, {Rocca-Volmerange}, {Rosen}, {Ruiz-Fuertes},
  {Russo}, {Sembay}, {Serraller Vizcaino}, {Short}, {Siebert}, {Silva},
  {Sinachopoulos}, {Slezak}, {Soffel}, {Sosnowska}, {Strai{\v{z}}ys}, {ter
  Linden}, {Terrell}, {Theil}, {Tiede}, {Troisi}, {Tsalmantza}, {Tur},
  {Vaccari}, {Vachier}, {Valles}, {Van Hamme}, {Veltz}, {Virtanen}, {Wallut},
  {Wichmann}, {Wilkinson}, {Ziaeepour}, \& {Zschocke}}]{2016A&A...595A...1G}
{Gaia Collaboration}, {Prusti}, T., {de Bruijne}, J.~H.~J., {et~al.} 2016,
  \aap, 595, A1, \dodoi{10.1051/0004-6361/201629272}

\bibitem[{{Gaia Collaboration} {et~al.}(2018){Gaia Collaboration}, {Brown},
  {Vallenari}, {Prusti}, {de Bruijne}, {Babusiaux}, {Bailer-Jones}, {Biermann},
  {Evans}, {Eyer}, {Jansen}, {Jordi}, {Klioner}, {Lammers}, {Lindegren},
  {Luri}, {Mignard}, {Panem}, {Pourbaix}, {Randich}, {Sartoretti}, {Siddiqui},
  {Soubiran}, {van Leeuwen}, {Walton}, {Arenou}, {Bastian}, {Cropper},
  {Drimmel}, {Katz}, {Lattanzi}, {Bakker}, {Cacciari}, {Casta{\~n}eda},
  {Chaoul}, {Cheek}, {De Angeli}, {Fabricius}, {Guerra}, {Holl}, {Masana},
  {Messineo}, {Mowlavi}, {Nienartowicz}, {Panuzzo}, {Portell}, {Riello},
  {Seabroke}, {Tanga}, {Th{\'e}venin}, {Gracia-Abril}, {Comoretto},
  {Garcia-Reinaldos}, {Teyssier}, {Altmann}, {Andrae}, {Audard},
  {Bellas-Velidis}, {Benson}, {Berthier}, {Blomme}, {Burgess}, {Busso},
  {Carry}, {Cellino}, {Clementini}, {Clotet}, {Creevey}, {Davidson}, {De
  Ridder}, {Delchambre}, {Dell'Oro}, {Ducourant},
  {Fern{\'a}ndez-Hern{\'a}ndez}, {Fouesneau}, {Fr{\'e}mat}, {Galluccio},
  {Garc{\'\i}a-Torres}, {Gonz{\'a}lez-N{\'u}{\~n}ez}, {Gonz{\'a}lez-Vidal},
  {Gosset}, {Guy}, {Halbwachs}, {Hambly}, {Harrison}, {Hern{\'a}ndez},
  {Hestroffer}, {Hodgkin}, {Hutton}, {Jasniewicz}, {Jean-Antoine-Piccolo},
  {Jordan}, {Korn}, {Krone-Martins}, {Lanzafame}, {Lebzelter}, {L{\"o}ffler},
  {Manteiga}, {Marrese}, {Mart{\'\i}n-Fleitas}, {Moitinho}, {Mora}, {Muinonen},
  {Osinde}, {Pancino}, {Pauwels}, {Petit}, {Recio-Blanco}, {Richards},
  {Rimoldini}, {Robin}, {Sarro}, {Siopis}, {Smith}, {Sozzetti}, {S{\"u}veges},
  {Torra}, {van Reeven}, {Abbas}, {Abreu Aramburu}, {Accart}, {Aerts},
  {Altavilla}, {{\'A}lvarez}, {Alvarez}, {Alves}, {Anderson}, {Andrei},
  {Anglada Varela}, {Antiche}, {Antoja}, {Arcay}, {Astraatmadja}, {Bach},
  {Baker}, {Balaguer-N{\'u}{\~n}ez}, {Balm}, {Barache}, {Barata}, {Barbato},
  {Barblan}, {Barklem}, {Barrado}, {Barros}, {Barstow}, {Bartholom{\'e}
  Mu{\~n}oz}, {Bassilana}, {Becciani}, {Bellazzini}, {Berihuete}, {Bertone},
  {Bianchi}, {Bienaym{\'e}}, {Blanco-Cuaresma}, {Boch}, {Boeche}, {Bombrun},
  {Borrachero}, {Bossini}, {Bouquillon}, {Bourda}, {Bragaglia}, {Bramante},
  {Breddels}, {Bressan}, {Brouillet}, {Br{\"u}semeister}, {Brugaletta},
  {Bucciarelli}, {Burlacu}, {Busonero}, {Butkevich}, {Buzzi}, {Caffau},
  {Cancelliere}, {Cannizzaro}, {Cantat-Gaudin}, {Carballo}, {Carlucci},
  {Carrasco}, {Casamiquela}, {Castellani}, {Castro-Ginard}, {Charlot},
  {Chemin}, {Chiavassa}, {Cocozza}, {Costigan}, {Cowell}, {Crifo}, {Crosta},
  {Crowley}, {Cuypers}, {Dafonte}, {Damerdji}, {Dapergolas}, {David}, {David},
  {de Laverny}, {De Luise}, {De March}, {de Martino}, {de Souza}, {de Torres},
  {Debosscher}, {del Pozo}, {Delbo}, {Delgado}, {Delgado}, {Di Matteo},
  {Diakite}, {Diener}, {Distefano}, {Dolding}, {Drazinos}, {Dur{\'a}n},
  {Edvardsson}, {Enke}, {Eriksson}, {Esquej}, {Eynard Bontemps}, {Fabre},
  {Fabrizio}, {Faigler}, {Falc{\~a}o}, {Farr{\`a}s Casas}, {Federici},
  {Fedorets}, {Fernique}, {Figueras}, {Filippi}, {Findeisen}, {Fonti},
  {Fraile}, {Fraser}, {Fr{\'e}zouls}, {Gai}, {Galleti}, {Garabato},
  {Garc{\'\i}a-Sedano}, {Garofalo}, {Garralda}, {Gavel}, {Gavras}, {Gerssen},
  {Geyer}, {Giacobbe}, {Gilmore}, {Girona}, {Giuffrida}, {Glass}, {Gomes},
  {Granvik}, {Gueguen}, {Guerrier}, {Guiraud}, {Guti{\'e}rrez-S{\'a}nchez},
  {Haigron}, {Hatzidimitriou}, {Hauser}, {Haywood}, {Heiter}, {Helmi}, {Heu},
  {Hilger}, {Hobbs}, {Hofmann}, {Holland}, {Huckle}, {Hypki}, {Icardi},
  {Jan{\ss}en}, {Jevardat de Fombelle}, {Jonker}, {Juh{\'a}sz}, {Julbe},
  {Karampelas}, {Kewley}, {Klar}, {Kochoska}, {Kohley}, {Kolenberg},
  {Kontizas}, {Kontizas}, {Koposov}, {Kordopatis}, {Kostrzewa-Rutkowska},
  {Koubsky}, {Lambert}, {Lanza}, {Lasne}, {Lavigne}, {Le Fustec}, {Le
  Poncin-Lafitte}, {Lebreton}, {Leccia}, {Leclerc}, {Lecoeur-Taibi},
  {Lenhardt}, {Leroux}, {Liao}, {Licata}, {Lindstr{\o}m}, {Lister}, {Livanou},
  {Lobel}, {L{\'o}pez}, {Managau}, {Mann}, {Mantelet}, {Marchal}, {Marchant},
  {Marconi}, {Marinoni}, {Marschalk{\'o}}, {Marshall}, {Martino}, {Marton},
  {Mary}, {Massari}, {Matijevi{\v{c}}}, {Mazeh}, {McMillan}, {Messina},
  {Michalik}, {Millar}, {Molina}, {Molinaro}, {Moln{\'a}r}, {Montegriffo},
  {Mor}, {Morbidelli}, {Morel}, {Morris}, {Mulone}, {Muraveva}, {Musella},
  {Nelemans}, {Nicastro}, {Noval}, {O'Mullane}, {Ord{\'e}novic},
  {Ord{\'o}{\~n}ez-Blanco}, {Osborne}, {Pagani}, {Pagano}, {Pailler},
  {Palacin}, {Palaversa}, {Panahi}, {Pawlak}, {Piersimoni}, {Pineau}, {Plachy},
  {Plum}, {Poggio}, {Poujoulet}, {Pr{\v{s}}a}, {Pulone}, {Racero}, {Ragaini},
  {Rambaux}, {Ramos-Lerate}, {Regibo}, {Reyl{\'e}}, {Riclet}, {Ripepi}, {Riva},
  {Rivard}, {Rixon}, {Roegiers}, {Roelens}, {Romero-G{\'o}mez}, {Rowell},
  {Royer}, {Ruiz-Dern}, {Sadowski}, {Sagrist{\`a} Sell{\'e}s}, {Sahlmann},
  {Salgado}, {Salguero}, {Sanna}, {Santana-Ros}, {Sarasso}, {Savietto},
  {Schultheis}, {Sciacca}, {Segol}, {Segovia}, {S{\'e}gransan}, {Shih},
  {Siltala}, {Silva}, {Smart}, {Smith}, {Solano}, {Solitro}, {Sordo}, {Soria
  Nieto}, {Souchay}, {Spagna}, {Spoto}, {Stampa}, {Steele},
  {Steidelm{\"u}ller}, {Stephenson}, {Stoev}, {Suess}, {Surdej}, {Szabados},
  {Szegedi-Elek}, {Tapiador}, {Taris}, {Tauran}, {Taylor}, {Teixeira},
  {Terrett}, {Teyssandier}, {Thuillot}, {Titarenko}, {Torra Clotet}, {Turon},
  {Ulla}, {Utrilla}, {Uzzi}, {Vaillant}, {Valentini}, {Valette}, {van Elteren},
  {Van Hemelryck}, {van Leeuwen}, {Vaschetto}, {Vecchiato}, {Veljanoski},
  {Viala}, {Vicente}, {Vogt}, {von Essen}, {Voss}, {Votruba}, {Voutsinas},
  {Walmsley}, {Weiler}, {Wertz}, {Wevers}, {Wyrzykowski}, {Yoldas},
  {{\v{Z}}erjal}, {Ziaeepour}, {Zorec}, {Zschocke}, {Zucker}, {Zurbach}, \&
  {Zwitter}}]{2018A&A...616A...1G}
{Gaia Collaboration}, {Brown}, A.~G.~A., {Vallenari}, A., {et~al.} 2018, \aap,
  616, A1, \dodoi{10.1051/0004-6361/201833051}

\bibitem[{{Gaia Collaboration} {et~al.}(2021){Gaia Collaboration}, {Brown},
  {Vallenari}, {Prusti}, {de Bruijne}, {Babusiaux}, {Biermann}, {Creevey},
  {Evans}, {Eyer}, {Hutton}, {Jansen}, {Jordi}, {Klioner}, {Lammers},
  {Lindegren}, {Luri}, {Mignard}, {Panem}, {Pourbaix}, {Randich}, {Sartoretti},
  {Soubiran}, {Walton}, {Arenou}, {Bailer-Jones}, {Bastian}, {Cropper},
  {Drimmel}, {Katz}, {Lattanzi}, {van Leeuwen}, {Bakker}, {Cacciari},
  {Casta{\~n}eda}, {De Angeli}, {Ducourant}, {Fabricius}, {Fouesneau},
  {Fr{\'e}mat}, {Guerra}, {Guerrier}, {Guiraud}, {Jean-Antoine Piccolo},
  {Masana}, {Messineo}, {Mowlavi}, {Nicolas}, {Nienartowicz}, {Pailler},
  {Panuzzo}, {Riclet}, {Roux}, {Seabroke}, {Sordo}, {Tanga}, {Th{\'e}venin},
  {Gracia-Abril}, {Portell}, {Teyssier}, {Altmann}, {Andrae}, {Bellas-Velidis},
  {Benson}, {Berthier}, {Blomme}, {Brugaletta}, {Burgess}, {Busso}, {Carry},
  {Cellino}, {Cheek}, {Clementini}, {Damerdji}, {Davidson}, {Delchambre},
  {Dell'Oro}, {Fern{\'a}ndez-Hern{\'a}ndez}, {Galluccio}, {Garc{\'\i}a-Lario},
  {Garcia-Reinaldos}, {Gonz{\'a}lez-N{\'u}{\~n}ez}, {Gosset}, {Haigron},
  {Halbwachs}, {Hambly}, {Harrison}, {Hatzidimitriou}, {Heiter},
  {Hern{\'a}ndez}, {Hestroffer}, {Hodgkin}, {Holl}, {Jan{\ss}en}, {Jevardat de
  Fombelle}, {Jordan}, {Krone-Martins}, {Lanzafame}, {L{\"o}ffler}, {Lorca},
  {Manteiga}, {Marchal}, {Marrese}, {Moitinho}, {Mora}, {Muinonen}, {Osborne},
  {Pancino}, {Pauwels}, {Petit}, {Recio-Blanco}, {Richards}, {Riello},
  {Rimoldini}, {Robin}, {Roegiers}, {Rybizki}, {Sarro}, {Siopis}, {Smith},
  {Sozzetti}, {Ulla}, {Utrilla}, {van Leeuwen}, {van Reeven}, {Abbas}, {Abreu
  Aramburu}, {Accart}, {Aerts}, {Aguado}, {Ajaj}, {Altavilla}, {{\'A}lvarez},
  {{\'A}lvarez Cid-Fuentes}, {Alves}, {Anderson}, {Anglada Varela}, {Antoja},
  {Audard}, {Baines}, {Baker}, {Balaguer-N{\'u}{\~n}ez}, {Balbinot}, {Balog},
  {Barache}, {Barbato}, {Barros}, {Barstow}, {Bartolom{\'e}}, {Bassilana},
  {Bauchet}, {Baudesson-Stella}, {Becciani}, {Bellazzini}, {Bernet}, {Bertone},
  {Bianchi}, {Blanco-Cuaresma}, {Boch}, {Bombrun}, {Bossini}, {Bouquillon},
  {Bragaglia}, {Bramante}, {Breedt}, {Bressan}, {Brouillet}, {Bucciarelli},
  {Burlacu}, {Busonero}, {Butkevich}, {Buzzi}, {Caffau}, {Cancelliere},
  {C{\'a}novas}, {Cantat-Gaudin}, {Carballo}, {Carlucci}, {Carnerero},
  {Carrasco}, {Casamiquela}, {Castellani}, {Castro-Ginard}, {Castro Sampol},
  {Chaoul}, {Charlot}, {Chemin}, {Chiavassa}, {Cioni}, {Comoretto}, {Cooper},
  {Cornez}, {Cowell}, {Crifo}, {Crosta}, {Crowley}, {Dafonte}, {Dapergolas},
  {David}, {David}, {de Laverny}, {De Luise}, {De March}, {De Ridder}, {de
  Souza}, {de Teodoro}, {de Torres}, {del Peloso}, {del Pozo}, {Delbo},
  {Delgado}, {Delgado}, {Delisle}, {Di Matteo}, {Diakite}, {Diener},
  {Distefano}, {Dolding}, {Eappachen}, {Edvardsson}, {Enke}, {Esquej}, {Fabre},
  {Fabrizio}, {Faigler}, {Fedorets}, {Fernique}, {Fienga}, {Figueras},
  {Fouron}, {Fragkoudi}, {Fraile}, {Franke}, {Gai}, {Garabato},
  {Garcia-Gutierrez}, {Garc{\'\i}a-Torres}, {Garofalo}, {Gavras}, {Gerlach},
  {Geyer}, {Giacobbe}, {Gilmore}, {Girona}, {Giuffrida}, {Gomel}, {Gomez},
  {Gonzalez-Santamaria}, {Gonz{\'a}lez-Vidal}, {Granvik},
  {Guti{\'e}rrez-S{\'a}nchez}, {Guy}, {Hauser}, {Haywood}, {Helmi}, {Hidalgo},
  {Hilger}, {H{\l}adczuk}, {Hobbs}, {Holland}, {Huckle}, {Jasniewicz},
  {Jonker}, {Juaristi Campillo}, {Julbe}, {Karbevska}, {Kervella}, {Khanna},
  {Kochoska}, {Kontizas}, {Kordopatis}, {Korn}, {Kostrzewa-Rutkowska},
  {Kruszy{\'n}ska}, {Lambert}, {Lanza}, {Lasne}, {Le Campion}, {Le Fustec},
  {Lebreton}, {Lebzelter}, {Leccia}, {Leclerc}, {Lecoeur-Taibi}, {Liao},
  {Licata}, {Lindstr{\o}m}, {Lister}, {Livanou}, {Lobel}, {Madrero Pardo},
  {Managau}, {Mann}, {Marchant}, {Marconi}, {Marcos Santos}, {Marinoni},
  {Marocco}, {Marshall}, {Martin Polo}, {Mart{\'\i}n-Fleitas}, {Masip},
  {Massari}, {Mastrobuono-Battisti}, {Mazeh}, {McMillan}, {Messina},
  {Michalik}, {Millar}, {Mints}, {Molina}, {Molinaro}, {Moln{\'a}r},
  {Montegriffo}, {Mor}, {Morbidelli}, {Morel}, {Morris}, {Mulone}, {Munoz},
  {Muraveva}, {Murphy}, {Musella}, {Noval}, {Ord{\'e}novic}, {Orr{\`u}},
  {Osinde}, {Pagani}, {Pagano}, {Palaversa}, {Palicio}, {Panahi}, {Pawlak},
  {Pe{\~n}alosa Esteller}, {Penttil{\"a}}, {Piersimoni}, {Pineau}, {Plachy},
  {Plum}, {Poggio}, {Poretti}, {Poujoulet}, {Pr{\v{s}}a}, {Pulone}, {Racero},
  {Ragaini}, {Rainer}, {Raiteri}, {Rambaux}, {Ramos}, {Ramos-Lerate}, {Re
  Fiorentin}, {Regibo}, {Reyl{\'e}}, {Ripepi}, {Riva}, {Rixon}, {Robichon},
  {Robin}, {Roelens}, {Rohrbasser}, {Romero-G{\'o}mez}, {Rowell}, {Royer},
  {Rybicki}, {Sadowski}, {Sagrist{\`a} Sell{\'e}s}, {Sahlmann}, {Salgado},
  {Salguero}, {Samaras}, {Sanchez Gimenez}, {Sanna}, {Santove{\~n}a},
  {Sarasso}, {Schultheis}, {Sciacca}, {Segol}, {Segovia}, {S{\'e}gransan},
  {Semeux}, {Shahaf}, {Siddiqui}, {Siebert}, {Siltala}, {Slezak}, {Smart},
  {Solano}, {Solitro}, {Souami}, {Souchay}, {Spagna}, {Spoto}, {Steele},
  {Steidelm{\"u}ller}, {Stephenson}, {S{\"u}veges}, {Szabados}, {Szegedi-Elek},
  {Taris}, {Tauran}, {Taylor}, {Teixeira}, {Thuillot}, {Tonello}, {Torra},
  {Torra}, {Turon}, {Unger}, {Vaillant}, {van Dillen}, {Vanel}, {Vecchiato},
  {Viala}, {Vicente}, {Voutsinas}, {Weiler}, {Wevers}, {Wyrzykowski}, {Yoldas},
  {Yvard}, {Zhao}, {Zorec}, {Zucker}, {Zurbach}, \&
  {Zwitter}}]{2021A&A...649A...1G}
---. 2021, \aap, 649, A1, \dodoi{10.1051/0004-6361/202039657}

\bibitem[{{Gebhardt} {et~al.}(2000){Gebhardt}, {Kormendy}, {Ho}, {Bender},
  {Bower}, {Dressler}, {Faber}, {Filippenko}, {Green}, {Grillmair}, {Lauer},
  {Magorrian}, {Pinkney}, {Richstone}, \& {Tremaine}}]{2000ApJ...543L...5G}
{Gebhardt}, K., {Kormendy}, J., {Ho}, L.~C., {et~al.} 2000, \apjl, 543, L5,
  \dodoi{10.1086/318174}

\bibitem[{{Gunn} {et~al.}(1998){Gunn}, {Carr}, {Rockosi}, {Sekiguchi}, {Berry},
  {Elms}, {de Haas}, {Ivezi{\'c}}, {Knapp}, {Lupton}, {Pauls}, {Simcoe},
  {Hirsch}, {Sanford}, {Wang}, {York}, {Harris}, {Annis}, {Bartozek},
  {Boroski}, {Bakken}, {Haldeman}, {Kent}, {Holm}, {Holmgren}, {Petravick},
  {Prosapio}, {Rechenmacher}, {Doi}, {Fukugita}, {Shimasaku}, {Okada}, {Hull},
  {Siegmund}, {Mannery}, {Blouke}, {Heidtman}, {Schneider}, {Lucinio}, \&
  {Brinkman}}]{1998AJ....116.3040G}
{Gunn}, J.~E., {Carr}, M., {Rockosi}, C., {et~al.} 1998, \aj, 116, 3040,
  \dodoi{10.1086/300645}

\bibitem[{Guo {et~al.}(2017)Guo, Gao, Liu, \& Yin}]{guo2017improved}
Guo, X., Gao, L., Liu, X., \& Yin, J. 2017, in Ijcai, 1753--1759

\bibitem[{Gwyn(2012)}]{gwyn2012}
Gwyn, S. D.~J. 2012, The Astronomical Journal, 143, 38,
  \dodoi{10.1088/0004-6256/143/2/38}

\bibitem[{{Hambly} {et~al.}(2008){Hambly}, {Collins}, {Cross}, {Mann}, {Read},
  {Sutorius}, {Bond}, {Bryant}, {Emerson}, {Lawrence}, {Rimoldini}, {Stewart},
  {Williams}, {Adamson}, {Hirst}, {Dye}, \& {Warren}}]{2008MNRAS.384..637H}
{Hambly}, N.~C., {Collins}, R.~S., {Cross}, N.~J.~G., {et~al.} 2008, \mnras,
  384, 637, \dodoi{10.1111/j.1365-2966.2007.12700.x}

\bibitem[{{Hewett} {et~al.}(2006){Hewett}, {Warren}, {Leggett}, \&
  {Hodgkin}}]{2006MNRAS.367..454H}
{Hewett}, P.~C., {Warren}, S.~J., {Leggett}, S.~K., \& {Hodgkin}, S.~T. 2006,
  \mnras, 367, 454, \dodoi{10.1111/j.1365-2966.2005.09969.x}

\bibitem[{{Hlo{\v{z}}ek} {et~al.}(2020){Hlo{\v{z}}ek}, {Ponder}, {Malz}, {Dai},
  {Narayan}, {Ishida}, {Allam}, {Bahmanyar}, {Biswas}, {Galbany}, {Jha},
  {Jones}, {Kessler}, {Lochner}, {Mahabal}, {Mandel}, {Mart{\'\i}nez-Galarza},
  {McEwen}, {Muthukrishna}, {Peiris}, {Peters}, \& {Setzer}}]{Hlozek2020}
{Hlo{\v{z}}ek}, R., {Ponder}, K.~A., {Malz}, A.~I., {et~al.} 2020, arXiv
  e-prints, arXiv:2012.12392.
\newblock \doarXiv{2012.12392}

\bibitem[{Ho(1995)}]{ho1995random}
Ho, T.~K. 1995, in Proceedings of 3rd international conference on document
  analysis and recognition, Vol.~1, IEEE, 278--282

\bibitem[{{Hodgkin} {et~al.}(2009){Hodgkin}, {Irwin}, {Hewett}, \&
  {Warren}}]{2009MNRAS.394..675H}
{Hodgkin}, S.~T., {Irwin}, M.~J., {Hewett}, P.~C., \& {Warren}, S.~J. 2009,
  \mnras, 394, 675, \dodoi{10.1111/j.1365-2966.2008.14387.x}

\bibitem[{Hunter(2007)}]{4160265}
Hunter, J.~D. 2007, Computing in Science Engineering, 9, 90,
  \dodoi{10.1109/MCSE.2007.55}

\bibitem[{{Ivezi{\'c}} {et~al.}(2007){Ivezi{\'c}}, {Smith}, {Miknaitis}, {Lin},
  {Tucker}, {Lupton}, {Gunn}, {Knapp}, {Strauss}, {Sesar}, {Doi}, {Tanaka},
  {Fukugita}, {Holtzman}, {Kent}, {Yanny}, {Schlegel}, {Finkbeiner},
  {Padmanabhan}, {Rockosi}, {Juri{\'c}}, {Bond}, {Lee}, {Stoughton}, {Jester},
  {Harris}, {Harding}, {Morrison}, {Brinkmann}, {Schneider}, \&
  {York}}]{2007AJ....134..973I}
{Ivezi{\'c}}, {\v{Z}}., {Smith}, J.~A., {Miknaitis}, G., {et~al.} 2007, \aj,
  134, 973, \dodoi{10.1086/519976}

\bibitem[{{Ivezi{\'c}} {et~al.}(2019){Ivezi{\'c}}, {Kahn}, {Tyson}, {Abel},
  {Acosta}, {Allsman}, {Alonso}, {AlSayyad}, {Anderson}, {Andrew}, {Angel},
  {Angeli}, {Ansari}, {Antilogus}, {Araujo}, {Armstrong}, {Arndt}, {Astier},
  {Aubourg}, {Auza}, {Axelrod}, {Bard}, {Barr}, {Barrau}, {Bartlett}, {Bauer},
  {Bauman}, {Baumont}, {Bechtol}, {Bechtol}, {Becker}, {Becla}, {Beldica},
  {Bellavia}, {Bianco}, {Biswas}, {Blanc}, {Blazek}, {Blandford}, {Bloom},
  {Bogart}, {Bond}, {Booth}, {Borgland}, {Borne}, {Bosch}, {Boutigny},
  {Brackett}, {Bradshaw}, {Brandt}, {Brown}, {Bullock}, {Burchat}, {Burke},
  {Cagnoli}, {Calabrese}, {Callahan}, {Callen}, {Carlin}, {Carlson},
  {Chandrasekharan}, {Charles-Emerson}, {Chesley}, {Cheu}, {Chiang}, {Chiang},
  {Chirino}, {Chow}, {Ciardi}, {Claver}, {Cohen-Tanugi}, {Cockrum}, {Coles},
  {Connolly}, {Cook}, {Cooray}, {Covey}, {Cribbs}, {Cui}, {Cutri}, {Daly},
  {Daniel}, {Daruich}, {Daubard}, {Daues}, {Dawson}, {Delgado}, {Dellapenna},
  {de Peyster}, {de Val-Borro}, {Digel}, {Doherty}, {Dubois},
  {Dubois-Felsmann}, {Durech}, {Economou}, {Eifler}, {Eracleous}, {Emmons},
  {Fausti Neto}, {Ferguson}, {Figueroa}, {Fisher-Levine}, {Focke}, {Foss},
  {Frank}, {Freemon}, {Gangler}, {Gawiser}, {Geary}, {Gee}, {Geha}, {Gessner},
  {Gibson}, {Gilmore}, {Glanzman}, {Glick}, {Goldina}, {Goldstein}, {Goodenow},
  {Graham}, {Gressler}, {Gris}, {Guy}, {Guyonnet}, {Haller}, {Harris},
  {Hascall}, {Haupt}, {Hernandez}, {Herrmann}, {Hileman}, {Hoblitt}, {Hodgson},
  {Hogan}, {Howard}, {Huang}, {Huffer}, {Ingraham}, {Innes}, {Jacoby}, {Jain},
  {Jammes}, {Jee}, {Jenness}, {Jernigan}, {Jevremovi{\'c}}, {Johns}, {Johnson},
  {Johnson}, {Jones}, {Juramy-Gilles}, {Juri{\'c}}, {Kalirai}, {Kallivayalil},
  {Kalmbach}, {Kantor}, {Karst}, {Kasliwal}, {Kelly}, {Kessler}, {Kinnison},
  {Kirkby}, {Knox}, {Kotov}, {Krabbendam}, {Krughoff}, {Kub{\'a}nek},
  {Kuczewski}, {Kulkarni}, {Ku}, {Kurita}, {Lage}, {Lambert}, {Lange},
  {Langton}, {Le Guillou}, {Levine}, {Liang}, {Lim}, {Lintott}, {Long},
  {Lopez}, {Lotz}, {Lupton}, {Lust}, {MacArthur}, {Mahabal}, {Mandelbaum},
  {Markiewicz}, {Marsh}, {Marshall}, {Marshall}, {May}, {McKercher}, {McQueen},
  {Meyers}, {Migliore}, {Miller}, {Mills}, {Miraval}, {Moeyens}, {Moolekamp},
  {Monet}, {Moniez}, {Monkewitz}, {Montgomery}, {Morrison}, {Mueller},
  {Muller}, {Mu{\~n}oz Arancibia}, {Neill}, {Newbry}, {Nief}, {Nomerotski},
  {Nordby}, {O'Connor}, {Oliver}, {Olivier}, {Olsen}, {O'Mullane}, {Ortiz},
  {Osier}, {Owen}, {Pain}, {Palecek}, {Parejko}, {Parsons}, {Pease},
  {Peterson}, {Peterson}, {Petravick}, {Libby Petrick}, {Petry},
  {Pierfederici}, {Pietrowicz}, {Pike}, {Pinto}, {Plante}, {Plate}, {Plutchak},
  {Price}, {Prouza}, {Radeka}, {Rajagopal}, {Rasmussen}, {Regnault}, {Reil},
  {Reiss}, {Reuter}, {Ridgway}, {Riot}, {Ritz}, {Robinson}, {Roby}, {Roodman},
  {Rosing}, {Roucelle}, {Rumore}, {Russo}, {Saha}, {Sassolas}, {Schalk},
  {Schellart}, {Schindler}, {Schmidt}, {Schneider}, {Schneider}, {Schoening},
  {Schumacher}, {Schwamb}, {Sebag}, {Selvy}, {Sembroski}, {Seppala}, {Serio},
  {Serrano}, {Shaw}, {Shipsey}, {Sick}, {Silvestri}, {Slater}, {Smith},
  {Smith}, {Sobhani}, {Soldahl}, {Storrie-Lombardi}, {Stover}, {Strauss},
  {Street}, {Stubbs}, {Sullivan}, {Sweeney}, {Swinbank}, {Szalay}, {Takacs},
  {Tether}, {Thaler}, {Thayer}, {Thomas}, {Thornton}, {Thukral}, {Tice},
  {Trilling}, {Turri}, {Van Berg}, {Vanden Berk}, {Vetter}, {Virieux},
  {Vucina}, {Wahl}, {Walkowicz}, {Walsh}, {Walter}, {Wang}, {Wang}, {Warner},
  {Wiecha}, {Willman}, {Winters}, {Wittman}, {Wolff}, {Wood-Vasey}, {Wu},
  {Xin}, {Yoachim}, \& {Zhan}}]{2019ApJ...873..111I}
{Ivezi{\'c}}, {\v{Z}}., {Kahn}, S.~M., {Tyson}, J.~A., {et~al.} 2019, \apj,
  873, 111, \dodoi{10.3847/1538-4357/ab042c}

\bibitem[{{Jankov} {et~al.}(2021){Jankov}, {Ili{\'c}}, \&
  {Kova{\v{c}}evi{\'c}}}]{2021POBeo.100..241J}
{Jankov}, I., {Ili{\'c}}, D., \& {Kova{\v{c}}evi{\'c}}, A. 2021, in XIX Serbian
  Astronomical Conference, Vol. 100, 241--246

\bibitem[{Jarvis {et~al.}(2012)Jarvis, Bonfield, Bruce, Geach, McAlpine,
  McLure, Gonz{\'{a} }lez-Solares, Irwin, Lewis, Yoldas, Andreon, Cross,
  Emerson, Dalton, Dunlop, Hodgkin, Le, Karouzos, Meisenheimer, Oliver,
  Rawlings, Simpson, Smail, Smith, Sullivan, Sutherland, White, \&
  Zwart}]{Jarvis_2012}
Jarvis, M.~J., Bonfield, D.~G., Bruce, V.~A., {et~al.} 2012, Monthly Notices of
  the Royal Astronomical Society, 428, 1281, \dodoi{10.1093/mnras/sts118}

\bibitem[{{Jiang} {et~al.}(2016){Jiang}, {McGreer}, {Fan}, {Strauss},
  {Ba{\~n}ados}, {Becker}, {Bian}, {Farnsworth}, {Shen}, {Wang}, {Wang},
  {Wang}, {White}, {Wu}, {Wu}, {Yang}, \& {Yang}}]{Jiang16}
{Jiang}, L., {McGreer}, I.~D., {Fan}, X., {et~al.} 2016, \apj, 833, 222,
  \dodoi{10.3847/1538-4357/833/2/222}

\bibitem[{{Jolliffe}(1986)}]{1986pca..book.....J}
{Jolliffe}, I.~T. 1986, {Principal component analysis}

\bibitem[{{Kaczmarczik} {et~al.}(2009){Kaczmarczik}, {Richards}, {Mehta}, \&
  {Schlegel}}]{2009AJ....138...19K}
{Kaczmarczik}, M.~C., {Richards}, G.~T., {Mehta}, S.~S., \& {Schlegel}, D.~J.
  2009, \aj, 138, 19, \dodoi{10.1088/0004-6256/138/1/19}

\bibitem[{Kasliwal {et~al.}(2017)Kasliwal, Vogeley, \& Richards}]{kasliwal2017}
Kasliwal, V.~P., Vogeley, M.~S., \& Richards, G.~T. 2017, Mon Not R Astron Soc,
  470, 3027, \dodoi{10.1093/mnras/stx1420}

\bibitem[{{Kelly} {et~al.}(2009){Kelly}, {Bechtold}, \&
  {Siemiginowska}}]{2009ApJ...698..895K}
{Kelly}, B.~C., {Bechtold}, J., \& {Siemiginowska}, A. 2009, \apj, 698, 895,
  \dodoi{10.1088/0004-637X/698/1/895}

\bibitem[{{Kim} {et~al.}(2011){Kim}, {Protopapas}, {Byun}, {Alcock}, {Khardon},
  \& {Trichas}}]{2011ApJ...735...68K}
{Kim}, D.-W., {Protopapas}, P., {Byun}, Y.-I., {et~al.} 2011, \apj, 735, 68,
  \dodoi{10.1088/0004-637X/735/2/68}

\bibitem[{Kluyver {et~al.}(2016)Kluyver, Ragan-Kelley, P{\'e}rez, Granger,
  Bussonnier, Frederic, Kelley, Hamrick, Grout, Corlay, Ivanov, Avila, Abdalla,
  Willing, \& development team}]{soton403913}
Kluyver, T., Ragan-Kelley, B., P{\'e}rez, F., {et~al.} 2016, in Positioning and
  Power in Academic Publishing: Players, Agents and Agendas, ed. F.~Loizides \&
  B.~Scmidt (IOS Press), 87--90.
\newblock \url{https://eprints.soton.ac.uk/403913/}

\bibitem[{{Koo} \& {Kron}(1982)}]{1982A&A...105..107K}
{Koo}, D.~C., \& {Kron}, R.~G. 1982, \aap, 105, 107

\bibitem[{{Koo} {et~al.}(1986){Koo}, {Kron}, \&
  {Cudworth}}]{1986PASP...98..285K}
{Koo}, D.~C., {Kron}, R.~G., \& {Cudworth}, K.~M. 1986, \pasp, 98, 285,
  \dodoi{10.1086/131756}

\bibitem[{{Kormendy} \& {Ho}(2013)}]{2013ARA&A..51..511K}
{Kormendy}, J., \& {Ho}, L.~C. 2013, \araa, 51, 511,
  \dodoi{10.1146/annurev-astro-082708-101811}

\bibitem[{{Kovacevic} {et~al.}(2022){Kovacevic}, {Radovic}, {Ilic}, {Popovic},
  {Assef}, {Sanchez-Saez}, {Nikutta}, {Raiteri}, {Yoon}, {Homayouni}, {Li},
  {Caplar}, {Czerny}, {Panda}, {Ricci}, {Jankov}, {Landt}, {Wolf},
  {Kovacevic-Dojcinovic}, {Lakicevic}, {Savic}, {Vince}, {Simic},
  {Cvorovic-Hajdinjak}, \& {Marceta-Mandic}}]{2022arXiv220806203K}
{Kovacevic}, A.~B., {Radovic}, V., {Ilic}, D., {et~al.} 2022, arXiv e-prints,
  arXiv:2208.06203.
\newblock \doarXiv{2208.06203}

\bibitem[{{Koz{\l}owski} {et~al.}(2016){Koz{\l}owski}, {Kochanek}, {Ashby},
  {Assef}, {Brodwin}, {Eisenhardt}, {Jannuzi}, \&
  {Stern}}]{2016ApJ...817..119K}
{Koz{\l}owski}, S., {Kochanek}, C.~S., {Ashby}, M. L.~N., {et~al.} 2016, \apj,
  817, 119, \dodoi{10.3847/0004-637X/817/2/119}

\bibitem[{{Koz{\l}owski} {et~al.}(2010{\natexlab{a}}){Koz{\l}owski},
  {Kochanek}, {Stern}, {Ashby}, {Assef}, {Bock}, {Borys}, {Brand}, {Brodwin},
  {Brown}, {Cool}, {Cooray}, {Croft}, {Dey}, {Eisenhardt}, {Gonzalez},
  {Gorjian}, {Griffith}, {Grogin}, {Ivison}, {Jacob}, {Jannuzi}, {Mainzer},
  {Moustakas}, {R{\"o}ttgering}, {Seymour}, {Smith}, {Stanford}, {Stauffer},
  {Sullivan}, {van Breugel}, {Willner}, \& {Wright}}]{2010ApJ...716..530K}
{Koz{\l}owski}, S., {Kochanek}, C.~S., {Stern}, D., {et~al.}
  2010{\natexlab{a}}, \apj, 716, 530, \dodoi{10.1088/0004-637X/716/1/530}

\bibitem[{{Koz{\l}owski} {et~al.}(2010{\natexlab{b}}){Koz{\l}owski},
  {Kochanek}, {Udalski}, {Wyrzykowski}, {Soszy{\'n}ski}, {Szyma{\'n}ski},
  {Kubiak}, {Pietrzy{\'n}ski}, {Szewczyk}, {Ulaczyk}, {Poleski}, \& {OGLE
  Collaboration}}]{2010ApJ...708..927K}
{Koz{\l}owski}, S., {Kochanek}, C.~S., {Udalski}, A., {et~al.}
  2010{\natexlab{b}}, \apj, 708, 927, \dodoi{10.1088/0004-637X/708/2/927}

\bibitem[{Kramer(1991)}]{Kramer1991NonlinearPC}
Kramer, M.~A. 1991, Aiche Journal, 37, 233

\bibitem[{{Kron} \& {Chiu}(1981)}]{1981PASP...93..397K}
{Kron}, R.~G., \& {Chiu}, L. T.~G. 1981, \pasp, 93, 397, \dodoi{10.1086/130845}

\bibitem[{{Lang} {et~al.}(2016){Lang}, {Hogg}, \&
  {Mykytyn}}]{2016ascl.soft04008L}
{Lang}, D., {Hogg}, D.~W., \& {Mykytyn}, D. 2016, {The Tractor: Probabilistic
  astronomical source detection and measurement}, Astrophysics Source Code
  Library, record ascl:1604.008.
\newblock \doeprint{1604.008}

\bibitem[{{Lawrence} {et~al.}(2007){Lawrence}, {Warren}, {Almaini}, {Edge},
  {Hambly}, {Jameson}, {Lucas}, {Casali}, {Adamson}, {Dye}, {Emerson},
  {Foucaud}, {Hewett}, {Hirst}, {Hodgkin}, {Irwin}, {Lodieu}, {McMahon},
  {Simpson}, {Smail}, {Mortlock}, \& {Folger}}]{Lawrence07}
{Lawrence}, A., {Warren}, S.~J., {Almaini}, O., {et~al.} 2007, \mnras, 379,
  1599, \dodoi{10.1111/j.1365-2966.2007.12040.x}

\bibitem[{{Lecun} {et~al.}(2015){Lecun}, {Bengio}, \&
  {Hinton}}]{2015Natur.521..436L}
{Lecun}, Y., {Bengio}, Y., \& {Hinton}, G. 2015, \nat, 521, 436,
  \dodoi{10.1038/nature14539}

\bibitem[{Lloyd(1982)}]{1056489}
Lloyd, S. 1982, IEEE Transactions on Information Theory, 28, 129,
  \dodoi{10.1109/TIT.1982.1056489}

\bibitem[{{Lochner} {et~al.}(2016){Lochner}, {McEwen}, {Peiris}, {Lahav}, \&
  {Winter}}]{2016ApJS..225...31L}
{Lochner}, M., {McEwen}, J.~D., {Peiris}, H.~V., {Lahav}, O., \& {Winter},
  M.~K. 2016, \apjs, 225, 31, \dodoi{10.3847/0067-0049/225/2/31}

\bibitem[{{LSST Science Collaboration} {et~al.}(2017){LSST Science
  Collaboration}, {Marshall}, {Anguita}, {Bianco}, {Bellm}, {Brandt},
  {Clarkson}, {Connolly}, {Gawiser}, {Ivezic}, {Jones}, {Lochner}, {Lund},
  {Mahabal}, {Nidever}, {Olsen}, {Ridgway}, {Rhodes}, {Shemmer}, {Trilling},
  {Vivas}, {Walkowicz}, {Willman}, {Yoachim}, {Anderson}, {Antilogus}, {Angus},
  {Arcavi}, {Awan}, {Biswas}, {Bell}, {Bennett}, {Britt}, {Buzasi},
  {Casetti-Dinescu}, {Chomiuk}, {Claver}, {Cook}, {Davenport}, {Debattista},
  {Digel}, {Doctor}, {Firth}, {Foley}, {Fong}, {Galbany}, {Giampapa}, {Gizis},
  {Graham}, {Grillmair}, {Gris}, {Haiman}, {Hartigan}, {Hawley}, {Hlozek},
  {Jha}, {Johns-Krull}, {Kanbur}, {Kalogera}, {Kashyap}, {Kasliwal}, {Kessler},
  {Kim}, {Kurczynski}, {Lahav}, {Liu}, {Malz}, {Margutti}, {Matheson},
  {McEwen}, {McGehee}, {Meibom}, {Meyers}, {Monet}, {Neilsen}, {Newman},
  {O'Dowd}, {Peiris}, {Penny}, {Peters}, {Poleski}, {Ponder}, {Richards},
  {Rho}, {Rubin}, {Schmidt}, {Schuhmann}, {Shporer}, {Slater}, {Smith},
  {Soares-Santos}, {Stassun}, {Strader}, {Strauss}, {Street}, {Stubbs},
  {Sullivan}, {Szkody}, {Trimble}, {Tyson}, {de Val-Borro}, {Valenti},
  {Wagoner}, {Wood-Vasey}, \& {Zauderer}}]{2017arXiv170804058L}
{LSST Science Collaboration}, {Marshall}, P., {Anguita}, T., {et~al.} 2017,
  arXiv e-prints, arXiv:1708.04058.
\newblock \doarXiv{1708.04058}

\bibitem[{{Luo} {et~al.}(2017){Luo}, {Brandt}, {Xue}, {Lehmer}, {Alexander},
  {Bauer}, {Vito}, {Yang}, {Basu-Zych}, {Comastri}, {Gilli}, {Gu},
  {Hornschemeier}, {Koekemoer}, {Liu}, {Mainieri}, {Paolillo}, {Ranalli},
  {Rosati}, {Schneider}, {Shemmer}, {Smail}, {Sun}, {Tozzi}, {Vignali}, \&
  {Wang}}]{2017ApJS..228....2L}
{Luo}, B., {Brandt}, W.~N., {Xue}, Y.~Q., {et~al.} 2017, \apjs, 228, 2,
  \dodoi{10.3847/1538-4365/228/1/2}

\bibitem[{{Macuga} {et~al.}(2019){Macuga}, {Martini}, {Miller}, {Brodwin},
  {Hayashi}, {Kodama}, {Koyama}, {Overzier}, {Shimakawa}, {Tadaki}, \&
  {Tanaka}}]{2019ApJ...874...54M}
{Macuga}, M., {Martini}, P., {Miller}, E.~D., {et~al.} 2019, \apj, 874, 54,
  \dodoi{10.3847/1538-4357/ab0746}

\bibitem[{{Mahabal} {et~al.}(2017){Mahabal}, {Sheth}, {Gieseke}, {Pai},
  {Djorgovski}, {Drake}, {Graham}, \& {the CSS/CRTS/PTF
  Collaboration}}]{2017arXiv170906257M}
{Mahabal}, A., {Sheth}, K., {Gieseke}, F., {et~al.} 2017, arXiv e-prints,
  arXiv:1709.06257.
\newblock \doarXiv{1709.06257}

\bibitem[{{Matsuoka} {et~al.}(2018){Matsuoka}, {Iwasawa}, {Onoue}, {Kashikawa},
  {Strauss}, {Lee}, {Imanishi}, {Nagao}, {Akiyama}, {Asami}, {Bosch},
  {Furusawa}, {Goto}, {Gunn}, {Harikane}, {Ikeda}, {Izumi}, {Kawaguchi},
  {Kato}, {Kikuta}, {Kohno}, {Komiyama}, {Lupton}, {Minezaki}, {Miyazaki},
  {Morokuma}, {Murayama}, {Niida}, {Nishizawa}, {Oguri}, {Ono}, {Ouchi},
  {Price}, {Sameshima}, {Schulze}, {Shirakata}, {Silverman}, {Sugiyama},
  {Tait}, {Takada}, {Takata}, {Tanaka}, {Tang}, {Toba}, {Utsumi}, {Wang}, \&
  {Yamashita}}]{Matsuoka18}
{Matsuoka}, Y., {Iwasawa}, K., {Onoue}, M., {et~al.} 2018, \apjs, 237, 5,
  \dodoi{10.3847/1538-4365/aac724}

\bibitem[{{Mazzucchelli} {et~al.}(2017){Mazzucchelli}, {Ba{\~n}ados},
  {Venemans}, {Decarli}, {Farina}, {Walter}, {Eilers}, {Rix}, {Simcoe},
  {Stern}, {Fan}, {Schlafly}, {De Rosa}, {Hennawi}, {Chambers}, {Greiner},
  {Burgett}, {Draper}, {Kaiser}, {Kudritzki}, {Magnier}, {Metcalfe}, {Waters},
  \& {Wainscoat}}]{Mazzucchelli17}
{Mazzucchelli}, C., {Ba{\~n}ados}, E., {Venemans}, B.~P., {et~al.} 2017, \apj,
  849, 91, \dodoi{10.3847/1538-4357/aa9185}

\bibitem[{{McGreer} {et~al.}(2013){McGreer}, {Jiang}, {Fan}, {Richards},
  {Strauss}, {Ross}, {White}, {Shen}, {Schneider}, {Myers}, {Brandt}, {DeGraf},
  {Glikman}, {Ge}, \& {Streblyanska}}]{McGreer13}
{McGreer}, I.~D., {Jiang}, L., {Fan}, X., {et~al.} 2013, \apj, 768, 105,
  \dodoi{10.1088/0004-637X/768/2/105}

\bibitem[{{McHardy} {et~al.}(2018){McHardy}, {Connolly}, {Horne}, {Cackett},
  {Gelbord}, {Peterson}, {Pahari}, {Gehrels}, {Goad}, {Lira}, {Arevalo},
  {Baldi}, {Brandt}, {Breedt}, {Chand}, {Dewangan}, {Done}, {Elvis},
  {Emmanoulopoulos}, {Fausnaugh}, {Kaspi}, {Kochanek}, {Korista}, {Papadakis},
  {Rao}, {Uttley}, {Vestergaard}, \& {Ward}}]{2018MNRAS.480.2881M}
{McHardy}, I.~M., {Connolly}, S.~D., {Horne}, K., {et~al.} 2018, \mnras, 480,
  2881, \dodoi{10.1093/mnras/sty1983}

\bibitem[{McKinney {et~al.}(2010)}]{mckinney2010data}
McKinney, W., {et~al.} 2010, in Proceedings of the 9th Python in Science
  Conference, Vol. 445, Austin, TX, 51--56

\bibitem[{{McMahon} {et~al.}(2013){McMahon}, {Banerji}, {Gonzalez}, {Koposov},
  {Bejar}, {Lodieu}, {Rebolo}, \& {VHS Collaboration}}]{Mcmahon13}
{McMahon}, R.~G., {Banerji}, M., {Gonzalez}, E., {et~al.} 2013, The Messenger,
  154, 35

\bibitem[{Moreno {et~al.}(2019)Moreno, Vogeley, Richards, \& Yu}]{moreno2019}
Moreno, J., Vogeley, M.~S., Richards, G.~T., \& Yu, W. 2019, PASP, 131, 063001,
  \dodoi{10.1088/1538-3873/ab1597}

\bibitem[{{Myers} {et~al.}(2015){Myers}, {Palanque-Delabrouille}, {Prakash},
  {P{\^a}ris}, {Yeche}, {Dawson}, {Bovy}, {Lang}, {Schlegel}, {Newman},
  {Petitjean}, {Kneib}, {Laurent}, {Percival}, {Ross}, {Seo}, {Tinker},
  {Armengaud}, {Brownstein}, {Burtin}, {Cai}, {Comparat}, {Kasliwal},
  {Kulkarni}, {Laher}, {Levitan}, {McBride}, {McGreer}, {Miller}, {Nugent},
  {Ofek}, {Rossi}, {Ruan}, {Schneider}, {Sesar}, {Streblyanska}, \&
  {Surace}}]{2015ApJS..221...27M}
{Myers}, A.~D., {Palanque-Delabrouille}, N., {Prakash}, A., {et~al.} 2015,
  \apjs, 221, 27, \dodoi{10.1088/0067-0049/221/2/27}

\bibitem[{{Netzer}(2015)}]{2015ARA&A..53..365N}
{Netzer}, H. 2015, \araa, 53, 365, \dodoi{10.1146/annurev-astro-082214-122302}

\bibitem[{{Ni} {et~al.}(2021){Ni}, {Brandt}, {Chen}, {Luo}, {Nyland}, {Yang},
  {Zou}, {Aird}, {Alexander}, {Bauer}, {Lacy}, {Lehmer}, {Mallick}, {Salvato},
  {Schneider}, {Tozzi}, {Traulsen}, {Vaccari}, {Vignali}, {Vito}, {Xue},
  {Banerji}, {Chow}, {Comastri}, {Del Moro}, {Gilli}, {Mullaney}, {Paolillo},
  {Schwope}, {Shemmer}, {Sun}, {Timlin}, \& {Trump}}]{2021ApJS..256...21N}
{Ni}, Q., {Brandt}, W.~N., {Chen}, C.-T., {et~al.} 2021, \apjs, 256, 21,
  \dodoi{10.3847/1538-4365/ac0dc6}

\bibitem[{{Nidever} {et~al.}(2021){Nidever}, {Dey}, {Fasbender}, {Juneau},
  {Meisner}, {Wishart}, {Scott}, {Matt}, {Nikutta}, \&
  {Pucha}}]{2021AJ....161..192N}
{Nidever}, D.~L., {Dey}, A., {Fasbender}, K., {et~al.} 2021, \aj, 161, 192,
  \dodoi{10.3847/1538-3881/abd6e1}

\bibitem[{{Nyland} {et~al.}(2017){Nyland}, {Lacy}, {Sajina}, {Pforr}, {Farrah},
  {Wilson}, {Surace}, {H{\"a}u{\ss}ler}, {Vaccari}, \&
  {Jarvis}}]{2017ApJS..230....9N}
{Nyland}, K., {Lacy}, M., {Sajina}, A., {et~al.} 2017, \apjs, 230, 9,
  \dodoi{10.3847/1538-4365/aa6fed}

\bibitem[{{Padovani} {et~al.}(2017){Padovani}, {Alexander}, {Assef}, {De
  Marco}, {Giommi}, {Hickox}, {Richards}, {Smol{\v{c}}i{\'c}},
  {Hatziminaoglou}, {Mainieri}, \& {Salvato}}]{2017A&ARv..25....2P}
{Padovani}, P., {Alexander}, D.~M., {Assef}, R.~J., {et~al.} 2017, \aapr, 25,
  2, \dodoi{10.1007/s00159-017-0102-9}

\bibitem[{{Panda} {et~al.}(2019){Panda}, {Mart{\'\i}nez-Aldama}, \&
  {Zaja{\v{c}}ek}}]{2019FrASS...6...75P}
{Panda}, S., {Mart{\'\i}nez-Aldama}, M.~L., \& {Zaja{\v{c}}ek}, M. 2019,
  Frontiers in Astronomy and Space Sciences, 6, 75,
  \dodoi{10.3389/fspas.2019.00075}

\bibitem[{Pedregosa {et~al.}(2011)Pedregosa, Varoquaux, Gramfort, Michel,
  Thirion, Grisel, Blondel, Prettenhofer, Weiss, Dubourg,
  {et~al.}}]{pedregosa2011scikit}
Pedregosa, F., Varoquaux, G., Gramfort, A., {et~al.} 2011, Journal of machine
  learning research, 12, 2825

\bibitem[{{Peters} {et~al.}(2015){Peters}, {Richards}, {Myers}, {Strauss},
  {Schmidt}, {Ivezi{\'c}}, {Ross}, {MacLeod}, \&
  {Riegel}}]{2015ApJ...811...95P}
{Peters}, C.~M., {Richards}, G.~T., {Myers}, A.~D., {et~al.} 2015, \apj, 811,
  95, \dodoi{10.1088/0004-637X/811/2/95}

\bibitem[{{Pierre} {et~al.}(2007){Pierre}, {Chiappetti}, {Pacaud}, {Gueguen},
  {Libbrecht}, {Altieri}, {Aussel}, {Gandhi}, {Garcet}, {Gosset}, {Paioro},
  {Ponman}, {Read}, {Refregier}, {Starck}, {Surdej}, {Valtchanov}, {Adami},
  {Alloin}, {Alshino}, {Andreon}, {Birkinshaw}, {Bremer}, {Detal}, {Duc},
  {Galaz}, {Jones}, {Le F{\`e}vre}, {Le F{\`e}vre}, {Maccagni}, {Mazure},
  {Quintana}, {R{\"o}ttgering}, {Sprimont}, {Tasse}, {Trinchieri}, \&
  {Willis}}]{2007MNRAS.382..279P}
{Pierre}, M., {Chiappetti}, L., {Pacaud}, F., {et~al.} 2007, \mnras, 382, 279,
  \dodoi{10.1111/j.1365-2966.2007.12354.x}

\bibitem[{{Poliszczuk} {et~al.}(2021){Poliszczuk}, {Pollo}, {Ma{\l}ek},
  {Durkalec}, {Pearson}, {Goto}, {Kim}, {Malkan}, {Oi}, {Ho}, {Shim},
  {Pearson}, {Hwang}, {Toba}, \& {Kim}}]{2021A&A...651A.108P}
{Poliszczuk}, A., {Pollo}, A., {Ma{\l}ek}, K., {et~al.} 2021, \aap, 651, A108,
  \dodoi{10.1051/0004-6361/202040219}

\bibitem[{{Poulain} {et~al.}(2020){Poulain}, {Paolillo}, {De Cicco}, {Brandt},
  {Bauer}, {Falocco}, {Vagnetti}, {Grado}, {Ragosta}, {Botticella},
  {Cappellaro}, {Pignata}, {Vaccari}, {Schipani}, {Covone}, {Longo}, \&
  {Napolitano}}]{2020A&A...634A..50P}
{Poulain}, M., {Paolillo}, M., {De Cicco}, D., {et~al.} 2020, \aap, 634, A50,
  \dodoi{10.1051/0004-6361/201937108}

\bibitem[{{Pozo Nu{\~n}ez} {et~al.}(2023){Pozo Nu{\~n}ez}, {Bruckmann},
  {Deesamutara}, {Czerny}, {Panda}, {Lobban}, {Pietrzy{\'n}ski}, \&
  {Polsterer}}]{2023MNRAS.522.2002P}
{Pozo Nu{\~n}ez}, F., {Bruckmann}, C., {Deesamutara}, S., {et~al.} 2023,
  \mnras, 522, 2002, \dodoi{10.1093/mnras/stad286}

\bibitem[{{Raiteri} {et~al.}(2022){Raiteri}, {Carnerero}, {Balmaverde},
  {Bellm}, {Clarkson}, {D'Ammando}, {Paolillo}, {Richards}, {Villata},
  {Yoachim}, \& {Yoon}}]{2022ApJS..258....3R}
{Raiteri}, C.~M., {Carnerero}, M.~I., {Balmaverde}, B., {et~al.} 2022, \apjs,
  258, 3, \dodoi{10.3847/1538-4365/ac3bb0}

\bibitem[{{Reed} {et~al.}(2017){Reed}, {McMahon}, {Martini}, {Banerji},
  {Auger}, {Hewett}, {Koposov}, {Gibbons}, {Gonzalez-Solares}, {Ostrovski},
  {Tie}, {Abdalla}, {Allam}, {Benoit-L{\'e}vy}, {Bertin}, {Brooks},
  {Buckley-Geer}, {Burke}, {Carnero Rosell}, {Carrasco Kind}, {Carretero}, {da
  Costa}, {DePoy}, {Desai}, {Diehl}, {Doel}, {Evrard}, {Finley}, {Flaugher},
  {Fosalba}, {Frieman}, {Garc{\'\i}a-Bellido}, {Gaztanaga}, {Goldstein},
  {Gruen}, {Gruendl}, {Gutierrez}, {James}, {Kuehn}, {Kuropatkin}, {Lahav},
  {Lima}, {Maia}, {Marshall}, {Melchior}, {Miller}, {Miquel}, {Nord}, {Ogando},
  {Plazas}, {Romer}, {Sanchez}, {Scarpine}, {Schubnell}, {Sevilla-Noarbe},
  {Smith}, {Sobreira}, {Suchyta}, {Swanson}, {Tarle}, {Tucker}, {Walker}, \&
  {Wester}}]{Reed17}
{Reed}, S.~L., {McMahon}, R.~G., {Martini}, P., {et~al.} 2017, \mnras, 468,
  4702, \dodoi{10.1093/mnras/stx728}

\bibitem[{Reynolds(2009)}]{Reynolds2009}
Reynolds, D. 2009, Gaussian Mixture Models (Boston, MA: Springer US), 659--663,
  \dodoi{10.1007/978-0-387-73003-5_196}

\bibitem[{{Richards} {et~al.}(2002){Richards}, {Fan}, {Newberg}, {Strauss},
  {Vanden Berk}, {Schneider}, {Yanny}, {Boucher}, {Burles}, {Frieman}, {Gunn},
  {Hall}, {Ivezi{\'c}}, {Kent}, {Loveday}, {Lupton}, {Rockosi}, {Schlegel},
  {Stoughton}, {SubbaRao}, \& {York}}]{2002AJ....123.2945R}
{Richards}, G.~T., {Fan}, X., {Newberg}, H.~J., {et~al.} 2002, \aj, 123, 2945,
  \dodoi{10.1086/340187}

\bibitem[{{Richards} {et~al.}(2011){Richards}, {Starr}, {Butler}, {Bloom},
  {Brewer}, {Crellin-Quick}, {Higgins}, {Kennedy}, \&
  {Rischard}}]{2011ApJ...733...10R}
{Richards}, J.~W., {Starr}, D.~L., {Butler}, N.~R., {et~al.} 2011, \apj, 733,
  10, \dodoi{10.1088/0004-637X/733/1/10}

\bibitem[{{Risaliti} \& {Lusso}(2019)}]{2019NatAs...3..272R}
{Risaliti}, G., \& {Lusso}, E. 2019, Nature Astronomy, 3, 272,
  \dodoi{10.1038/s41550-018-0657-z}

\bibitem[{{Salpeter}(1964)}]{1964ApJ...140..796S}
{Salpeter}, E.~E. 1964, \apj, 140, 796, \dodoi{10.1086/147973}

\bibitem[{{Salvato} {et~al.}(2022){Salvato}, {Wolf}, {Dwelly}, {Georgakakis},
  {Brusa}, {Merloni}, {Liu}, {Toba}, {Nandra}, {Lamer}, {Buchner}, {Schneider},
  {Freund}, {Rau}, {Schwope}, {Nishizawa}, {Klein}, {Arcodia}, {Comparat},
  {Musiimenta}, {Nagao}, {Brunner}, {Malyali}, {Finoguenov}, {Anderson},
  {Shen}, {Ibarra-Medel}, {Trump}, {Brandt}, {Urry}, {Rivera}, {Krumpe},
  {Urrutia}, {Miyaji}, {Ichikawa}, {Schneider}, {Fresco}, {Boller}, {Haase},
  {Brownstein}, {Lane}, {Bizyaev}, \& {Nitschelm}}]{2022A&A...661A...3S}
{Salvato}, M., {Wolf}, J., {Dwelly}, T., {et~al.} 2022, \aap, 661, A3,
  \dodoi{10.1051/0004-6361/202141631}

\bibitem[{{S{\'a}nchez} {et~al.}(2020){S{\'a}nchez}, {Walter}, {Awan},
  {Chiang}, {Daniel}, {Gawiser}, {Glanzman}, {Kirkby}, {Mandelbaum}, {Slosar},
  {Wood-Vasey}, {AlSayyad}, {Burke}, {Digel}, {Jarvis}, {Johnson}, {Kelly},
  {Krughoff}, {Lupton}, {Marshall}, {Peterson}, {Price}, {Sembroski}, {Van
  Klaveren}, {Wiesner}, {Xin}, \& {LSST Dark Energy Science
  Collaboration}}]{Sanchez2020}
{S{\'a}nchez}, J., {Walter}, C.~W., {Awan}, H., {et~al.} 2020, \mnras, 497,
  210, \dodoi{10.1093/mnras/staa1957}

\bibitem[{{Sandage} \& {Luyten}(1967)}]{1967ApJ...148..767S}
{Sandage}, A., \& {Luyten}, W.~J. 1967, \apj, 148, 767, \dodoi{10.1086/149200}

\bibitem[{{Schmidt} {et~al.}(2010){Schmidt}, {Marshall}, {Rix}, {Jester},
  {Hennawi}, \& {Dobler}}]{2010ApJ...714.1194S}
{Schmidt}, K.~B., {Marshall}, P.~J., {Rix}, H.-W., {et~al.} 2010, \apj, 714,
  1194, \dodoi{10.1088/0004-637X/714/2/1194}

\bibitem[{Shirley {et~al.}(2021)Shirley, Duncan, Campos Varillas, Hurley,
  Małek, Roehlly, Smith, Aussel, Bakx, Buat, Burgarella, Christopher,
  Duivenvoorden, Eales, Efstathiou, González Solares, Griffin, Jarvis, Faro,
  Marchetti, McCheyne, Papadopoulos, Penner, Pons, Prescott, Rigby, Rottgering,
  Saxena, Scudder, Vaccari, Wang, \& Oliver}]{10.1093/mnras/stab1526}
Shirley, R., Duncan, K., Campos Varillas, M.~C., {et~al.} 2021, Monthly
  Notices of the Royal Astronomical Society, 507, 129,
  \dodoi{10.1093/mnras/stab1526}

\bibitem[{{Shy} {et~al.}(2022){Shy}, {Tak}, {Feigelson}, {Timlin}, \&
  {Babu}}]{2022AJ....164....6S}
{Shy}, S., {Tak}, H., {Feigelson}, E.~D., {Timlin}, J.~D., \& {Babu}, G.~J.
  2022, \aj, 164, 6, \dodoi{10.3847/1538-3881/ac6e64}

\bibitem[{{Smith} \& {Geach}(2022)}]{2022arXiv221103796S}
{Smith}, M.~J., \& {Geach}, J.~E. 2022, arXiv e-prints, arXiv:2211.03796.
\newblock \doarXiv{2211.03796}

\bibitem[{Stone(1974)}]{stone1974}
Stone, M. 1974, Journal of the Royal Statistical Society: Series B
  (Methodological), 36, 111,
  \dodoi{https://doi.org/10.1111/j.2517-6161.1974.tb00994.x}

\bibitem[{{Suberlak} {et~al.}(2021){Suberlak}, {Ivezi{\'c}}, \&
  {MacLeod}}]{2021ApJ...907...96S}
{Suberlak}, K.~L., {Ivezi{\'c}}, {\v{Z}}., \& {MacLeod}, C. 2021, \apj, 907,
  96, \dodoi{10.3847/1538-4357/abc698}

\bibitem[{{Tan} {et~al.}(2018){Tan}, {Sun}, {Kong}, {Zhang}, {Yang}, \&
  {Liu}}]{2018arXiv180801974T}
{Tan}, C., {Sun}, F., {Kong}, T., {et~al.} 2018, arXiv e-prints,
  arXiv:1808.01974.
\newblock \doarXiv{1808.01974}

\bibitem[{{Temple} {et~al.}(2021){Temple}, {Hewett}, \&
  {Banerji}}]{2021MNRAS.508..737T}
{Temple}, M.~J., {Hewett}, P.~C., \& {Banerji}, M. 2021, \mnras, 508, 737,
  \dodoi{10.1093/mnras/stab2586}

\bibitem[{{Trevese} {et~al.}(1989){Trevese}, {Pittella}, {Kron}, {Koo}, \&
  {Bershady}}]{1989AJ.....98..108T}
{Trevese}, D., {Pittella}, G., {Kron}, R.~G., {Koo}, D.~C., \& {Bershady}, M.
  1989, \aj, 98, 108, \dodoi{10.1086/115129}

\bibitem[{{Uttley} {et~al.}(2002){Uttley}, {McHardy}, \&
  {Papadakis}}]{2002MNRAS.332..231U}
{Uttley}, P., {McHardy}, I.~M., \& {Papadakis}, I.~E. 2002, \mnras, 332, 231,
  \dodoi{10.1046/j.1365-8711.2002.05298.x}

\bibitem[{van~der Maaten \& Hinton(2008)}]{vanDerMaaten2008}
van~der Maaten, L., \& Hinton, G. 2008, Journal of Machine Learning Research,
  9, 2579.
\newblock \url{http://www.jmlr.org/papers/v9/vandermaaten08a.html}

\bibitem[{{van der Walt} {et~al.}(2011){van der Walt}, {Colbert}, \&
  {Varoquaux}}]{2011CSE....13b..22V}
{van der Walt}, S., {Colbert}, S.~C., \& {Varoquaux}, G. 2011, Computing in
  Science and Engineering, 13, 22, \dodoi{10.1109/MCSE.2011.37}

\bibitem[{Van~Rossum \& Drake~Jr(1995)}]{van1995python}
Van~Rossum, G., \& Drake~Jr, F.~L. 1995, Python reference manual (Centrum voor
  Wiskunde en Informatica Amsterdam)

\bibitem[{{Venemans} {et~al.}(2015){Venemans}, {Ba{\~n}ados}, {Decarli},
  {Farina}, {Walter}, {Chambers}, {Fan}, {Rix}, {Schlafly}, {McMahon},
  {Simcoe}, {Stern}, {Burgett}, {Draper}, {Flewelling}, {Hodapp}, {Kaiser},
  {Magnier}, {Metcalfe}, {Morgan}, {Price}, {Tonry}, {Waters}, {AlSayyad},
  {Banerji}, {Chen}, {Gonz{\'a}lez-Solares}, {Greiner}, {Mazzucchelli},
  {McGreer}, {Miller}, {Reed}, \& {Sullivan}}]{Venemans15}
{Venemans}, B.~P., {Ba{\~n}ados}, E., {Decarli}, R., {et~al.} 2015, \apjl, 801,
  L11, \dodoi{10.1088/2041-8205/801/1/L11}

\bibitem[{{Wang} {et~al.}(2016){Wang}, {Wu}, {Fan}, {Yang}, {Yi}, {Bian},
  {McGreer}, {Yang}, {Ai}, {Dong}, {Zuo}, {Jiang}, {Green}, {Wang}, {Cai},
  {Wang}, \& {Yue}}]{Wang16}
{Wang}, F., {Wu}, X.-B., {Fan}, X., {et~al.} 2016, \apj, 819, 24,
  \dodoi{10.3847/0004-637X/819/1/24}

\bibitem[{{Wang} {et~al.}(2019){Wang}, {Yang}, {Fan}, {Wu}, {Yue}, {Li},
  {Bian}, {Jiang}, {Ba{\~n}ados}, {Schindler}, {Findlay}, {Davies}, {Decarli},
  {Farina}, {Green}, {Hennawi}, {Huang}, {Mazzuccheli}, {McGreer}, {Venemans},
  {Walter}, {Dye}, {Lyke}, {Myers}, \& {Nunez}}]{Wang19}
{Wang}, F., {Yang}, J., {Fan}, X., {et~al.} 2019, \apj, 884, 30,
  \dodoi{10.3847/1538-4357/ab2be5}

\bibitem[{{Warren} {et~al.}(1991){Warren}, {Hewett}, {Irwin}, \&
  {Osmer}}]{1991ApJS...76....1W}
{Warren}, S.~J., {Hewett}, P.~C., {Irwin}, M.~J., \& {Osmer}, P.~S. 1991,
  \apjs, 76, 1, \dodoi{10.1086/191563}

\bibitem[{Waskom {et~al.}(2017)Waskom, Botvinnik, O'Kane, Hobson, Lukauskas,
  Gemperline, Augspurger, Halchenko, Cole, Warmenhoven, de~Ruiter, Pye, Hoyer,
  Vanderplas, Villalba, Kunter, Quintero, Bachant, Martin, Meyer, Miles, Ram,
  Yarkoni, Williams, Evans, Fitzgerald, Brian, Fonnesbeck, Lee, \&
  Qalieh}]{michael_waskom_2017_883859}
Waskom, M., Botvinnik, O., O'Kane, D., {et~al.} 2017, mwaskom/seaborn: v0.8.1
  (September 2017), v0.8.1,  Zenodo, \dodoi{10.5281/zenodo.883859}

\bibitem[{{Willott} {et~al.}(2010){Willott}, {Delorme}, {Reyl{\'e}}, {Albert},
  {Bergeron}, {Crampton}, {Delfosse}, {Forveille}, {Hutchings}, {McLure},
  {Omont}, \& {Schade}}]{Willott10}
{Willott}, C.~J., {Delorme}, P., {Reyl{\'e}}, C., {et~al.} 2010, \aj, 139, 906,
  \dodoi{10.1088/0004-6256/139/3/906}

\bibitem[{{Wright} {et~al.}(2010){Wright}, {Eisenhardt}, {Mainzer}, {Ressler},
  {Cutri}, {Jarrett}, {Kirkpatrick}, {Padgett}, {McMillan}, {Skrutskie},
  {Stanford}, {Cohen}, {Walker}, {Mather}, {Leisawitz}, {Gautier}, {McLean},
  {Benford}, {Lonsdale}, {Blain}, {Mendez}, {Irace}, {Duval}, {Liu}, {Royer},
  {Heinrichsen}, {Howard}, {Shannon}, {Kendall}, {Walsh}, {Larsen}, {Cardon},
  {Schick}, {Schwalm}, {Abid}, {Fabinsky}, {Naes}, \& {Tsai}}]{Wright10}
{Wright}, E.~L., {Eisenhardt}, P. R.~M., {Mainzer}, A.~K., {et~al.} 2010, \aj,
  140, 1868, \dodoi{10.1088/0004-6256/140/6/1868}

\bibitem[{{Xie} {et~al.}(2015){Xie}, {Girshick}, \&
  {Farhadi}}]{2015arXiv151106335X}
{Xie}, J., {Girshick}, R., \& {Farhadi}, A. 2015, arXiv e-prints,
  arXiv:1511.06335.
\newblock \doarXiv{1511.06335}

\bibitem[{{Yang} {et~al.}(2019{\natexlab{a}}){Yang}, {Wang}, {Fan}, {Wu},
  {Bian}, {Ba{\~n}ados}, {Yue}, {Schindler}, {Yang}, {Jiang}, {McGreer},
  {Green}, \& {Dye}}]{Yang19a}
{Yang}, J., {Wang}, F., {Fan}, X., {et~al.} 2019{\natexlab{a}}, \apj, 871, 199,
  \dodoi{10.3847/1538-4357/aaf858}

\bibitem[{{Yang} {et~al.}(2019{\natexlab{b}}){Yang}, {Wang}, {Fan}, {Yue},
  {Wu}, {Li}, {Bian}, {Jiang}, {Ba{\~n}ados}, \& {Beletsky}}]{Yang19b}
---. 2019{\natexlab{b}}, \aj, 157, 236, \dodoi{10.3847/1538-3881/ab1be1}

\bibitem[{{York} {et~al.}(2000{\natexlab{a}}){York}, {Adelman}, {Anderson},
  {Anderson}, {Annis}, {Bahcall}, {Bakken}, {Barkhouser}, {Bastian}, {Berman},
  {Boroski}, {Bracker}, {Briegel}, {Briggs}, {Brinkmann}, {Brunner}, {Burles},
  {Carey}, {Carr}, {Castander}, {Chen}, {Colestock}, {Connolly}, {Crocker},
  {Csabai}, {Czarapata}, {Davis}, {Doi}, {Dombeck}, {Eisenstein}, {Ellman},
  {Elms}, {Evans}, {Fan}, {Federwitz}, {Fiscelli}, {Friedman}, {Frieman},
  {Fukugita}, {Gillespie}, {Gunn}, {Gurbani}, {de Haas}, {Haldeman}, {Harris},
  {Hayes}, {Heckman}, {Hennessy}, {Hindsley}, {Holm}, {Holmgren}, {Huang},
  {Hull}, {Husby}, {Ichikawa}, {Ichikawa}, {Ivezi{\'c}}, {Kent}, {Kim},
  {Kinney}, {Klaene}, {Kleinman}, {Kleinman}, {Knapp}, {Korienek}, {Kron},
  {Kunszt}, {Lamb}, {Lee}, {Leger}, {Limmongkol}, {Lindenmeyer}, {Long},
  {Loomis}, {Loveday}, {Lucinio}, {Lupton}, {MacKinnon}, {Mannery}, {Mantsch},
  {Margon}, {McGehee}, {McKay}, {Meiksin}, {Merelli}, {Monet}, {Munn},
  {Narayanan}, {Nash}, {Neilsen}, {Neswold}, {Newberg}, {Nichol}, {Nicinski},
  {Nonino}, {Okada}, {Okamura}, {Ostriker}, {Owen}, {Pauls}, {Peoples},
  {Peterson}, {Petravick}, {Pier}, {Pope}, {Pordes}, {Prosapio},
  {Rechenmacher}, {Quinn}, {Richards}, {Richmond}, {Rivetta}, {Rockosi},
  {Ruthmansdorfer}, {Sandford}, {Schlegel}, {Schneider}, {Sekiguchi}, {Sergey},
  {Shimasaku}, {Siegmund}, {Smee}, {Smith}, {Snedden}, {Stone}, {Stoughton},
  {Strauss}, {Stubbs}, {SubbaRao}, {Szalay}, {Szapudi}, {Szokoly}, {Thakar},
  {Tremonti}, {Tucker}, {Uomoto}, {Vanden Berk}, {Vogeley}, {Waddell}, {Wang},
  {Watanabe}, {Weinberg}, {Yanny}, {Yasuda}, \& {SDSS
  Collaboration}}]{2000AJ....120.1579Y}
{York}, D.~G., {Adelman}, J., {Anderson}, John~E., J., {et~al.}
  2000{\natexlab{a}}, \aj, 120, 1579, \dodoi{10.1086/301513}

\bibitem[{{York} {et~al.}(2000{\natexlab{b}}){York}, {Adelman}, {Anderson},
  {Anderson}, {Annis}, {Bahcall}, {Bakken}, {Barkhouser}, {Bastian}, {Berman},
  {Boroski}, {Bracker}, {Briegel}, {Briggs}, {Brinkmann}, {Brunner}, {Burles},
  {Carey}, {Carr}, {Castander}, {Chen}, {Colestock}, {Connolly}, {Crocker},
  {Csabai}, {Czarapata}, {Davis}, {Doi}, {Dombeck}, {Eisenstein}, {Ellman},
  {Elms}, {Evans}, {Fan}, {Federwitz}, {Fiscelli}, {Friedman}, {Frieman},
  {Fukugita}, {Gillespie}, {Gunn}, {Gurbani}, {de Haas}, {Haldeman}, {Harris},
  {Hayes}, {Heckman}, {Hennessy}, {Hindsley}, {Holm}, {Holmgren}, {Huang},
  {Hull}, {Husby}, {Ichikawa}, {Ichikawa}, {Ivezi{\'c}}, {Kent}, {Kim},
  {Kinney}, {Klaene}, {Kleinman}, {Kleinman}, {Knapp}, {Korienek}, {Kron},
  {Kunszt}, {Lamb}, {Lee}, {Leger}, {Limmongkol}, {Lindenmeyer}, {Long},
  {Loomis}, {Loveday}, {Lucinio}, {Lupton}, {MacKinnon}, {Mannery}, {Mantsch},
  {Margon}, {McGehee}, {McKay}, {Meiksin}, {Merelli}, {Monet}, {Munn},
  {Narayanan}, {Nash}, {Neilsen}, {Neswold}, {Newberg}, {Nichol}, {Nicinski},
  {Nonino}, {Okada}, {Okamura}, {Ostriker}, {Owen}, {Pauls}, {Peoples},
  {Peterson}, {Petravick}, {Pier}, {Pope}, {Pordes}, {Prosapio},
  {Rechenmacher}, {Quinn}, {Richards}, {Richmond}, {Rivetta}, {Rockosi},
  {Ruthmansdorfer}, {Sandford}, {Schlegel}, {Schneider}, {Sekiguchi}, {Sergey},
  {Shimasaku}, {Siegmund}, {Smee}, {Smith}, {Snedden}, {Stone}, {Stoughton},
  {Strauss}, {Stubbs}, {SubbaRao}, {Szalay}, {Szapudi}, {Szokoly}, {Thakar},
  {Tremonti}, {Tucker}, {Uomoto}, {Vanden Berk}, {Vogeley}, {Waddell}, {Wang},
  {Watanabe}, {Weinberg}, {Yanny}, {Yasuda}, \& {SDSS Collaboration}}]{York00}
---. 2000{\natexlab{b}}, \aj, 120, 1579, \dodoi{10.1086/301513}

\bibitem[{Yu \& Richards(2021)}]{yu2021}
Yu, W., \& Richards, G.~T. 2021, {{LSSTC AGN Data Challenge}} 2021, Tech. rep.,
  {GitHub}, \dodoi{10.17918/AGN_DataChallenge}

\bibitem[{{Yu} \& {Richards}(2022)}]{Yu2022a}
{Yu}, W., \& {Richards}, G.~T. 2022, {EzTao: Easier CARMA Modeling}.
\newblock \doeprint{2201.001}

\bibitem[{{Yu} {et~al.}(2022){Yu}, {Richards}, {Vogeley}, {Moreno}, \&
  {Graham}}]{Yu2022b}
{Yu}, W., {Richards}, G.~T., {Vogeley}, M.~S., {Moreno}, J., \& {Graham}, M.~J.
  2022, \apj, 936, 132, \dodoi{10.3847/1538-4357/ac8351}

\bibitem[{{Yu} {et~al.}(2020){Yu}, {Richards}, {Yoachim}, \&
  {Peters}}]{2020RNAAS...4..252Y}
{Yu}, W., {Richards}, G.~T., {Yoachim}, P., \& {Peters}, C. 2020, Research
  Notes of the American Astronomical Society, 4, 252,
  \dodoi{10.3847/2515-5172/abd6e2}

\bibitem[{Yu {et~al.}(2022)Yu, Richards, Buat, Brandt, Banerji, Ni, Shirley,
  Temple, Wang, \& Yang}]{weixiang_yu_2022_6878414}
Yu, W., Richards, G., Buat, V., {et~al.} 2022, LSSTC AGN Data Challenge 2021,
  1.1,  Zenodo, \dodoi{10.5281/zenodo.6878414}

\bibitem[{{Zebrun} {et~al.}(2001){Zebrun}, {Soszynski}, {Wozniak}, {Udalski},
  {Kubiak}, {Szymanski}, {Pietrzynski}, {Szewczyk}, \&
  {Wyrzykowski}}]{2001AcA....51..317Z}
{Zebrun}, K., {Soszynski}, I., {Wozniak}, P.~R., {et~al.} 2001, \actaa, 51,
  317.
\newblock \doarXiv{astro-ph/0110623}

\bibitem[{{Zel'dovich} \& {Novikov}(1964)}]{1964SPhD....9..246Z}
{Zel'dovich}, Y.~B., \& {Novikov}, I.~D. 1964, Soviet Physics Doklady, 9, 246

\bibitem[{{Zu} {et~al.}(2013){Zu}, {Kochanek}, {Koz{\l}owski}, \&
  {Udalski}}]{2013ApJ...765..106Z}
{Zu}, Y., {Kochanek}, C.~S., {Koz{\l}owski}, S., \& {Udalski}, A. 2013, \apj,
  765, 106, \dodoi{10.1088/0004-637X/765/2/106}

\end{thebibliography}
\bibliographystyle{aasjournal}



\end{document}